\documentclass[12pt]{article}
\usepackage{epsfig}
\usepackage{amsfonts}
\usepackage{latexsym}
\usepackage{amsmath,amssymb}
\usepackage{verbatim}
\usepackage{setspace}
\usepackage[small]{caption}
\usepackage{graphicx}
\usepackage{color}
\definecolor{darkgreen}{rgb}{0,0.5,0}
\definecolor{darkblue}{rgb}{0,0,0.6}
\definecolor{purple}{rgb}{0.4,0.15,0.21}
\usepackage[colorlinks=true,citecolor=darkgreen,linkcolor=purple,urlcolor=darkblue]{hyperref}
\usepackage[textheight=9in, textwidth=6.5in, letterpaper]{geometry}

\numberwithin{equation}{section}

\def\tilde{\widetilde}
\newcommand{\bea}{\begin{eqnarray}}
\newcommand{\eea}{\end{eqnarray}}
\newcommand{\be}{\begin{equation}}
\newcommand{\ee}{\end{equation}}
\newcommand{\ba}{\begin{align}}
\newcommand{\ea}{\end{align}}

\newcommand{\sech}{\mbox{sech}}

\renewcommand{\epsilon}{\varepsilon}

  \makeatletter
  \let\over=\@@over \let\overwithdelims=\@@overwithdelims
  \let\atop=\@@atop \let\atopwithdelims=\@@atopwithdelims
  \let\above=\@@above \let\abovewithdelims=\@@abovewithdelims
\renewcommand\section{\@startsection {section}{1}{\z@}%
                                   {-3.5ex \@plus -1ex \@minus -.2ex}
                                   {2.3ex \@plus.2ex}%
                                   {\normalfont\large\bfseries}}

\renewcommand\subsection{\@startsection{subsection}{2}{\z@}%
                                     {-3.25ex\@plus -1ex \@minus -.2ex}%
                                     {1.5ex \@plus .2ex}%
                                     {\normalfont\bfseries}}

\def\Or[#1]{{\text{O}}\left({#1}\right)}
\def\dotl[#1,#2]{\left\langle #1, #2 \right\rangle}
\def\dotlb[#1,#2]{[ #1, #2 ]}
\def\dotp[#1,#2]{(#1) \cdot (#2)}
\def\aff[#1,#2]{\hat{#1}(#2)}
\def\n4sym{{\cal N}=4 SYM}
\def\>{\rangle}
\def\<{\langle}
\def\weight[#1,#2,#3]{\{(#1),#2,#3\}}
\def\ads[#1]{$\text{AdS}_{#1}$}

\begin{document}

\unitlength = 1mm
\ \\

\begin{center}

{ \LARGE   \textsc{Higher Spin de Sitter Holography from Functional Determinants} }

\vspace{0.8cm}

Dionysios Anninos$^1$, Frederik Denef$^{2}$, George Konstantinidis$^1$ and Edgar Shaghoulian$^1$

\vspace{1cm}

\vspace{0.5cm}

$^1$ {\it Stanford Institute for Theoretical Physics, Stanford University}\\
$^2$ {\it Institute for Theoretical Physics, University of Leuven}\\

\vspace{1.0cm}

\end{center}

\begin{abstract}

We discuss further aspects of the higher spin dS/CFT correspondence. Using a recent result of Dunne and Kirsten, it is shown how to numerically compute the partition function of the free $Sp(N)$ model for a large class of $SO(3)$ preserving deformations of the flat/round metric on $\mathbb{R}^3/S^3$ and the source of the spin-zero single-trace operator dual to the bulk scalar. We interpret this partition function as a Hartle-Hawking wavefunctional. It has a local maximum about the pure de Sitter vacuum. Restricting to $SO(3)$ preserving deformations, other local maxima (which exceed the one near the de Sitter vacuum) can peak at inhomogeneous and anisotropic values of the late time metric and scalar profile.  
Numerical experiments suggest the remarkable observation that, upon fixing a certain average of the bulk scalar profile at $\mathcal{I}^+$, the wavefunction becomes normalizable in all the other (infinite) directions of the deformation. We elucidate the meaning of double trace deformations in the context of dS/CFT as a change of basis and as a convolution. Finally, we discuss possible extensions of higher spin de Sitter holography by coupling the free theory to a Chern-Simons term.


\end{abstract}

\pagebreak 
\setcounter{page}{1}
\pagestyle{plain}

\setcounter{tocdepth}{1}

\tableofcontents

\section{Introduction}

A natural object to consider when studying an asymptotically (approximately) de Sitter spacetime, such as the universe during the inflationary era, is the wavefunction \cite{Hartle:1983ai} as a function of small fluctuations of the bulk fields. For a free massless scalar $\phi$ (such as the inflaton in slow roll inflation) in the Bunch-Davies vacuum state\footnote{It might be more appropriate to call it the Bunch-Davies-Hartle-Hawking-Euclidean-Schomblond-Spindel-Chernikov-Tagirov-Mottola-Allen-Sasaki-Tanaka-Yamamoto-Critchley-Dowker-Candelas-Raine-Boerner-Duerr vacuum state.} $| E \rangle$ \cite{Bunch:1978yq,Mottola:1984ar,Allen:1985ux,Chernikov:1968zm,Schomblond:1976xc,Sasaki:1994yt,Dowker:1975tf,Candelas:1975du,Boerner:1969ff} in a fixed four-dimensional de Sitter background:
\begin{equation}\label{massless}
ds^2 = \frac{\ell^2}{\eta^2} \left( - d\eta^2 + d\vec{x}^2 \right)~, \quad \eta \in (-\infty,0)~,
\end{equation}
one finds the late time Hartle-Hawking Gaussian wavefunctional  at $\eta = \eta_c \to 0$ \cite{Maldacena:2002vr}:
\begin{equation}
 \lim_{\eta_c \to 0} | \Psi_{HH} \left[ \varphi(\vec{x}),\eta_c \right] | \sim \exp \left[{- \frac{{\ell^2}}{2} \int \frac{d^{3} k}{(2\pi)^3} \; k^3 \;  \varphi_{\vec{k}} \; \varphi_{-\vec{k}} } \right]~,
\end{equation}
where $\varphi_{\vec{k}}$ are the Fourier components of the late time profile $\varphi(\vec{x})$. Such a Gaussian wavefunction gives rise to the scale-invariant fluctuations of the cosmic background radiation. Understanding the behavior of such a wavefunction for a large range of values for its arguments, which include the metric, and with the inclusion of quantum corrections is a basic problem in quantum cosmology.

The perturbative Bunch-Davies wavefunction (\ref{massless}) was noticed to be a simple analytic continuation \cite{Maldacena:2002vr} of the partition function of a free massless field in a fixed Euclidean anti-de Sitter space. These observations, coupled with the correspondence between anti-de Sitter space and conformal field theory, motivate the proposal that at late times (or large spatial volume) $\Psi_{HH}$ is computed by a statistical (and hence Euclidean) conformal field theory, in what has come to be known as the dS/CFT conjecture \cite{Maldacena:2002vr,Strominger:2001pn,Witten:2001kn}.\footnote{See \cite{Anninos:2012qw} for a discussion of several aspects of de Sitter space. Other proposals include \cite{Alishahiha:2004md,Dong:2010pm,Dong:2011uf,McFadden:2009fg,Banks:2011qf,Banks:2008ep,Freivogel:2006xu,Anninos:2011kh,Roberts:2012jw,Harlow:2012dd,Garriga:2008ks,Parikh:2004wh,Anninos:2009yc,Anninos:2010gh,Anninos:2011zn,Anninos:2011af,Hertog:2011ky}.} In its weakest form dS/CFT conjectures that the Taylor coefficients of the logarithm of the late time Hartle-Hawking wavefunctional expanded about the empty de Sitter vacuum at large $N \sim \left( \ell/\ell_{p} \right)^{\#}$ are the correlation functions of such a non-unitary CFT. Namely, at some late time cutoff $\eta = \eta_c \to 0$ we have:
\begin{equation}\label{pertdscft}
\log \Psi_{HH}[\varphi(\vec{x}),\eta_c] = \sum_{n=1}^{\infty} \frac{1}{n !} \left( \int  d^3 {x}_1 \ldots \int d^3 {x}_n  \; \varphi(\vec{x}_1) \ldots \varphi(\vec{x}_n) \; \langle \mathcal{O}(\vec{x}_1) \ldots  \mathcal{O}(\vec{x}_n) \rangle_{CFT} \right)~.
\end{equation}
The correlators $\langle \mathcal{O}(\vec{x}_1) \ldots  \mathcal{O}(\vec{x}_n) \rangle_{CFT}$, where the operator $\mathcal{O}$ has been rescaled by an appropriate $\eta_c$ dependent factor, compute late time bulk correlation functions with future boundary conditions \cite{Anninos:2011jp}. The bulk late time profiles $\varphi(\vec{x})$ are taken to be infinitesimal such that $\Psi_{HH}[\varphi(\vec{x}),\eta_c]$ is merely a generating function of late time correlators about $\varphi(\vec{x})=0$. In its strongest form, the claim is that the CFT is a non-perturbative definition of $\Psi_{HH}$ for finite deviations away from the pure de Sitter vacuum and at finite $N$. Particularly, $\Psi_{HH}$ is computed by the partition function of the putative CFT with sources turned on.  The single-trace operators are dual to the bulk de Sitter fields. Abstractly speaking, if we could write down a complete basis $\mathcal{B}$ (which may include information about topology, geometry and matter) for the Hilbert space of our theory, $\Psi_{HH}$ would be computing the overlap of the Hartle-Hawking state $| E \rangle$ with a particular state $|\beta\rangle \in \mathcal{B}$. One could also consider computing the partition function with sources for more general multi-trace operators turned on.  As we shall see this computes the overlap of $| E \rangle$ with states which are not sharp eigenstates of the field operator or its conjugate momentum. More dramatically, if there are single-trace operators which are irrelevant they may correspond to an exit from the de Sitter phase (see for instance \cite{Bzowski:2012ih}).

de Sitter space arises as a non-linear classical solution to four-dimensional higher spin gravity \cite{Vasiliev:1990en,Vasiliev:1999ba,Iazeolla:2007wt,Vasiliev:1986td,Vasiliev:1992av,Vasiliev:1995dn}. This theory has a tower of light particles with increasing spin, including a spinless bulk scalar with mass $m^2\ell^2 = +2$ and a spin-two graviton. The scalar potential has a minimum about the pure de Sitter solution and the kinetic terms of the higher spin particles carry the right signs and are canonical. Hence the de Sitter vacuum is perturbatively stable and free of tachyons and ghosts in this theory. Beyond perturbation theory, the late time Hartle-Hawking wavefunctional of asymptotically de Sitter space in higher spin gravity (at least when the topology at $\mathcal{I}^+$ is sufficiently simple) is conjectured to be computed by the partition function of the $Sp(N)$ model \cite{arXiv:1108.5735} (see also \cite{Das:2012dt,Ng:2012xp}), with $N \sim (\ell/\ell_p)^2~$. This model comes in two flavors, either a free theory of $N$ anti-commuting scalar fields transforming as vectors under the $Sp(N)$ symmetry or as a critical theory obtained from the $Sp(N)$ model by a double trace deformation \cite{LeClair:2007iy}. The partition function of the critical model (at least at large $N$) as a functional of the sources of the single-trace operators computes the wavefunction in the ordinary field basis. On the other hand, the free theory computes the wavefunction in a slightly modified basis. The quantum mechanical analogue of this basis is given by eigenstates of the Hermitian operator $\hat{\varsigma} = \left( \beta \hat{x} - \alpha \hat{p} \right)$ with $\alpha$, $\beta \in \mathbb{R}$. Given the wavefunction in the coordinate basis $\psi(x)$ we can compute in the $\hat{\varsigma}$-basis by performing the transform:
\begin{equation}\label{tranqminv}
\psi(\varsigma) = \frac{1}{\sqrt{2\pi\alpha}}\int dx \; e^{-\frac{i}{\alpha}  \; \left( {\frac{{\beta}{x^2}}{2} -  \varsigma { x } } \right)} \psi(x)~, \quad \psi(x) = \frac{1}{\sqrt{2\pi \alpha}}\; e^{\frac{ i \beta x^2 }{2 \alpha } }\int \; d\varsigma \; e^{-\frac{i \varsigma x}{ \alpha}} \psi(\varsigma).
\end{equation}
Normalizability in the $\hat{x}$-basis implies normalizability in the $\hat{\varsigma}$-basis and vice-versa. However, a node-less $\psi(x)$ will not necessarily give a node-less $\psi(\varsigma)$. 

The free $Sp(N)$ model  computes the late time Hartle-Hawking wavefunction of the bulk scalar in the eigenbasis of the late time operator \cite{Anninos:2012ft}:
\begin{equation}\label{convention2}
 {\sqrt{N} \; \eta_c^2} \; \hat{\sigma} = \hat{\phi} - \eta_c^3 \; \hat{\pi}_{\phi} ~, \quad \quad \hat{\pi}_\phi \equiv - \frac{i}{\sqrt{\det{g_{ij}}}} \frac{\delta}{\delta \phi}~,
\end{equation}
where $\hat{\phi}$ is the bulk scalar field operator and $\hat{\pi}_{\phi}$ is the field momentum density operator.
We have taken $|\eta_c| \ll 1 $ as a late time cutoff and the combination $g_{ij}/\eta_c^2$ represents the spatial metric at this time in Fefferman-Graham gauge. It is somewhat remarkable that there exists a basis for which the wavefunctional is computed by a free dual theory given that in the ordinary field basis it is computed by a strongly coupled theory. The partition function of the free $Sp(N)$ theory on an $\mathbb{R}^3$ topology is an explicit resummation of the correlation functions and there are no non-perturbative phenomena. The remaining coordinates of the wavefunctional, which are late time profiles of the higher spin fields including the bulk graviton, are computed in the ordinary field basis for both the free and critical models.

The wavefunctionals for the bulk scalar in the two different bases are related by a functional version of the transform (\ref{tranqminv}). As shown in \cite{Anninos:2012ft}, in the large $N$ limit performing the transform amounts to finding the saddles of a functional equation. One of these saddles (not necessarily the dominant one) gives the field basis wavefunction whose perturbative expansion agrees with that computed perturbatively in the bulk about the pure de Sitter solution.  In most of what follows, we will perform computations in the $\hat{\sigma}$-basis, which amounts to the calculation of a functional determinant. This is a considerably hard object to compute and we will limit ourselves to situations where we have conditioned all bulk fields except the metric $g_{ij}$ and the scalar $\phi$ to have vanishing late time profiles. We should emphasize that this amounts to a very sharp conditioning and ultimately the situation could be altered by allowing higher spin deformations.\footnote{It might be worth mentioning that non-linear bulk solutions exist in higher spin gravity for which {\it only} the bulk metric and scalar are turned on \cite{Sezgin:2005hf}.}

The main body of this paper is dedicated to the study of the partition function of the $Sp(N)$ model for a large class of $SO(3)$ preserving deformations of the bulk scalar and graviton. That is to say we study the late time wavefunction for bulk graviton and scalar configurations with late time profile on an $\mathbb{R}^3$ topology:
\begin{equation}
ds^2 = dr^2 + f(r)^2 \; r^2 \; d\Omega^2~, \quad \phi = \phi(r)~.
\end{equation}
where $d\Omega^2_2$ is the round metric on an $S^2$ whose $SO(3)$ symmetry is the one preserved. We also study the analogous problem on an $S^3$ topology, which amounts to a simple conformal transformation of the $\mathbb{R}^3$ case, where the late time profiles take the form:
\begin{equation}
ds^2 = d\psi^2 + f(\psi)^2 \; \sin^2\psi \; d\Omega^2~, \quad \phi = \phi(\psi)~.
\end{equation}
This allows us to examine the behavior of the wavefunction of higher spin de Sitter space for inhomogeneous and anisotropic deformations which extend the rather uniform and homogenous deformations studied in \cite{Anninos:2012ft}. In \cite{Anninos:2012ft} it was found that the wavefunction in the $\hat{\sigma}$-basis diverges as a function of uniform mode of the bulk scalar on the round metric on $S^3$. One of the questions we would like to understand is whether such non-normalizabilities persist for less `global' late time configurations such as the above $SO(3)$ deformations. The above metric deformations are conformally equivalent to the flat/round metric on $\mathbb{R}^3/S^3$. We find particularly striking numerical evidence, discussed in section \ref{secconj}, that upon fixing the uniform mode of the bulk scalar on $S^3$ (in the conformal frame where it is endowed with the standard metric) all other directions of a general $SO(3)$ late time deformation are normalizable.  We have also analyzed a different geometric deformation, this time homogeneous but anisotropic, which does not keep the metric in the same conformal class. This is a squashing deformation $\alpha$ of the round metric on $S^3$ expressed as an $S^1$ fiber over an $S^2$:
\begin{equation}\label{squashedsphere}
ds^2 = \frac{1}{4} \left( d\theta^2 + \cos^2\theta d\phi^2 + \frac{1}{1+\alpha}\left( d\psi + \sin\theta d\phi \right)^2 \right)~,
\end{equation}
along with a uniform late time profile for the bulk scalar. When $\alpha = 0 $ the metric (\ref{squashedsphere}) reduces to the standard metric on $S^3$. Once again, so long as the scalar profile is kept fixed we find that the wavefunction is bounded in the $\alpha$ direction.

We begin by briefly reviewing the $Sp(N)$ model in section \ref{sectwo}. In section \ref{secfour}, using technology developed in \cite{Dunne:2006ct}, we compute the wavefunction (in the $\hat{\sigma}$-basis) for some Gaussian radial deformations of the bulk scalar. This amounts to computing a functional determinant of a scalar field with a radially dependent mass term. In section \ref{balloonsec} we compute the wavefunction for a radial deformation of the geometry, in the presence of a radial mass deformation, that takes it from the flat metric on $\mathbb{R}^3$ to a general form $ds^2 = dr^2 + r^2 f(r)^2 d\Omega^2_2\;$. In section \ref{secconj} we compute the wavefunction for several harmonics on the three-sphere and linear combinations thereof and note that the wavefunction seems to diverge only when the zero harmonic becomes large and negative. We then discuss the behavior of the wavefunction on a squashed sphere with a uniform profile of the late time bulk scalar in section \ref{secthree}. In section \ref{secthreehalf} we make some general remarks about double trace deformations. We end by speculating on possible extensions of higher spin holography in section \ref{secsix}. Most of our calculations can carry over to the $O(N)$ model and its AdS$_4$ dual in higher spin gravity \cite{Klebanov:2002ja,Sezgin:2002rt,hep-th/0103247,arXiv:0912.3462}.

\section{Wavefunctionals and the free $Sp(N)$ model}\label{sectwo}

We wish to study the Hartle-Hawking wavefunctional of an asymptotically de Sitter higher spin gravity for deformations of the bulk scalar and graviton away from the pure de Sitter solution. We will restrict the topology of space to be $\mathbb{R}^3$ or $S^3$ and allow only deformations that decay sufficiently fast at infinity. In this section, we remind the reader of the $Sp(N)$ theory and discuss how to compute its functional determinant for certain $SO(3)$ invariant radial deformations. One motivation to do so is to understand the behavior of the wavefunction of higher spin de Sitter space for mass deformations that are more `localized' than those studied in \cite{Anninos:2012ft} which were uniform over the entire $S^3$. It is also worth noting that computing analogous pieces of the Hartle-Hawking wavefunction for a simple toy model of Einstein gravity coupled to a scalar field with a simple potential would require significant numerical work even in the classical limit, let alone at finite $N$. One would have to find a complex solution of the Euclidean equations of motion that caps off smoothly in the interior and has the prescribed boundary values at large volume, and compute its on-shell action.

\subsection{Wavefunctional}\label{wf}

Recall that the action of the free $Sp(N)$ model on a curved background $g_{ij}$ with a source, $m(x^i)$, turned on for the $J^{(0)} = \Omega_{AB} \chi^A \chi^B \equiv \chi \cdot \chi$ operator (dual to the bulk scalar) is given by:
\begin{equation}
S = \frac{1}{2} \int d^3 x \sqrt{g} \; \Omega_{AB} \left( \partial_i \chi^A \partial_j \chi^B g^{ij} + \frac{R[g]}{8} \chi^A \chi^B  + m(x^i) \chi^A \chi^B \right)~, \quad \{ A, B\} = 1,2,\ldots,N~,
\end{equation}
where $N$ should be even. The fields $\chi^A$ are anti-commuting scalars that transform as $Sp(N)$ vectors and $\Omega_{AB}$ is the symplectic form. Notice that due to the presence of the conformal coupling, the action is invariant under a local Weyl transformation of the metric: $g_{ij} \to e^{2W(x^i)} g_{ij}$, so long as we also rescale the source as $m(x^i) \to e^{-2W(x^i)} m(x^i)$. From the bulk point of view this amounts to performing a coordinate transformation $\eta =  e^{-W(x^i)} \eta$, as can be seen by studying the Starobinski-Fefferman-Graham expansion \cite{Starobinsky:1982mr,fg,Anninos:2010zf} near $\eta = 0$:
\begin{equation}\label{fg}
\frac{ds^2}{\ell^2} = -\frac{d\eta^2}{\eta^2} + \frac{1}{\eta^2} \; g_{ij}(x^i) dx^i dx^j +\ldots~, \quad \phi = \eta \; \nu({x^i}) +  \eta^2 \; \mu(x^i)  + \ldots \; .
\end{equation}
Notice that $\mu(x^i) \equiv \sqrt{N} m(x^i)$ is {\it not} the coefficient of the slowest falling power of $\eta$ for bulk scalar. At the linearized level, the $\nu(x^i)$ profile is dual to the vev of the $J^{(0)}$ operator in the presence of an infinitesimal $m(x^i)$ source.

Computing the late time Hartle-Hawking wavefunctional in the $\hat{\sigma}$-basis with $\sigma = m(x^i)$ amounts to computing the partition function of the $Sp(N)$ theory with finite sources turned on. Given that it is a Gaussian theory, we can integrate out the anti-commuting $\chi^A$ fields and find:\
\begin{equation}
\lim_{\eta_c\to 0}  \Psi_{HH} \left[g_{ij}, m(x^i), \eta_c \right] = Z_{free} [g_{ij}, m(x^i)] = \left( \det \left[ - \nabla_g^2 + \frac{R[g]}{8} + m(x^i) \right] \right)^{N/2}~,
\end{equation}
where:
\begin{equation}
Z_{free} [g_{ij}, m(x^i)] \equiv \prod_{A=1}^N \int \mathcal{D} \chi^A \; e^{-S[\chi^A,m(x^i)]}~.
\end{equation}
In the case of a metric $g_{ij} = e^{2W(x^i)} \delta_{ij}$ that is conformally equivalent to the flat metric on $\mathbb{R}^3$, it is convenient to compute the functional determinant in the conformal frame where $g_{ij}$ is the flat metric. This amounts to rescaling the source to:
\begin{equation}
\hat{m}(x^i) = e^{2W(x^i)} m(x^i)~.
\end{equation}
We will use this fact in section \ref{balloonsec}.

\subsection{Functional determinant for radial deformations}\label{radialwf}

We have seen that for conformally flat metrics our problem reduces to computing a functional determinant:
\begin{equation}
\det \left[ - \nabla_{\mathbb{R}^3}^2 + \hat{m}(x^i) \right]~,
\end{equation}
where $\nabla_{\mathbb{R}^3}^2$ is the Laplacian of the round metric on $\mathbb{R}^3$, namely $ds^2 = dr^2 + r^2 d\Omega^2_2$. The above object is badly divergent unless we regulate it somehow. We will regulate it using a heat kernel or zeta function approach, both of which give the same answer. In fact, this precise problem has been studied by Dunne and Kirsten in \cite{Dunne:2006ct} for functions $\hat{m}(x^i)$ which only depend on the radial coordinate, i.e. $\hat{m}(x^i) = \hat{m}(r)$, and which vanish sufficiently fast at infinity. It was shown that the zeta function regulated determinant is given by the following sum:
\begin{equation}\label{dunne}
\log \left( \frac{\det \left[ - \nabla^2 + \mu^2 + \hat{m}(r) \right] }{\det \left[ - \nabla^2 + \mu^2  \right] } \right) = \sum_{l=0}^\infty (2l+1) \left( \log T^{(l)}(\infty) - \frac{ \int_0^\infty dr \; r \; \hat{m}(r)  }{2 l+1} \right)~.
\end{equation}
In the above, the factor $(2l +1)$ originates from the degeneracy of eigenfunctions on a two-sphere and $T^{(l)}(r)$ solves the equation:
\begin{equation}
- \frac{d^2}{dr^2} T^{(l)} (r) - 2 \left[ \frac{(1+l)}{r} + \mu \frac{ I_{3/2+l}(\mu r) }{ I_{1/2+l}(\mu r) } \right] \frac{d}{dr} T^{(l)} (r) + \hat{m}(r) T^{(l)}(r) = 0~,
\end{equation}
with boundary conditions $T^{(l)}(0) = 1$ and $d T^{(l)}(0) / dr = 0$. The parameter $\mu^2 \in \mathbb{R}$ is a constant mass parameter that we will set to zero. The derivation of the above formula employs the Gelfand-Yaglom theorem \cite{Gelfand:1959nq}, which expresses the regulated functional determinant of a one-dimensional Schr\"{o}dinger operator in terms of a single boundary value problem. The problem of computing the logarithm of a ratio of functional determinants for purely radial operators reduces to an infinite number of Gelfand-Yaglom problems, one for each $l$, whose solutions need to be summed (this is the first piece on the right hand side of (\ref{dunne})) and regularized (this is the second piece on the right hand side of (\ref{dunne})). The applicability of the formula requires $\hat{m}(r)$ to vanish faster than $r^{-2}$ at infinity, and these are the only types of deformations for which we will compute the wavefunction in the latter sections. When implementing the above formula we must sum up to a certain cutoff $l = l_{max}$ which we take to be $l_{max} = 45$. A discussion of how the error decreases with $l_{max}$ is given in appendix \ref{r3s3}.

\section{Simple examples of radial deformations}\label{secfour}

The purpose of this section is to exploit the general formula (\ref{dunne}) for a simple set of radial functions. By studying $\Psi_{HH}$ as a functional of $\hat{m}(r)$ we can identify some qualitative features already observed in \cite{Anninos:2012ft}, such as regions where the wavefunction oscillates and grows exponentially, as well as some new ones. Furthermore, we can study its dependence on more detailed features of the localized deformation. The zeroes of the wavefunction in the $\hat{\sigma}$-basis occur only when the effective potential $V_{eff}(r) = l(l+1)/r^2 +\hat{m}(r)$ of the differential operator $-\nabla^2_{\mathbb{R}^3} + \hat{m}(r)$ is negative for some range of $r$. If the effective potential were positive for all $r$ it could not have vanishing eigenvalues, and hence the wavefunction could not vanish. Thus, we expect all oscillations of the wavefunction in the $\hat{\sigma}$-basis to occur in directions where $\hat{m}(r)$ is negative for some range of $r$. 

Assessing the magnitude of the wavefunction as a functional of $\hat{m}(r)$ is a more complicated task. We observe that the wavefunctional acquires increasingly high local maxima between its oscillations only in regions where the quantity $I_{\hat m} \equiv \int_0^\infty dr \;r \; \hat{m}(r)$ appearing in (\ref{dunne}) becomes large and negative. 

It is important to note that because we are working with the flat metric on $\mathbb{R}^3$, which has no scale, our functional determinants will have an associated scaling symmetry given by $r \to r/\lambda$ and $\hat{m}(r) \to \hat{m}(r/\lambda)/\lambda^2$. 
We should thus fix the scaling when studying the functional determinant/wavefunction.
\begin{figure}
\begin{center}
{ \includegraphics[height=1.7in]{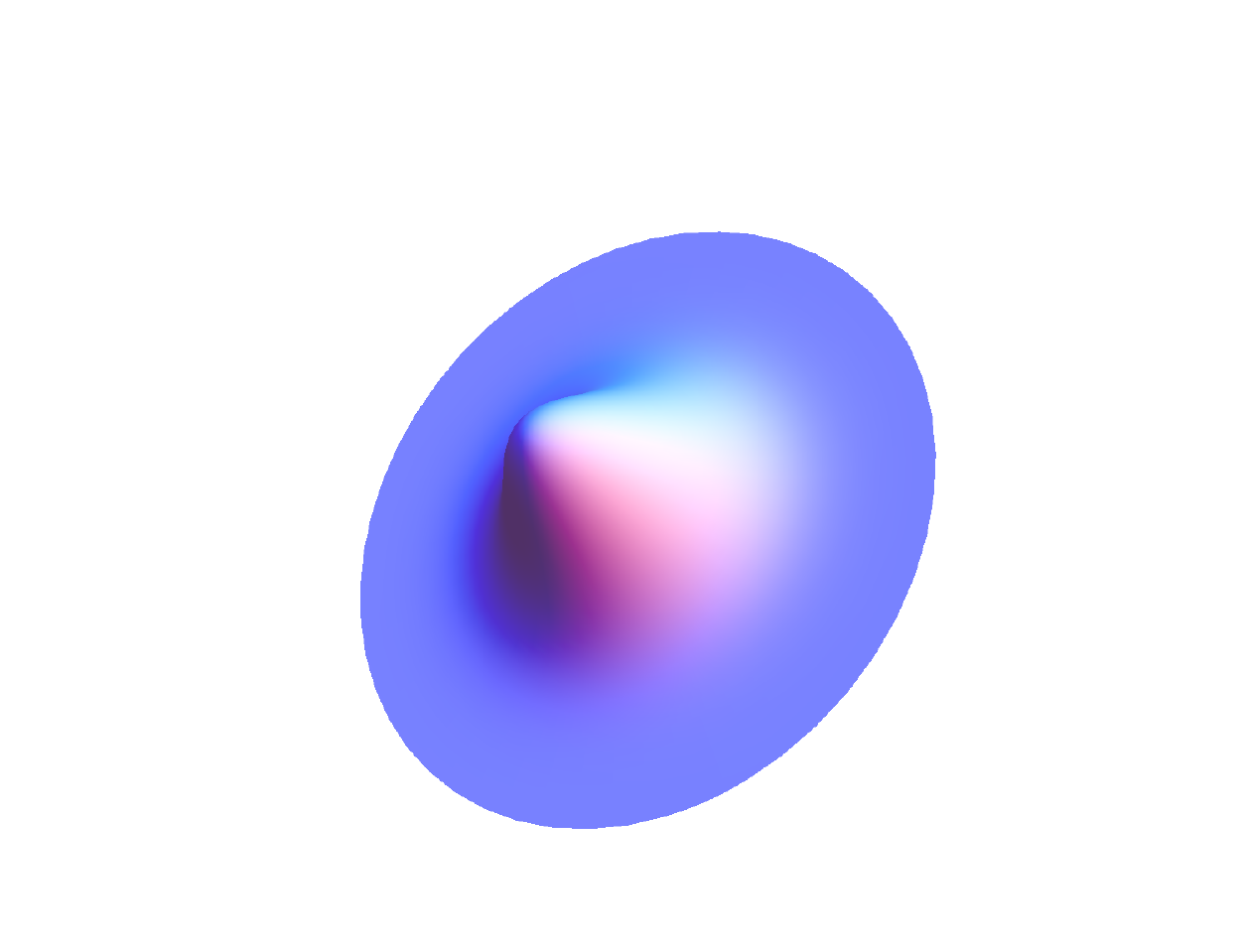}  \includegraphics[height=1.7in]{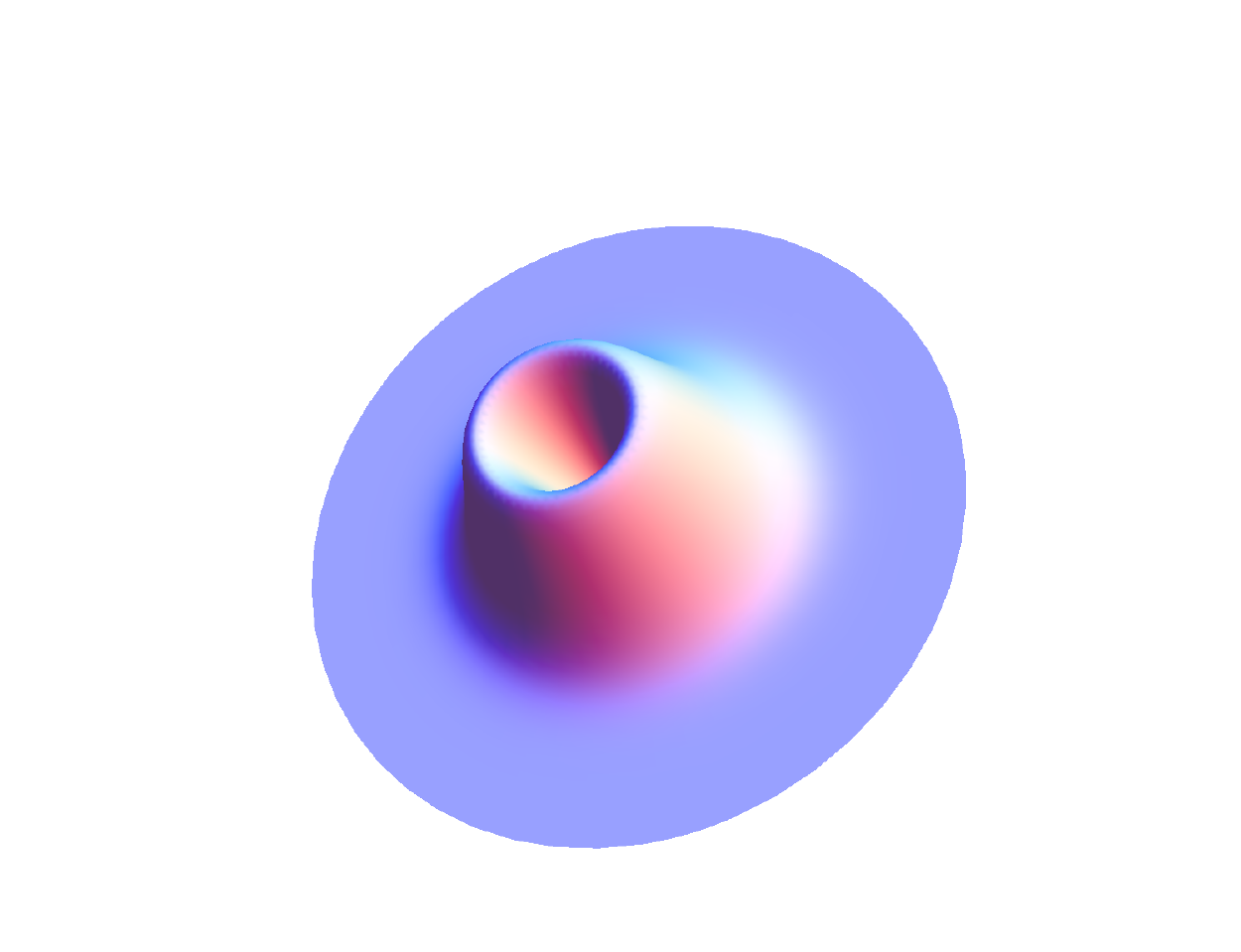}  \includegraphics[height=1.54in]{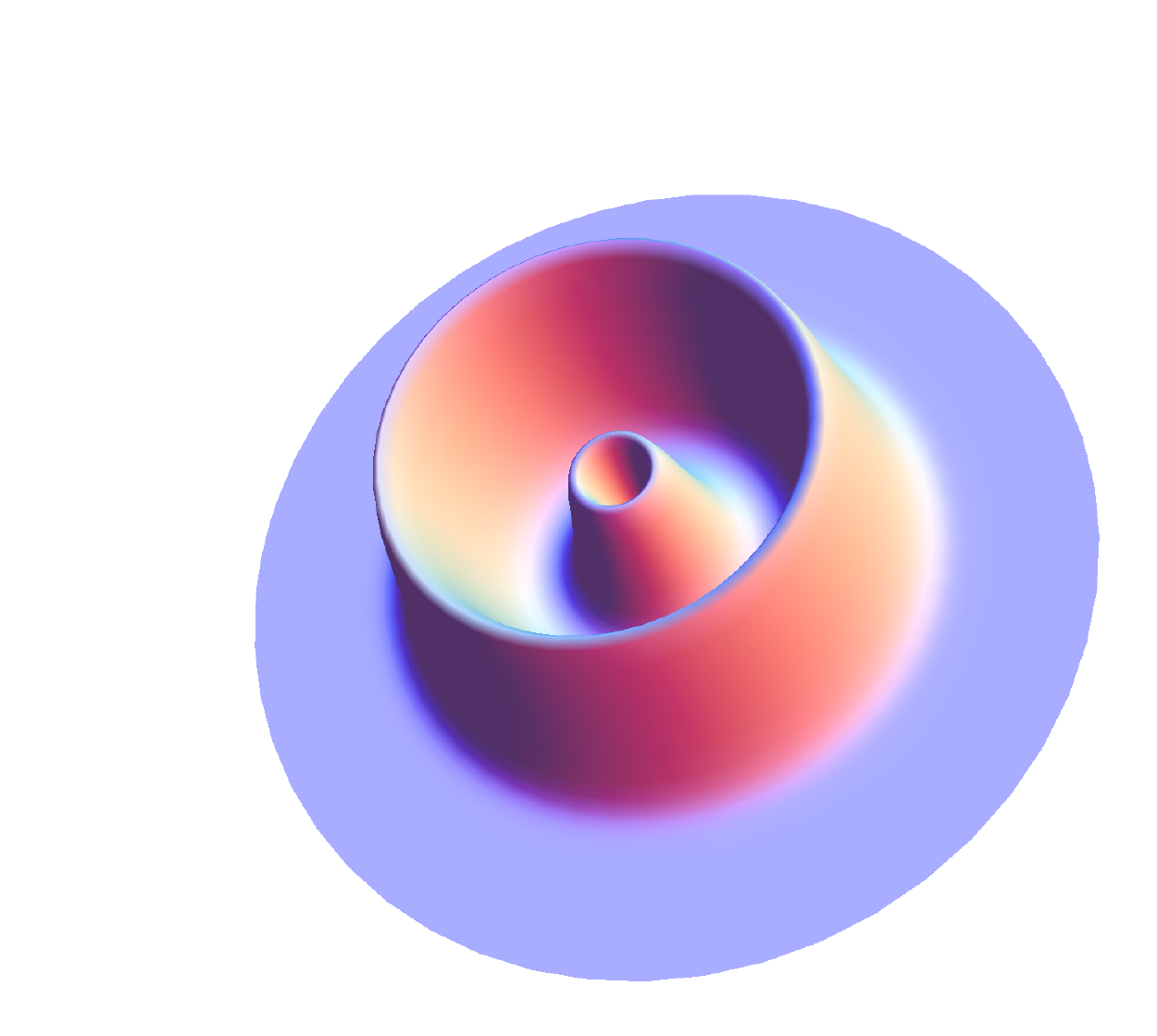}   }
\caption{Examples of the radial deformations (\ref{gauss}) on the left, (\ref{r2gauss}) in the middle and (\ref{dbgauss}) on the right. We have suppressed the polar coordinate  $\theta$ of the $S^2$ but kept the azimuthal direction.} \label{gaussians}
\end{center}
\end{figure} 


\subsection{Single Gaussian}

\begin{figure}
\begin{center}
{ {\includegraphics[height=2.2in]{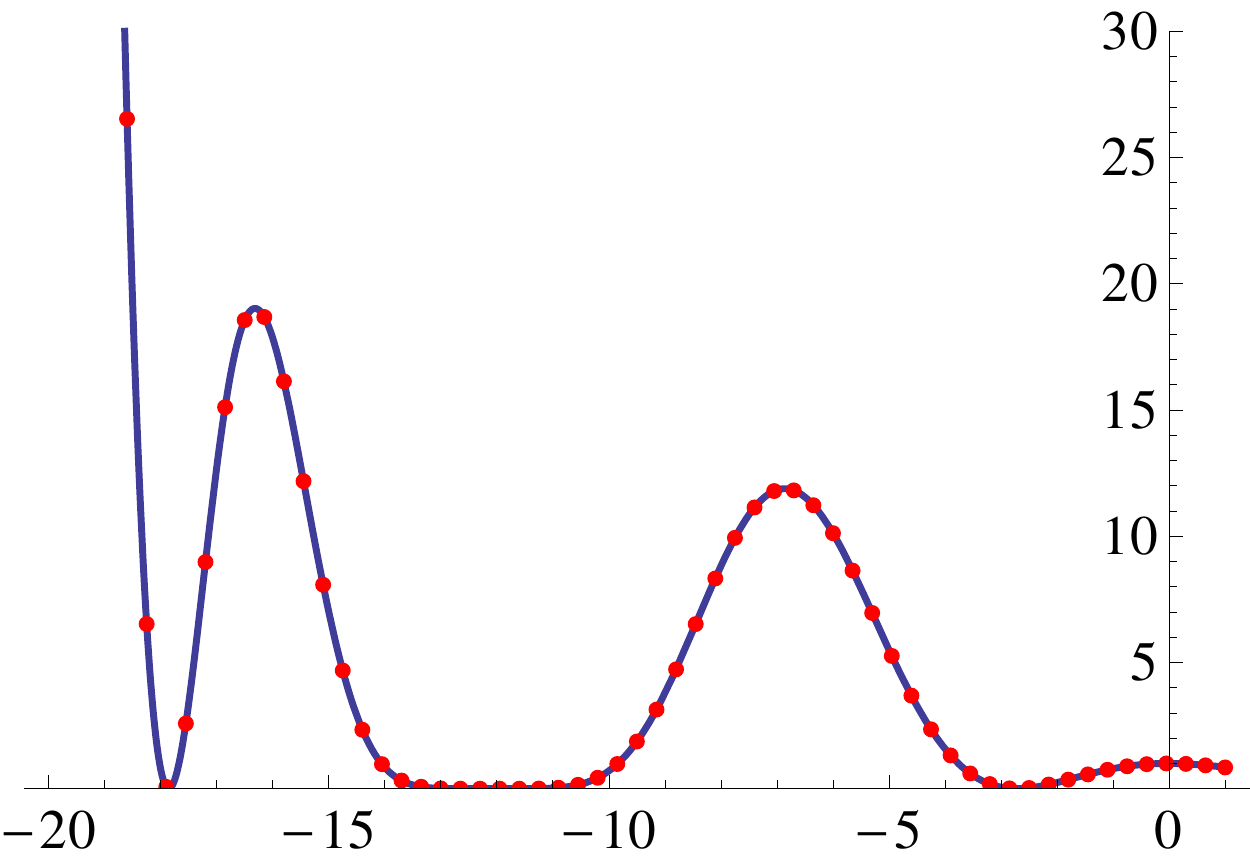}}}
\caption{Plot of $|\Psi_{HH}(\lambda,A)|^2$ for $N=2$ for the Gaussian profile (\ref{gauss}) with $\lambda=1$ using $l_{max} = 45$. The solid blue line is an interpolation of the numerically determined points (shown in red). The wavefunction grows and oscillates in the negative $A$ direction.}\label{singlegauss}
\end{center}
\end{figure}

We first consider $\hat{m}(r)$ to be given by a general single Gaussian profile:
\begin{equation}\label{gauss}
\hat{m}(r) = A \; \frac{e^{-r^2/\lambda^2}}{\lambda^2}~, \quad \quad \int_0^\infty dr \; r \; \hat{m}(r) = \frac{A}{2}~.
\end{equation}
See the left panel of figure \ref{gaussians} for an illustration of this deformation. Using equation (\ref{dunne}) we can explore $|\Psi_{HH}(\lambda,A)|$, where from now on whenever we write $\Psi_{HH}$ it is implied as a late time wavefunctional. Using the scaling relation $r \to r/\tilde{\lambda}$ and $\hat{m}(r) \to \hat{m}(r/\tilde{\lambda})/\tilde{\lambda}^2$ we can set $\lambda = 1$. In figure \ref{singlegauss} we show a plot of the functional determinant. We immediately notice the same qualitative feature that was present for the constant mass deformation on a round $S^3$ (displayed later on in figure \ref{sigmafig}). 
Namely, it oscillates and grows exponentially in the negative $A$ direction. This is somewhat expected since our deformation is qualitatively similar to the mass deformation on the flat metric on $\mathbb{R}^3$ one gets by the conformal transformation of a constant mass on $S^3$ (see appendix \ref{r3s3}). In particular, all oscillations occur for $A<0$ and the magnitude of the local maxima increases for increasing $|A|$ for fixed $\lambda$. 
\begin{figure}
\begin{center}
{\includegraphics[height=2.8in]{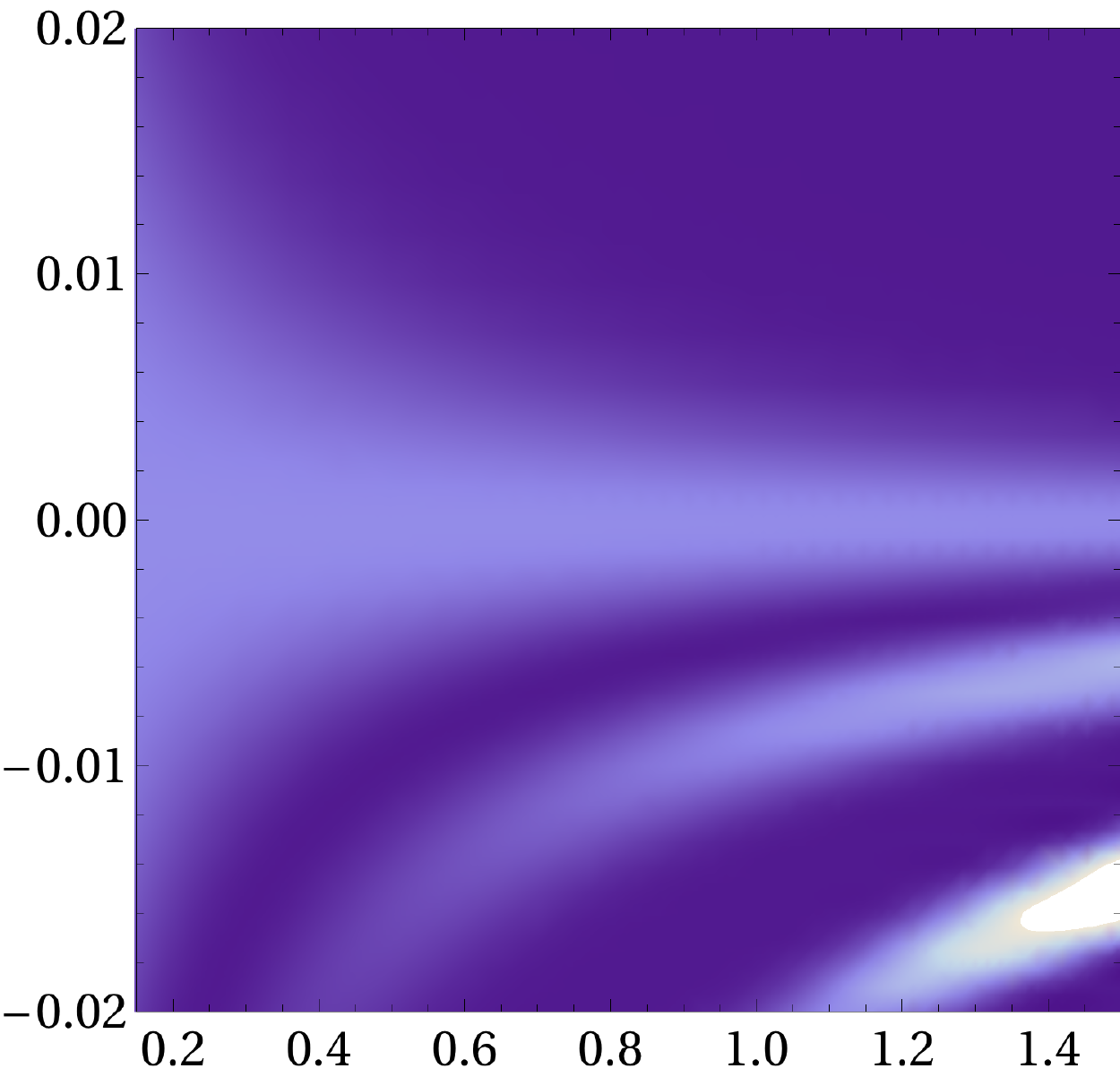} \quad
\includegraphics[width=2.8in]{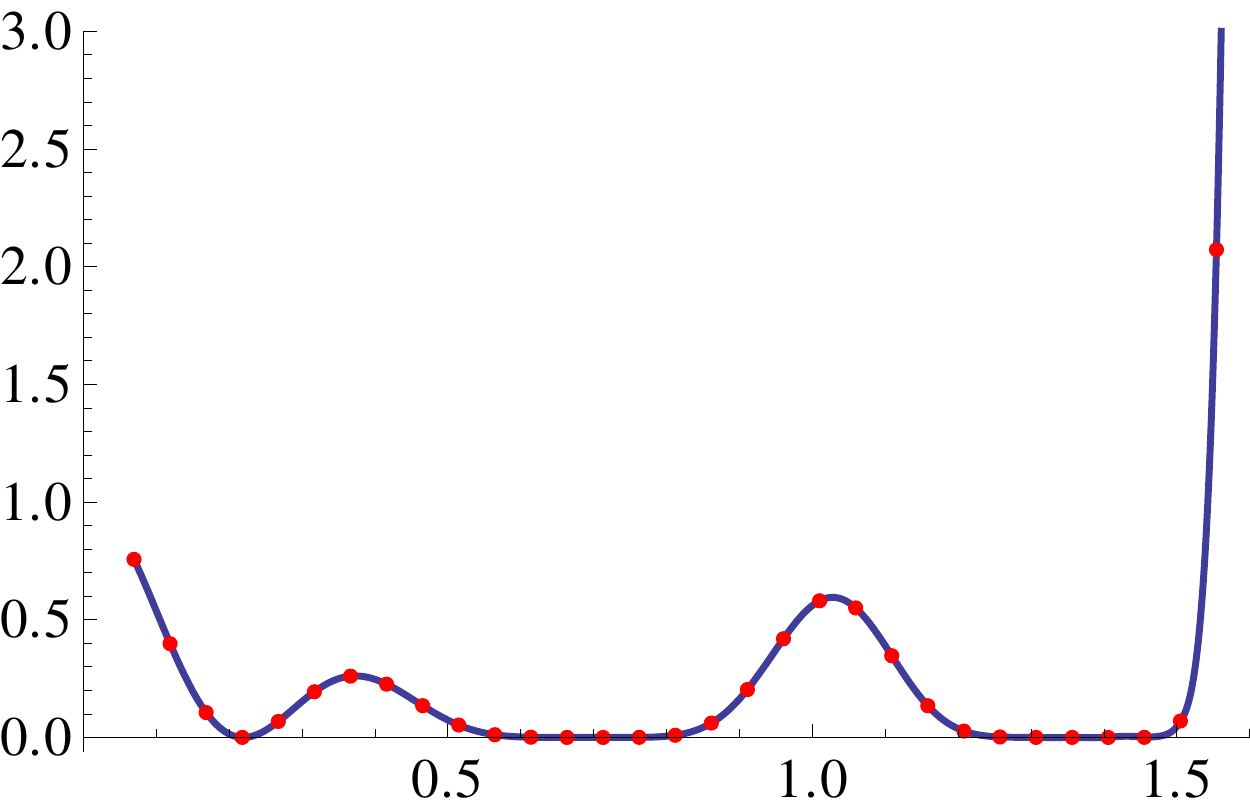}}\caption{Left: Density plot of $|\Psi_{HH}(\lambda,a,A)|^2$ for $N=2$ for the profile (\ref{r2gauss}) as a function of A (vertical) and $\lambda$ (horizontal) for $a=5$ using $l_{max} = 45$. Again, the wavefunction grows and oscillates in the negative $A$ and positive $\lambda$ directions. Right: Plot of $|\Psi_{HH}(\lambda,a,A)|^2$ for the profile (\ref{r2gauss}) as a function of $\lambda$ for $A=-0.022$ and $a=5$.}\label{singlegaussii}
\end{center}
\end{figure} 

\subsection{Gaussian Ring}

We can also study the functional determinant of a profile of the type:
\begin{multline}\label{r2gauss}
\hat{m}(r) = A \; e^{-(r-a)^2/\lambda^2} \; r^2~, \\  \int_0^\infty dr \; r \; \hat{m}(r) = \frac{A \; \lambda}{4}  \left[2 e^{-{a^2}/{\lambda ^2}} \lambda  \left(a^2+\lambda^2\right)+a \sqrt{\pi } \left(2 a^2+3 \lambda ^2\right) \left(1+\text{Erf}\left[\frac{a}{\lambda }\right]\right)\right],
\end{multline}
which describes a Gaussian-like ring peaked around $r \sim 1/2 \left(a+\sqrt{a^2+4 \lambda ^2}\right)$. $\text{Erf}[x]$ denotes the error function. See the middle panel of figure \ref{gaussians} for an illustration of this deformation. The factor of $r^2$ is included to ensure that the profile is continuously differentiable near the origin. Again, using the scaling relation $r \to r/\tilde{\lambda}$ and $\hat{m}(r) \to \hat{m}(r/\tilde{\lambda})/\tilde{\lambda}^2$ we either fix the value of $\lambda$, $|a|$ or $|A|$. We show an example in figure \ref{singlegaussii} where we have fixed the value of $a$.


\subsection{Double Gaussian}

\begin{figure}
\begin{center}
{\includegraphics[height=2.9in]{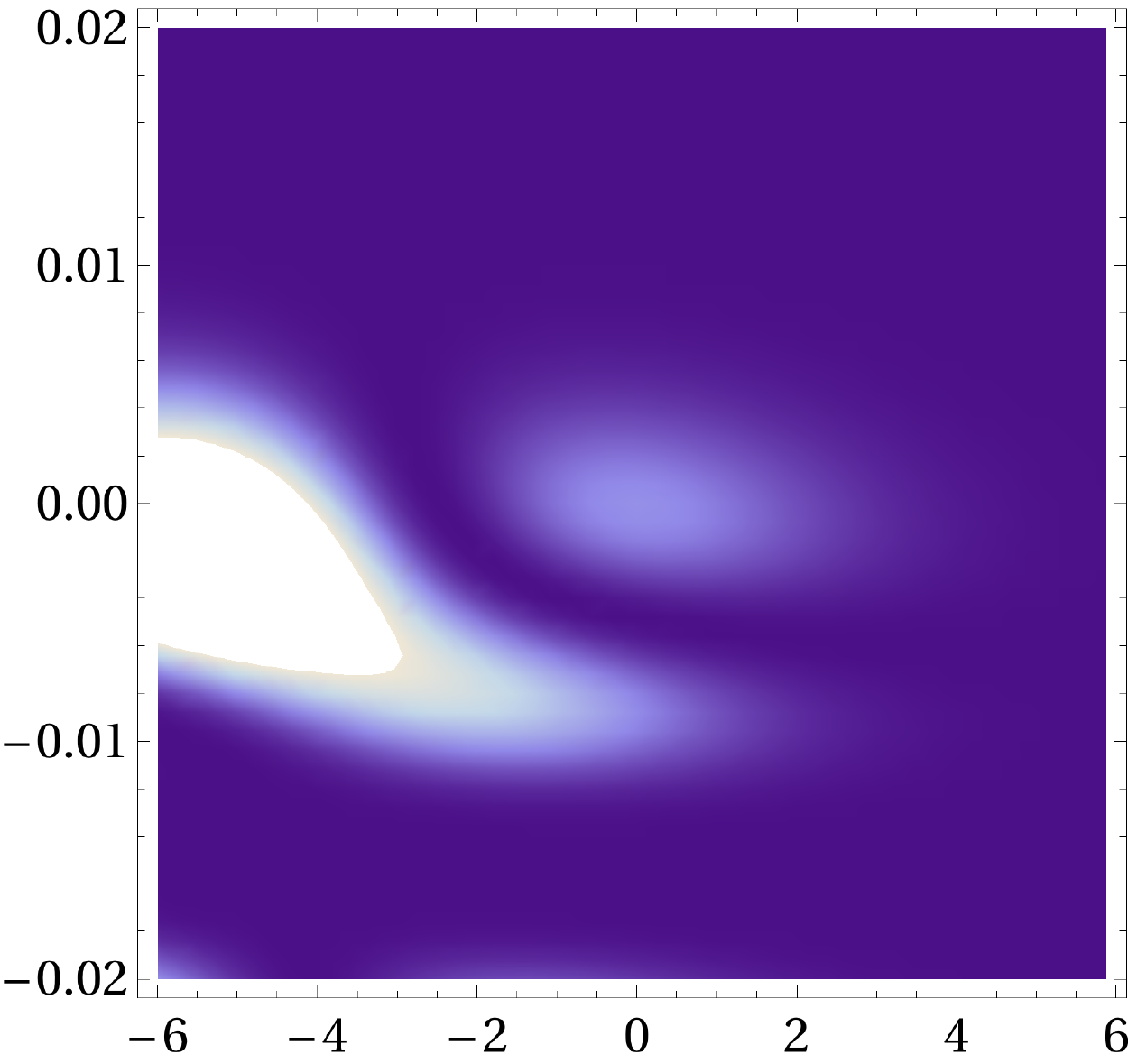} \quad \includegraphics[height=2.8in]{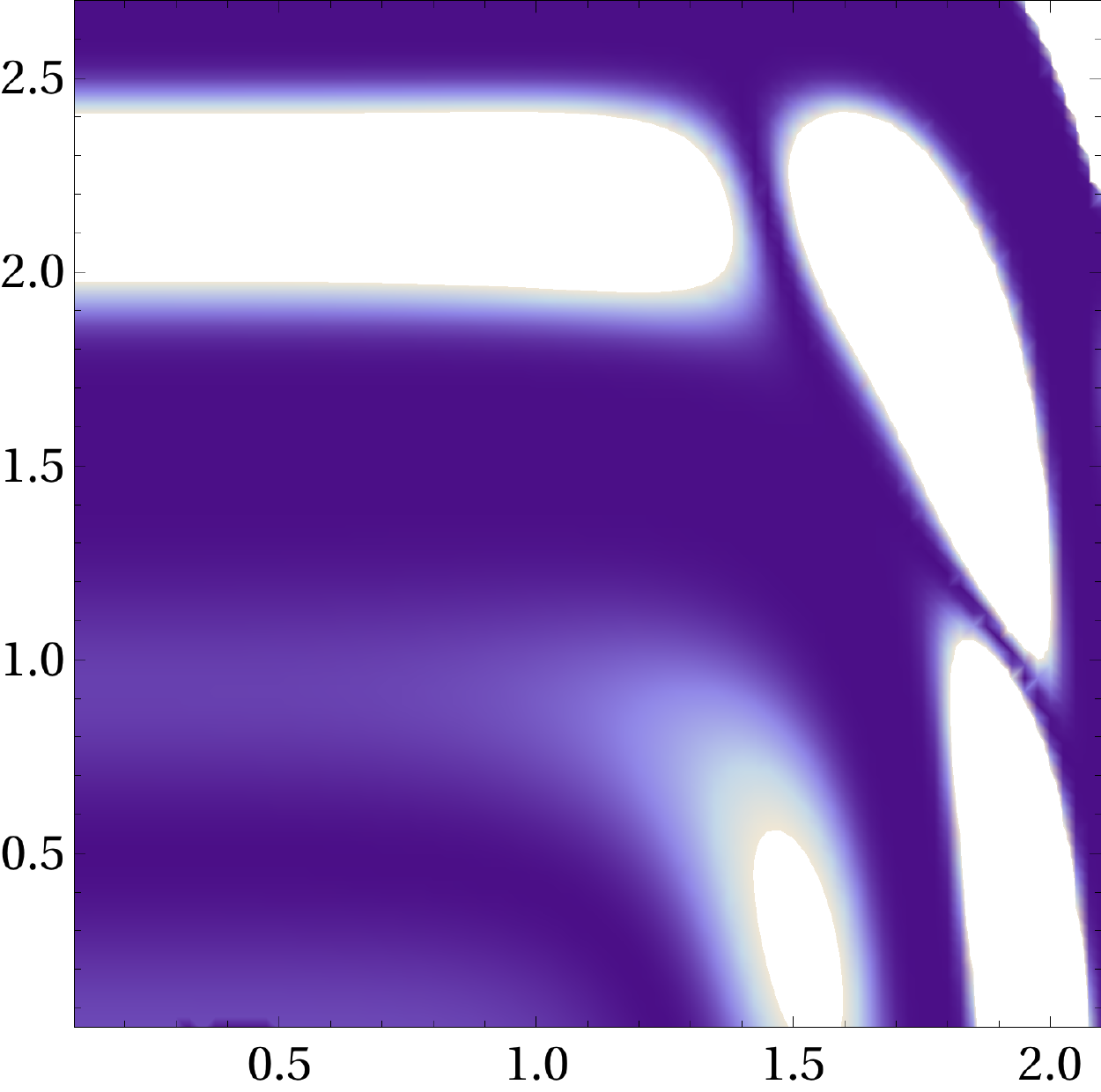} }
\caption{Left: Density plot of $|\Psi_{HH}(\lambda_i,a_i,A_i)|^2$ for $N=2$ for the double Gaussian profile (\ref{dbgauss}) as a function of $A_1$ ($x$-axis) and $A_2$ ($y$-axis) for $a_1=0$, $a_2=5$, $\lambda_1=\lambda_2=1$ using $l_{max} = 45$. The wavefunction grows and oscillates for negative $A_1$ and $A_2$. Right: Density plot of $|\Psi_{HH}(\lambda_i,a_i,A_i)|^2$ for the double Gaussian profile (\ref{dbgauss}) as a function of $\lambda_1$ ($x$-axis) and $\lambda_2$ ($y$-axis) with $A_1 = -1$, $A_2 = -1/100$, $a_1 = 0$ and $a_2 = 5$.}\label{doublegauss}
\end{center}
\end{figure}

As a third example we consider a double Gaussian profile:
\begin{equation}\label{dbgauss}
\hat{m}(r) = r^2 \left( A_1 \; e^{-(r-a_1)^2/\lambda_1^2} + A_2 \; e^{-(r-a_2)^2/\lambda_2^2} \right)~.
\end{equation}
See the right panel of figure \ref{gaussians} for an illustration of this deformation. An example of $|\Psi_{HH}(\lambda_i,a_i,A_i)|^2$ with $a_1 = 0$ is shown in figure \ref{doublegauss}. Once again we observe a pattern of maxima encircled by regions where the wavefunction squared vanishes identically. Furthermore, the wavefunction grows for increasingly negative values of $A_1$ and $A_2$.

\section{Radial deformations of flat $\mathbb{R}^3$ and pinching limits}\label{balloonsec}

In this section, we introduce and study a class of $SO(3)$ preserving deformations of the flat metric on $\mathbb{R}^3$. We show that they are conformally equivalent to the flat metric on $\mathbb{R}^3$. Thus, the wavefunction can only depend on such deformations of the metric if we also turn on a radial mass $m_b(r)$. Turning on such a mass, we can then perform the symmetry transformation discussed at the end of section \ref{wf} to get a mass deformation $\hat{m}(r)$ on the flat metric on $\mathbb{R}^3$. 
We will pick a functional form that is related by the symmetry transformations of \ref{wf} to a constant mass deformation on a deformed three-sphere geometry (see appendix \ref{r3s3} for details). Depending on the sign of a parameter, the deformed three-sphere will look either like a peanut or an inverse peanut, i.e. like bulbous pears inverted relative to one another and conjoined on their fatter ends. The partition function that we compute can then also be understood as the answer for the partition function on this deformed three-sphere with a constant mass deformation. The pinching limit will be when the waist of the peanut-shaped geometry vanishes. 

We emphasize that how we perceive these deformations of the late time metric depends greatly on what we decide are natural constant time slices, since there always exists a conformal frame where the late time metric is the flat one. Ideally, it would be useful to analyze a qualitatively similar geometric deformation that would take the original geometry outside its conformal class, but we must restrict to the former case in this section since we will be constrained by considering $SO(3)$ preserving deformations. Section \ref{secthree} will go beyond this restriction by considering a new conformal class.

\subsection{Balloon Geometry} 

Consider the following class of $SO(3)$ preserving metrics defined on $\mathbb{R}^3$:
\begin{equation}\label{evap}
ds^2 = dr^2 + r^2 f_\zeta(r)^2 d\Omega^2_2~, \quad d\Omega^2_2 \equiv d\theta^2 + \sin^2\theta d\phi^2~,
\end{equation}
with $r \in [0,\infty)$. Consider a family of smooth functions $f_\zeta(r)$ with $\zeta\leq\zeta^* $ for positive $\zeta^*$ that 
tend to unity both at large $r$ and near $r=0$.  We require that $f_{\zeta}(r)$ vanishes at some $r = r^*$ for the critical value $\zeta = \zeta^*$. We furthermore impose that:
\begin{equation}\label{noconical}
\lim_{\zeta \to \zeta^*} \left. \frac{d^2}{dr^2} \left( r^2 f_\zeta(r)^2 \right) \right\vert_{r=r^*} = 2~.
\end{equation}
\begin{figure}
\begin{center}
{ \includegraphics[height=2in]{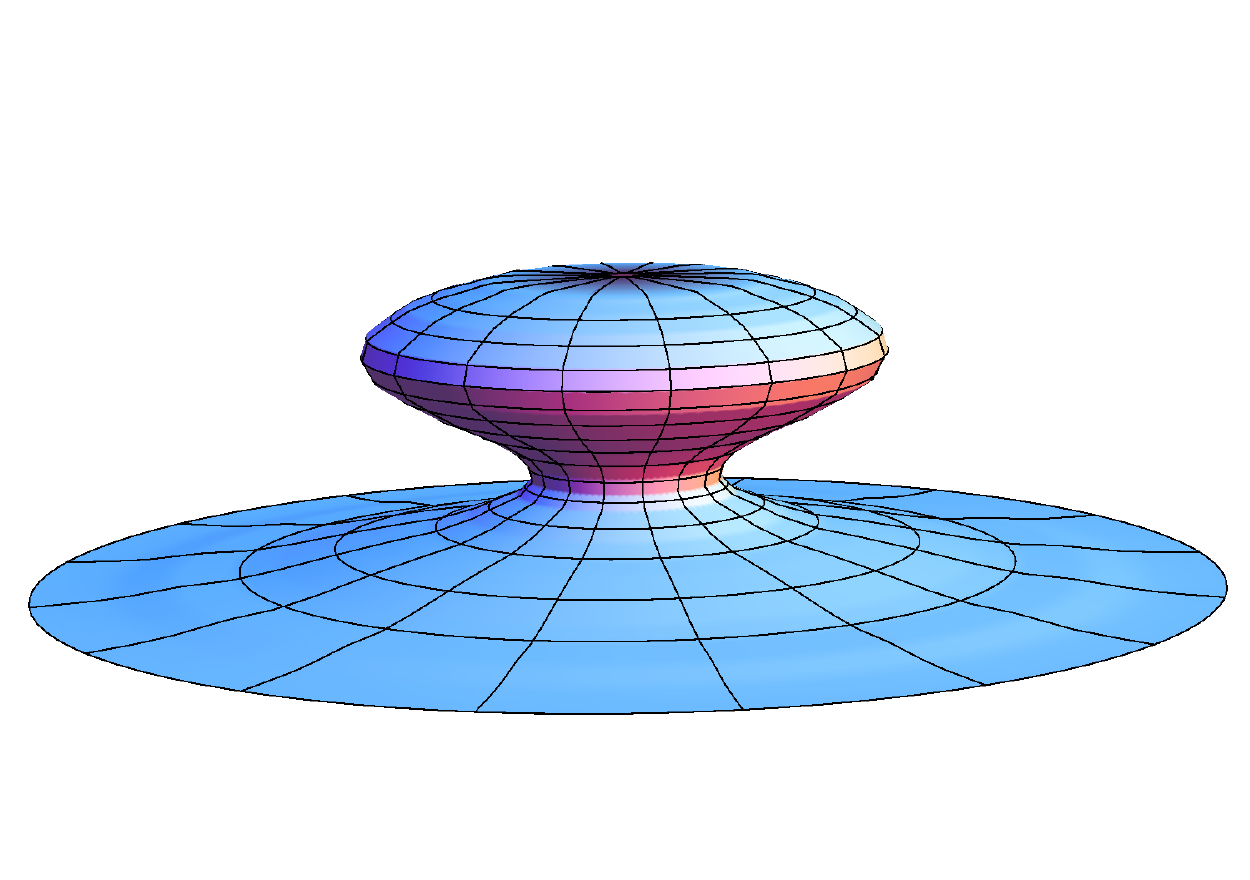}   }
\caption{The ``balloon" deformation of $\mathbb{R}^3$, defined by (\ref{evap}) and (\ref{balloonfunction}), represented schematically for positive $\zeta$. }\label{schematic}
\end{center} 
\end{figure}
For positive $\zeta$, the geometries described by (\ref{evap}) can be pictured as a two-sphere whose size at some finite $\zeta < \zeta^*$ grows, then shrinks and subsequently grows again. If $f_\zeta(r) = 1$ the geometry is of course nothing more than the flat metric on $\mathbb{R}^3$. As we approach $\zeta = \zeta^*$, the size of the two-sphere tends to vanish at $r = r^*$ eventually pinching the geometry into a warped three-sphere and a slightly deformed metric on $\mathbb{R}^3$. The choice (\ref{noconical}) ensures there are no conical singularities at the pinching point. It would be of interest in and of itself to study geometries with conical singularities (see for example \cite{Bordag:1996fw}). As a concrete example, we will take:
\begin{equation}\label{balloonfunction}
f_\zeta(r)^2 =  1 -  \zeta \left( r^2 + \frac{1}{\zeta^*} (\gamma r)^4  \right)   e^{- (r - a)^2/\lambda^2} ~.
\end{equation}
The parameters $a$ and $\lambda$ are chosen and $\gamma$ is tuned to obey the condition (\ref{noconical}). Though $\zeta^*$ is not an independent parameter it is useful to isolate in the expression. A schematic representation of this deformation, for positive $\zeta$, is presented in figure \ref{schematic}.
 
\subsection{Conformal Flatness of Balloon Geometry}

\begin{figure}
\begin{center}
{\includegraphics[width=3in]{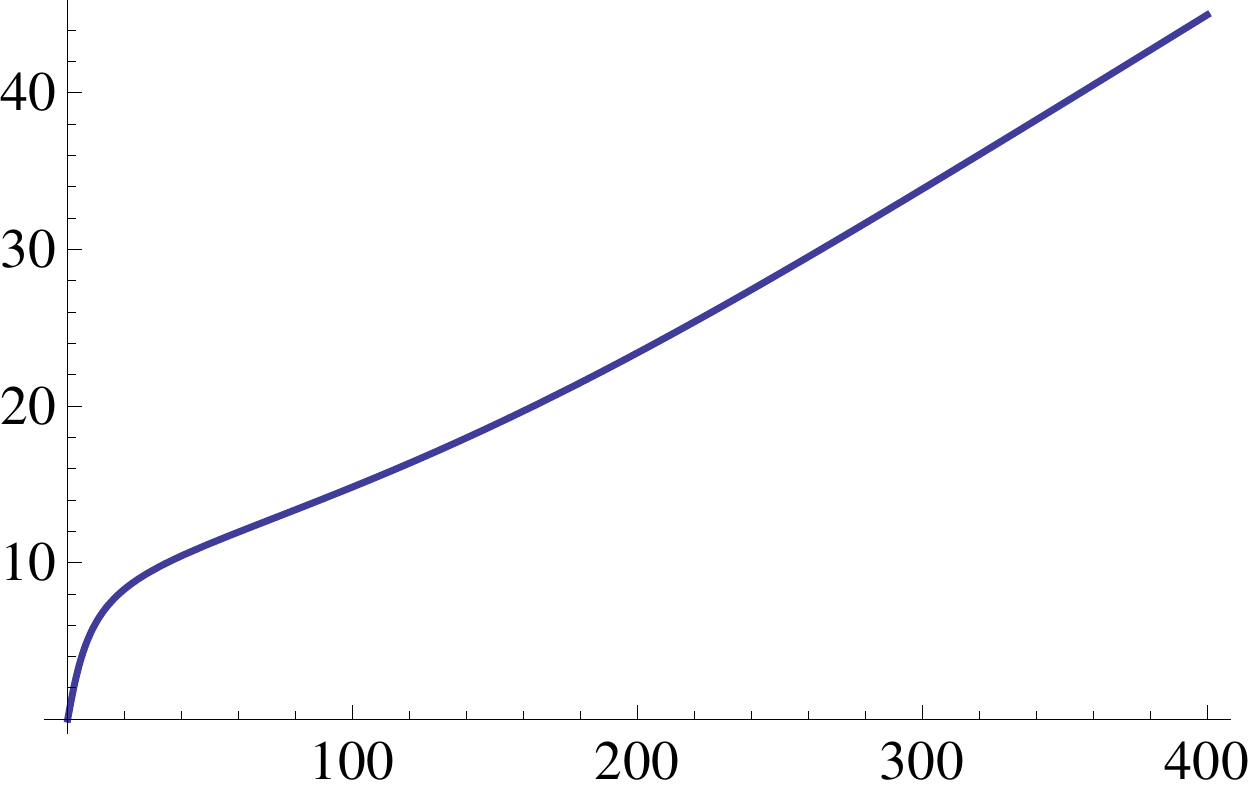} \quad \includegraphics[width=3in]{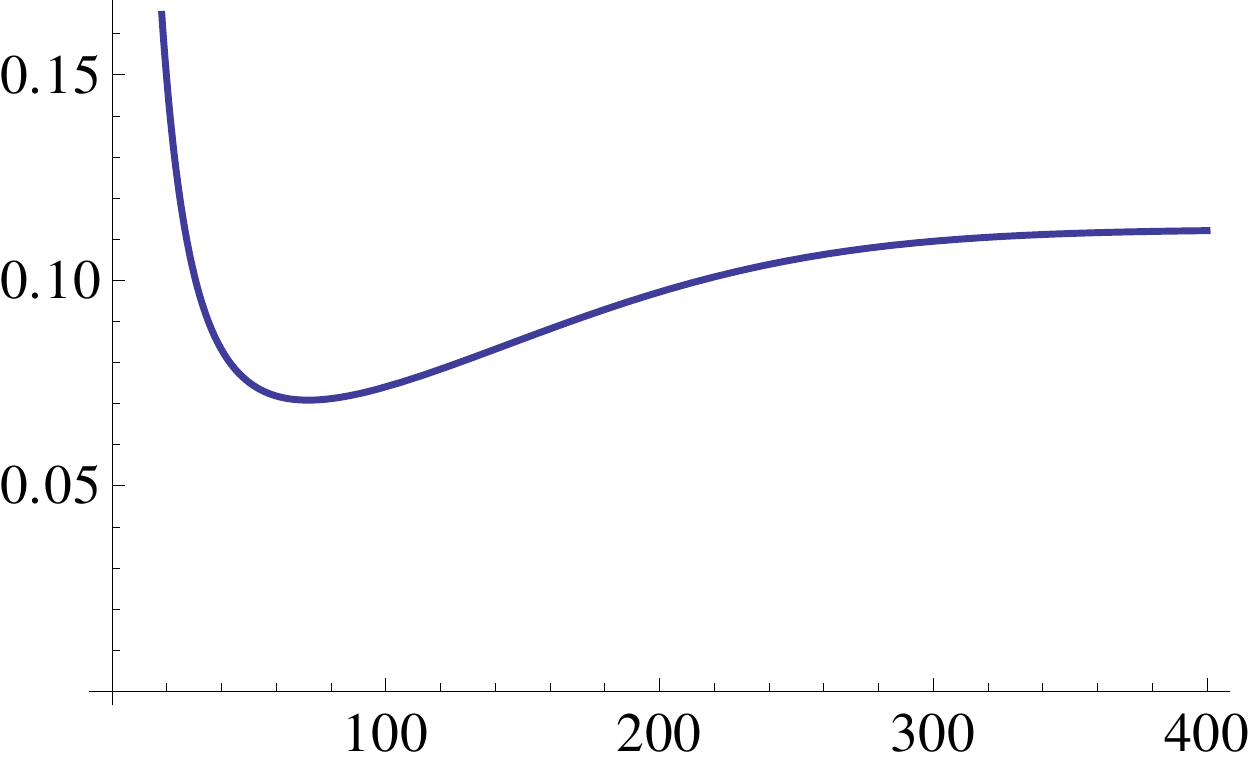}}
\caption{Plot of $g(x)$ (left) and $dg(x)/dx$ (right) as obtained by numerically solving equation \ref{conformal_ode} ($\zeta^*=1741.51$, $\zeta=9\zeta^*/10$, $\gamma=1.36612$, $a=-95$, $\lambda=30$).}\label{gexample}
\end{center}
\end{figure}

It is important to note that the geometry (\ref{evap}) is conformally flat. This can be shown in a straightforward fashion. Consider a coordinate transformation $r = g(x)$. It immediately follows that if the following ordinary differential equation:
\begin{equation}\label{conformal_ode}
x \frac{d g(x)}{dx}  = g(x) f_\zeta(g(x))~,
\end{equation} 
has a smooth solution for $g(x)$ whose derivative is positive for all $x > 0$ then our metric becomes:
\begin{equation}\label{conformal}
ds^2 = \left( \frac{d g(x)}{dx}  \right)^2 \left( dx^2 + x^2 d\Omega_2^2 \right)~.
\end{equation}
Though we cannot solve the non-linear o.d.e analytically, we can easily evaluate it numerically and confirm for several cases that $g(x)$ satisfies the necessary requirements. Hence, our metric (\ref{evap}) is indeed conformally equivalent to the flat metric on $\mathbb{R}^3$.
In figure \ref{gexample} we give a numerical example of this.

This result is already of some interest even for the case of ordinary Einstein gravity. It informs us that upon conditioning that all other fields vanish at late times, the absolute value of the late time Hartle-Hawking wavefunction, $|\Psi_{HH}[g_{ij}]|$, is independent of any radial $SO(3)$ preserving deformation of the late time metric. Indeed, from the bulk perspective a smooth conformal transformation of the late time three-metric can be induced by a time diffeomorphism that preserves the Starobinski-Fefferman-Graham form. From a holographic perspective, the late time wavefunction is computed by the partition function of a three-dimensional conformal field theory and thus only depends on the conformal metric (recall there are no conformal anomalies in three dimensions).

\subsection{Wavefunctions and balloon geometries}

\begin{figure}
\begin{center}
\includegraphics[width=3in]{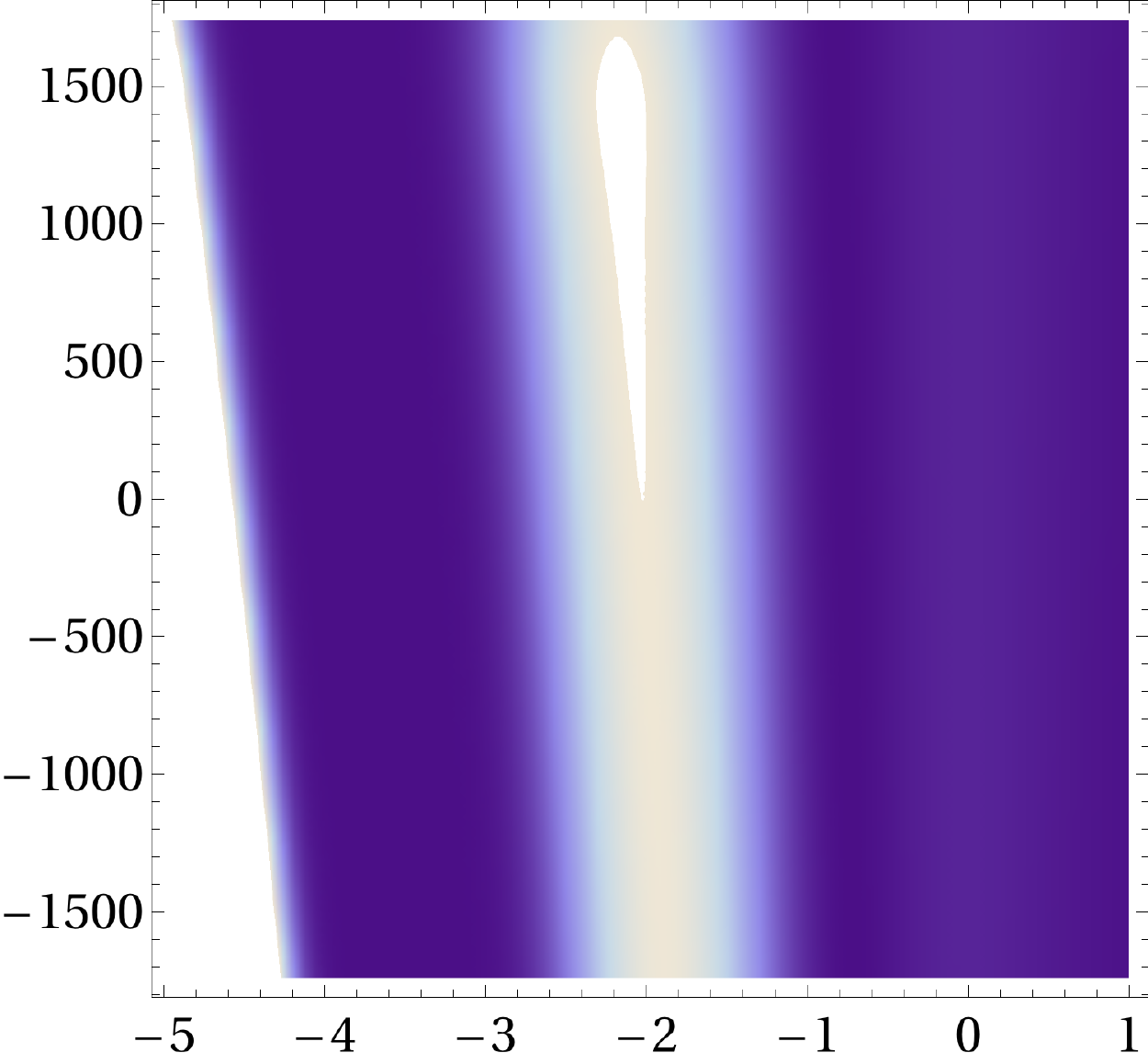}
\end{center}
\caption{Density plot of the $|\Psi_{HH}(\zeta,m)|^2$ for $N=2$ as a function of the pinching parameter $\zeta$ (vertical axis) and an overall mass deformation $m$ (horizontal axis) using $l_{max}=45$, $\zeta_*=1741.51$, $\gamma=1.36612$, $a=-95$ and $\lambda=30$.} 
\begin{center}\label{balloonfigi}
{\includegraphics[width=3in]{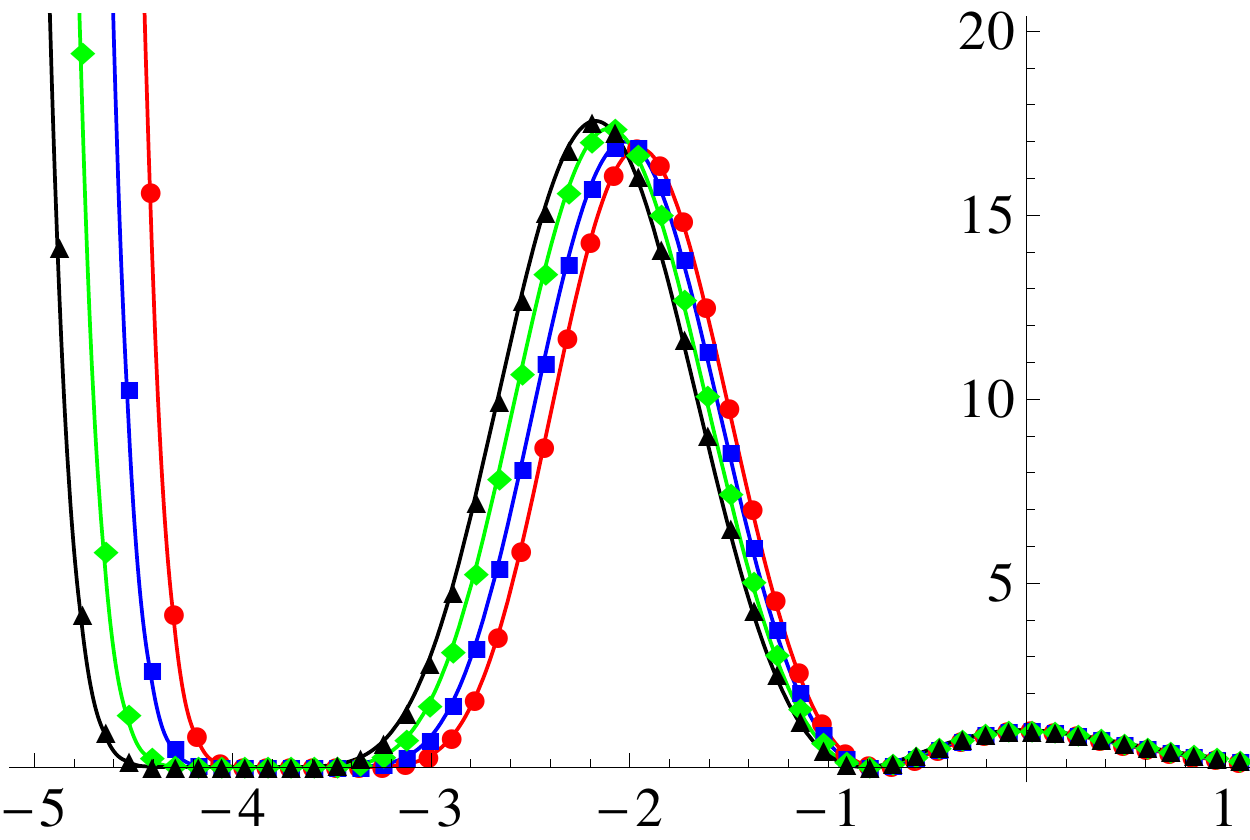} \quad \includegraphics[width=3in]{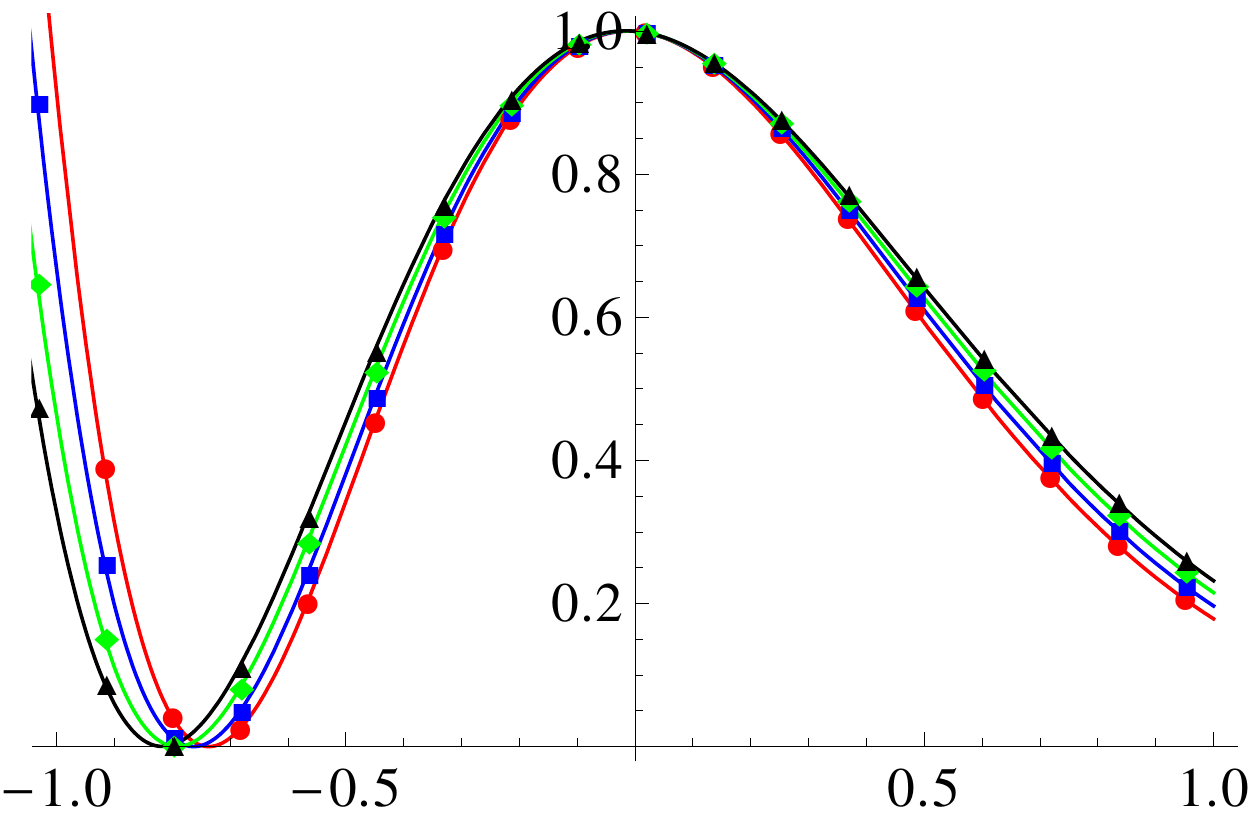}}
\caption{Left: $|\Psi_{HH}(\zeta,m)|^2$ for $N=2$ as a function of $m$ for $\zeta=-\zeta^*/2$ (red dots), $\zeta=0$ (blue squares), $\zeta=\zeta^*/2$ (green diamonds) and $\zeta=9 \zeta^*/10$ (black triangles) using $l_{max}=45$, $\zeta_*=1741.51$, $\gamma=1.36612$, $a=-95$ and $\lambda=30$. Right: Same as left but for different plot range.}\label{balloonfigii}
\end{center}
\end{figure}

We now examine what happens to the functional determinant as we vary the waist parameter $\zeta$ for an example. We will also turn on a mass that would correspond to a uniform mass $m$ on the deformed three-sphere (as discussed further in appendix \ref{balloonapp}), which upon the conformal transformation discussed there becomes:
\begin{equation}
m_b(r) = \left(\frac{2}{1+r^2}\right)^2 m~.
\end{equation}
This is the mass deformation on the balloon geometry. The final deformation $\hat{m}(x)$, to be used in the Dunne-Kirsten formula, is obtained by performing a conformal rescaling of the balloon geometry to the flat metric on $\mathbb{R}^3$: $ds^2=dx^2+x^2\, d\Omega_2^2$. This requires a conformal rescaling of $m_b(r)$ to:
\begin{equation}
\hat{m}(x)=\left(\frac{dg(x)}{dx}\right)^2 \left(\frac{2}{1+r(x)^2}\right)^2 m~.
\end{equation}
Thus, we will study the functional determinant as a function of $m$ and $\zeta$. In figures \ref{balloonfigi} and \ref{balloonfigii} we display our numerical results. As expected, at $m = 0$, nothing changes as we vary $\zeta$ since the balloon geometries are conformally flat. However, when we turn on $m \neq 0$ the wavefunction becomes sensitive to changes in $\zeta$. Interestingly, decreasing the girth of the throat while keeping everything else fixed is favored, at least near $m = 0$. Thus, though supressed exponentially with respect to the local maximum of the wavefunction at $m = 0$, the wavefunction does not vanish in the pinching limit. It is tempting to speculate that such pieces of the wavefunction might be connected to the fragmentation picture of \cite{Bousso:1998bn}.

\section{Spherical harmonics and a conjecture}\label{secconj}

In this section, we present numerical evidence that when mapping the problem back to the three-sphere (using the discussion in appendix \ref{r3s3}), all profiles give a normalizable wavefunction upon fixing their average value over the whole three-sphere. Thus, it is conceivable that the only divergence of the wavefunction occurs precisely for large and negative values of a uniform profile over the whole three-sphere \cite{Anninos:2012ft}; a single direction in an infinite dimensional configuration space! For instance, as we shall show below, by mapping the Gaussian profile (\ref{gauss}) to the three-sphere and removing the zero mode from its expansion in terms of three-sphere harmonics, the resultant profile produces a normalizable wavefunction as a function of its amplitude.

\subsection{Three-sphere harmonics}

We now study some examples of $SO(3)$ invariant deformations which correspond to harmonics of the three-sphere conformally mapped back to $\mathbb{R}^3$. These harmonics are the eigenfunctions of the Klein-Gordon operator on the three-sphere with metric $ds^2 = d\psi^2 + \sin^2\psi \; d\Omega_2^2$. 
The $k^{th}$ harmonic (independent of the $S^2$ coordinates) is given by:
\begin{equation}\label{sphericalharmonic}
F_k(\psi) = c_k ~\text{csc}(\psi) ~\text{sin}[\,(1+k) \psi \,]~, \quad k = 0,1,2,\ldots~.
\end{equation}
As explained in appendix \ref{r3s3}, to evaluate the partition function for this deformation using the method of Dunne and Kirsten we must first perform the coordinate transformation $\psi(r) = 2 \cot^{-1} r^{-1}$, and then scale the deformation by the inverse of the conformal factor that maps the three-sphere metric to the metric on $\mathbb{R}^3$. The final radial deformation which is to be used in the Dunne-Kirsten formula is:
\begin{equation}\label{sphericalharmonicR3}
\hat m_k(r) = c_k \frac{2 ~\text{sin}[\,2(1+k) \text{tan}^{-1}r\,]}{ \left(r+r^3\right)}~.
\end{equation}
Taking $k=0$ corresponds to the zero mode which has been previously studied in \cite{Anninos:2012ft} and was found to be oscillatory and divergent as the coefficient $c_k$ goes to large negative values. In figure \ref{harmonicsPlots} we plot the partition functions for higher harmonics as a function of the coefficient $c_k$. We notice that they are all well-behaved and normalizable, at least in the range we have explored.
This motivates us to consider deformations which are linear combinations of spherical harmonics. Having looked at a deformation that is the linear combination of the zero mode with the first harmonic, as well as a deformation that is the linear combination of the first harmonic with the second harmonic, we notice that the partition function is not divergent so long as the coefficient of the zero mode is kept fixed.




\begin{figure}[p!]
\begin{center}
{ \includegraphics[height=1.7in]{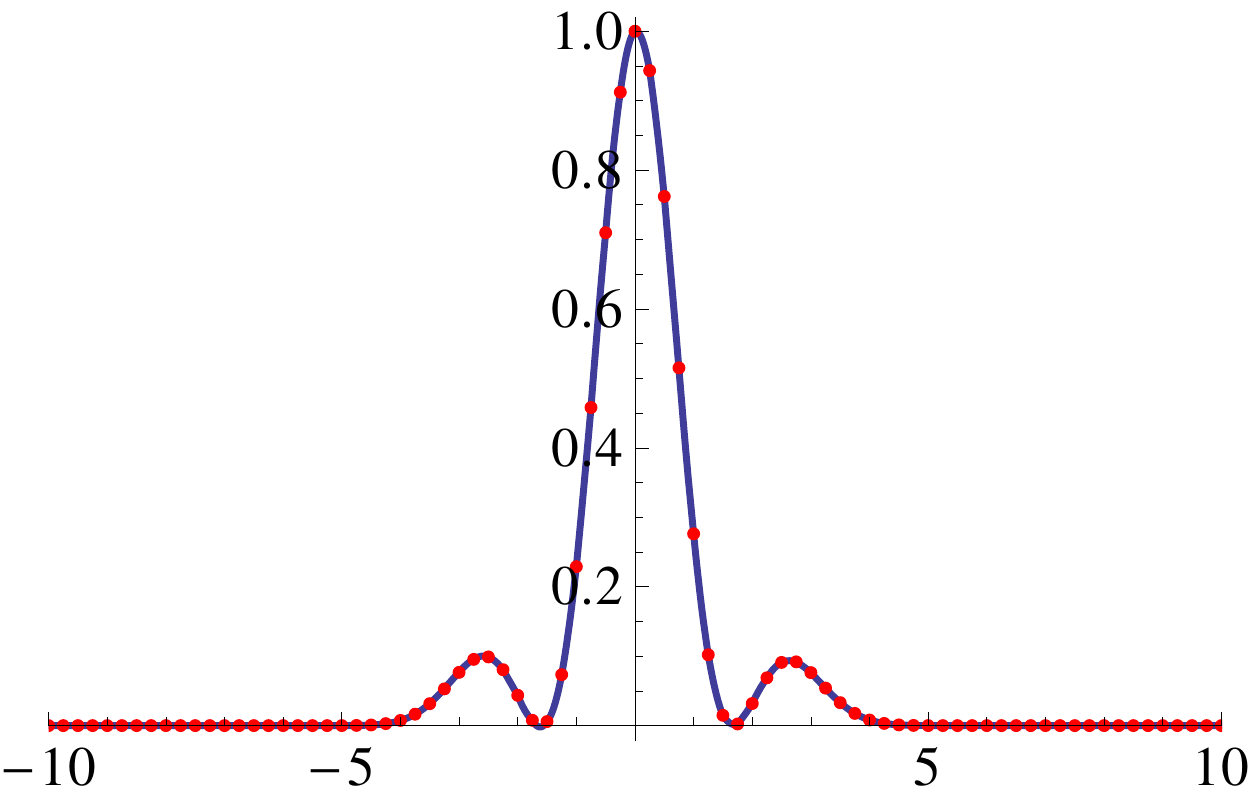} \;\; \quad  \includegraphics[height=1.7in]{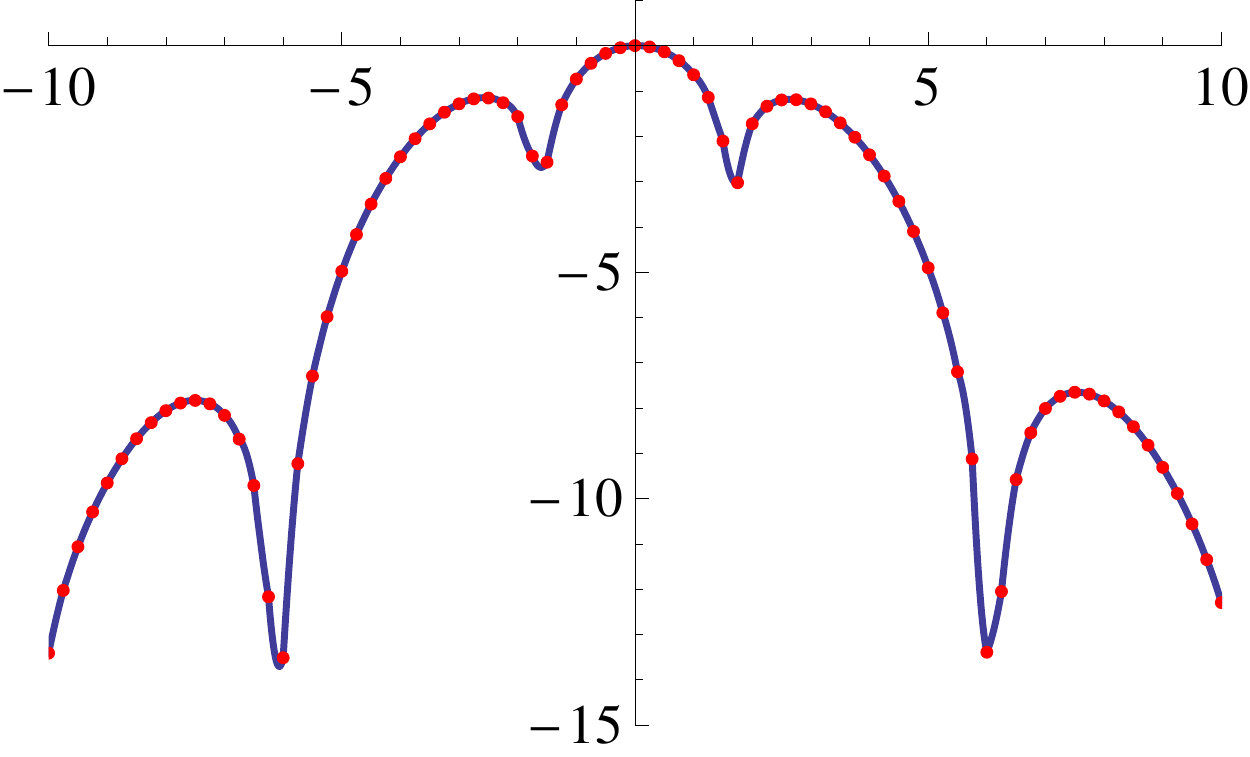}  }
\caption{Left: Plot of $|\Psi_{HH} (c_1) |^2$ for $N=2$ for the first harmonic mapped to $\mathbb{R}^3$ given in (\ref{sphericalharmonicR3}) using $l_{max} = 45$. Right: Plot of $\log |\Psi_{HH} (c_1) |^2$ for $N=2$ using $l_{max} = 45$.\newline\newline}\label{harmonicsPlots}
{ \includegraphics[height=1.27in]{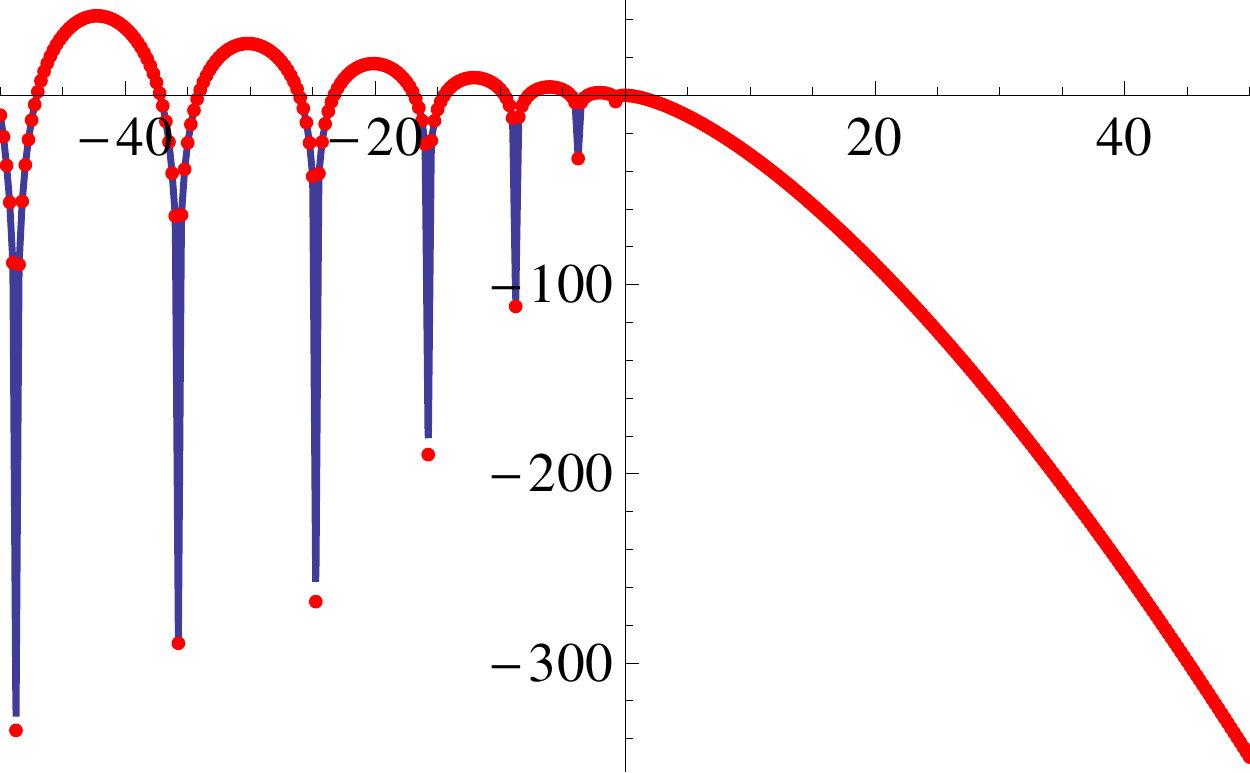} \;\; \includegraphics[height=1.27in]{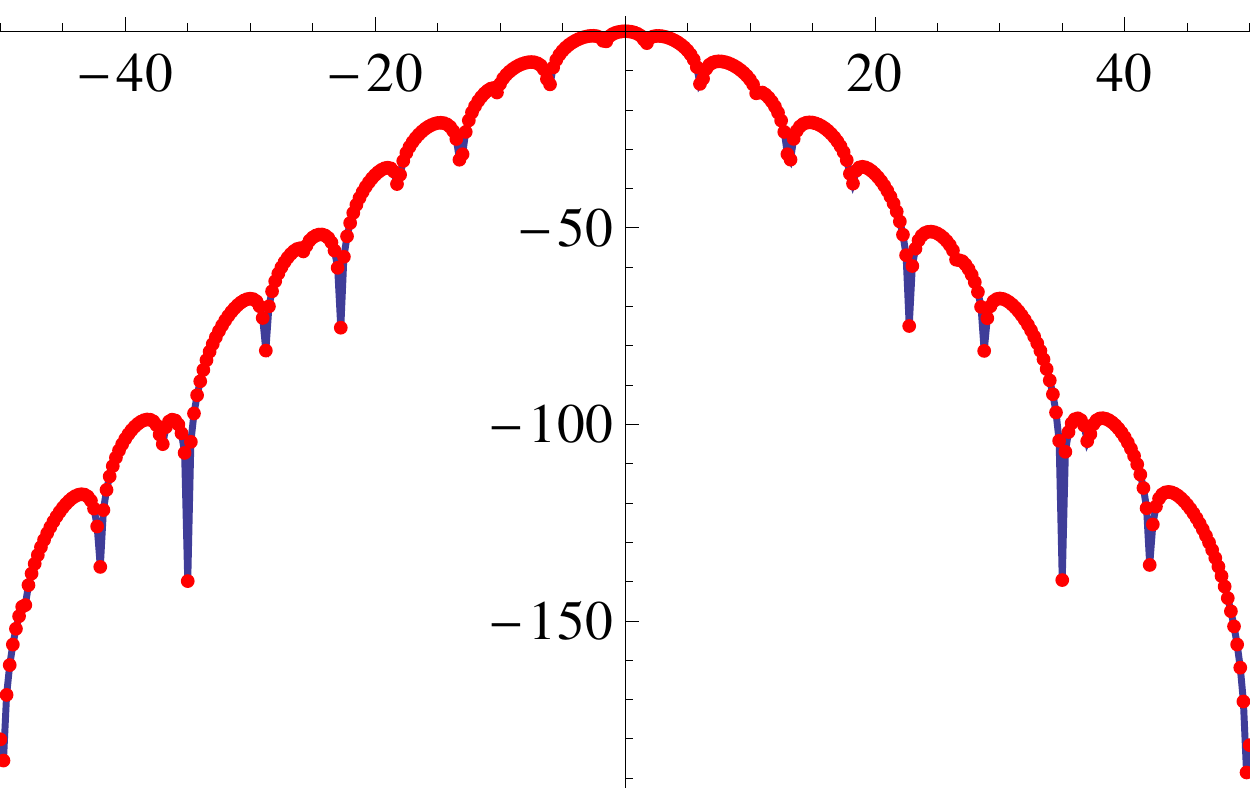} \;\; \includegraphics[height=1.27in]{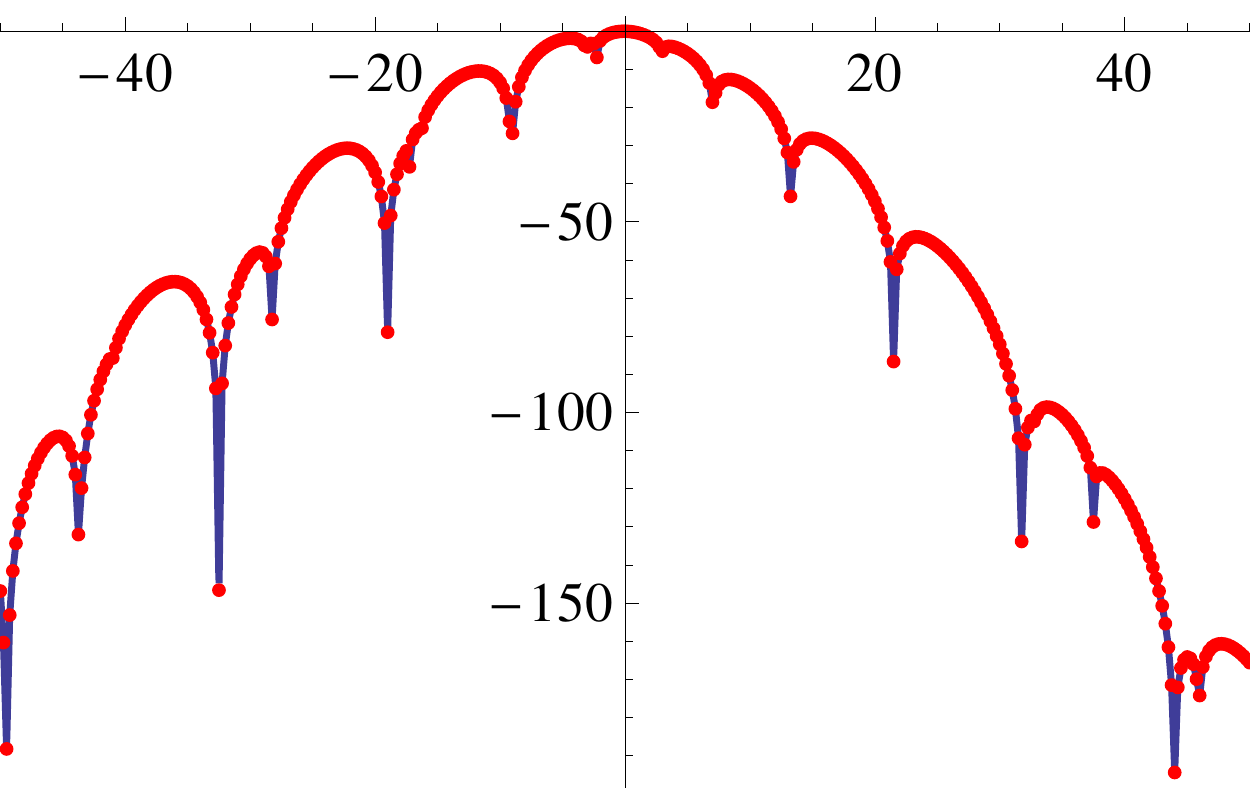} \;\; \includegraphics[height=1.27in]{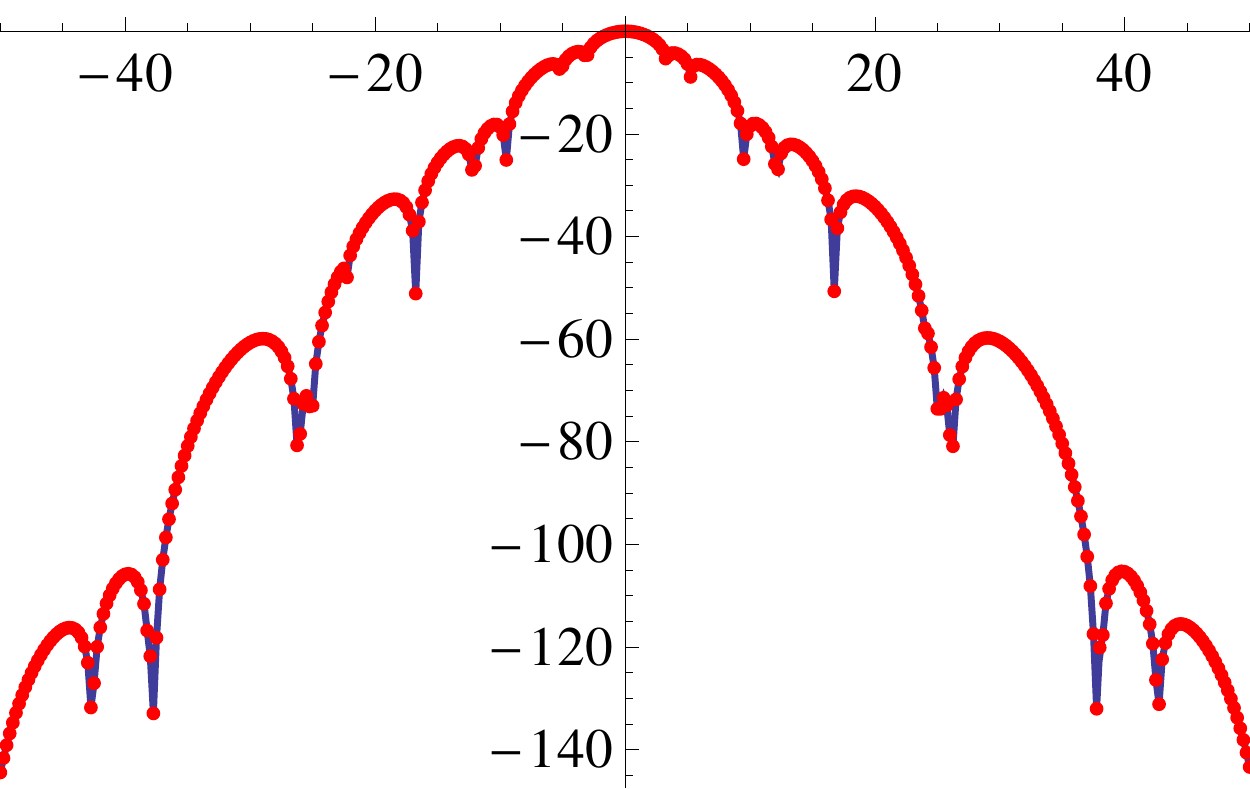}  \;\; \includegraphics[height=1.27in]{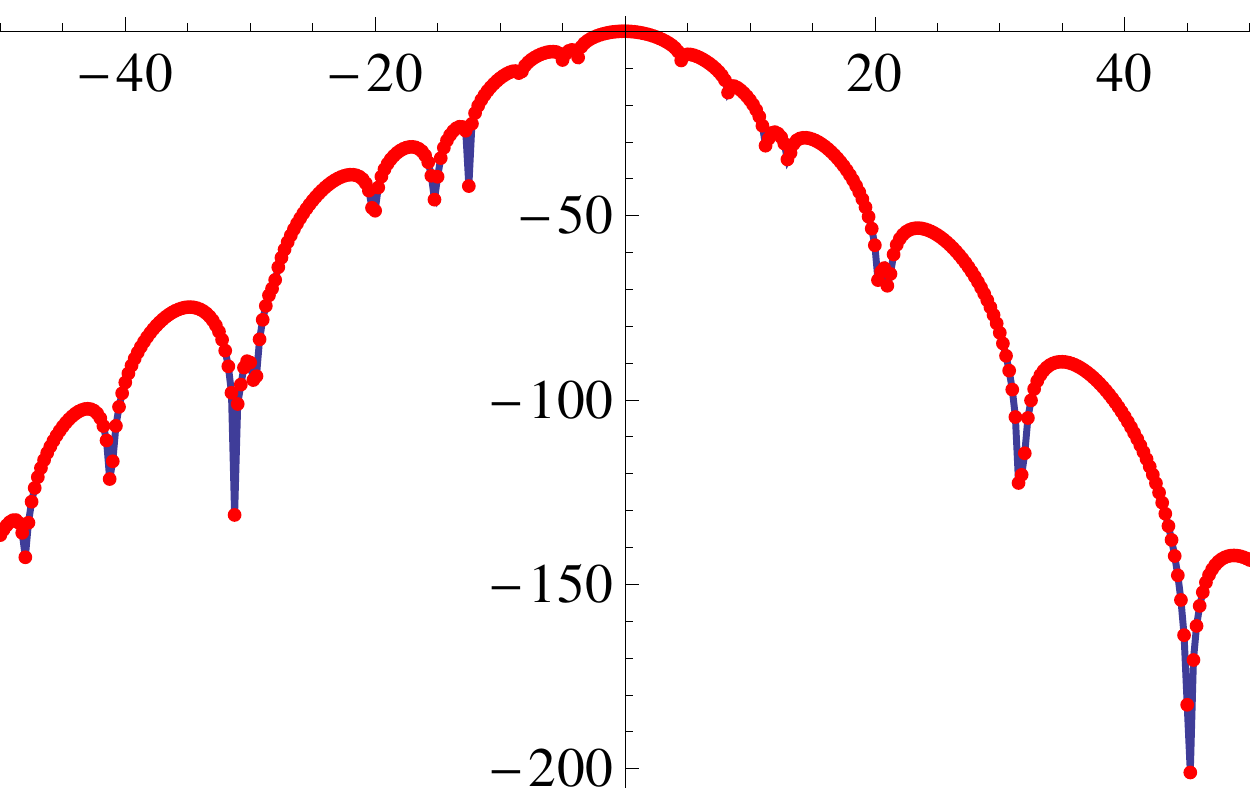}  \;\;\includegraphics[height=1.27in]{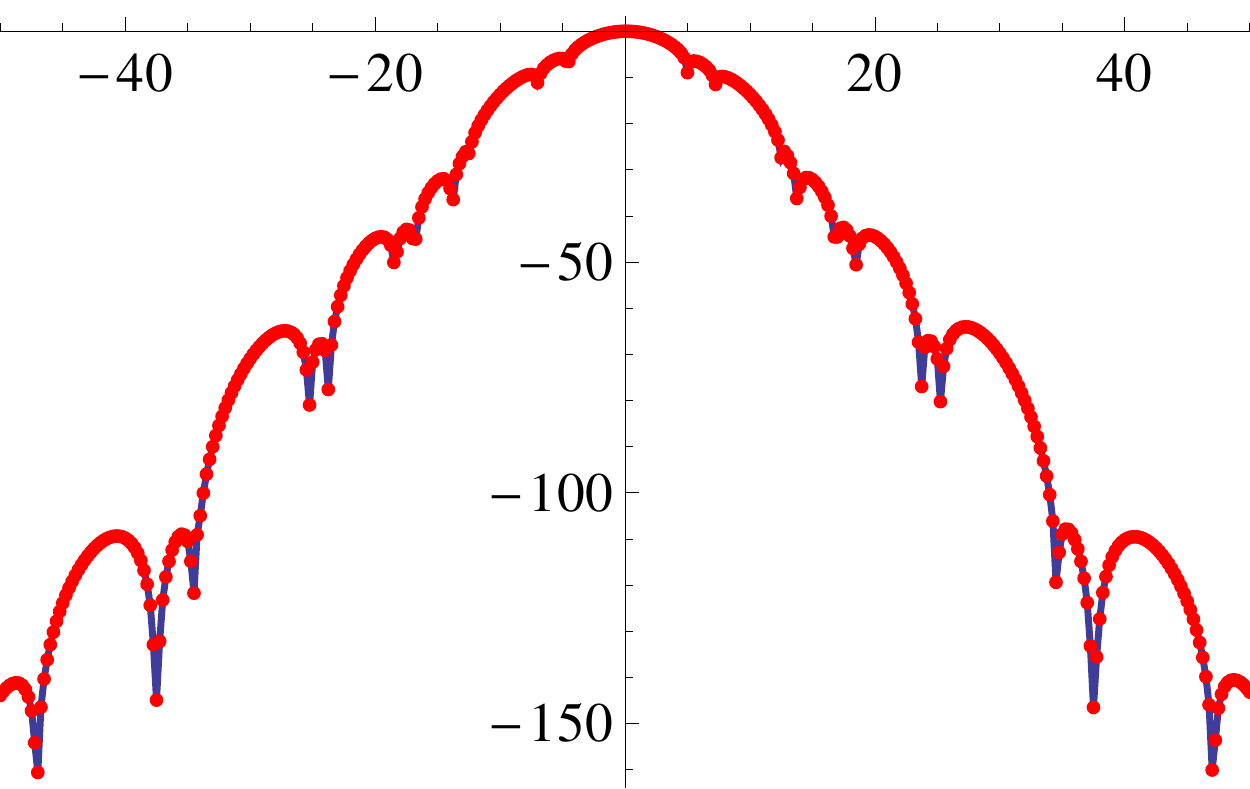} \;\; \includegraphics[height=1.27in]{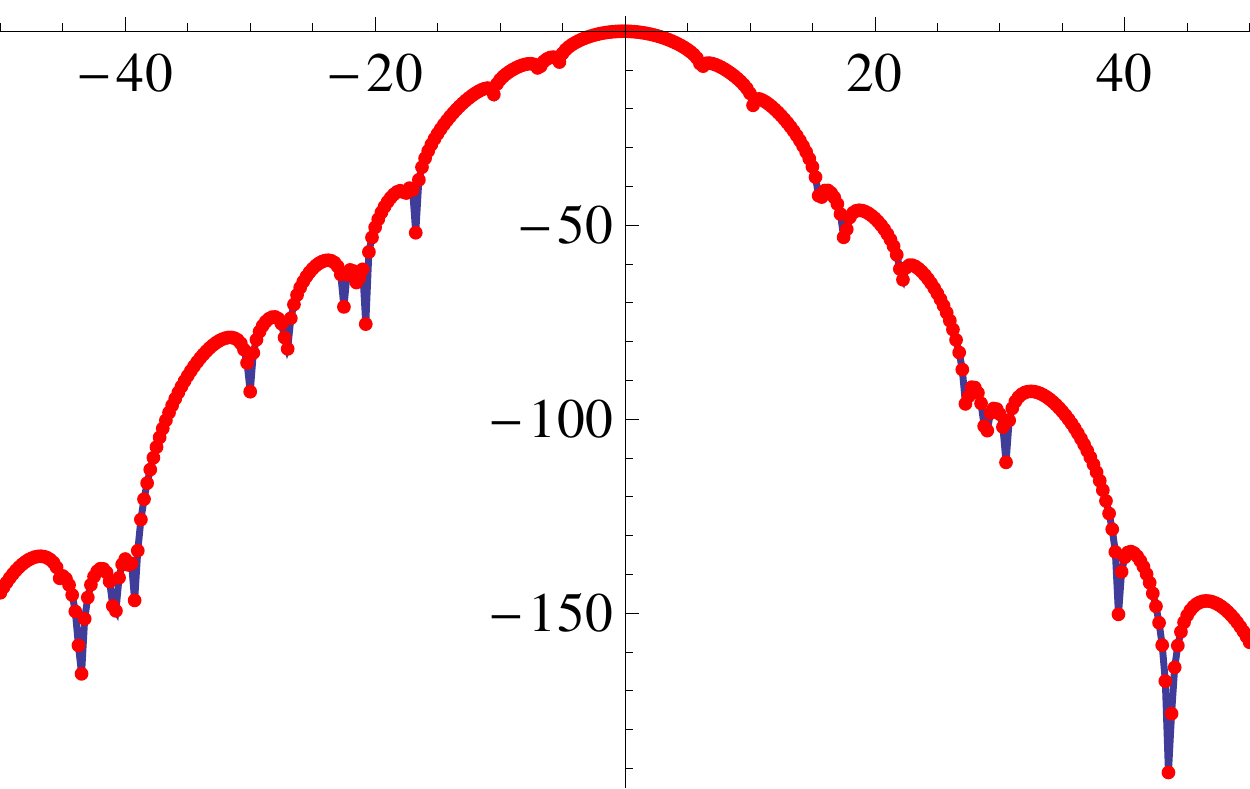} \;\; \includegraphics[height=1.27in]{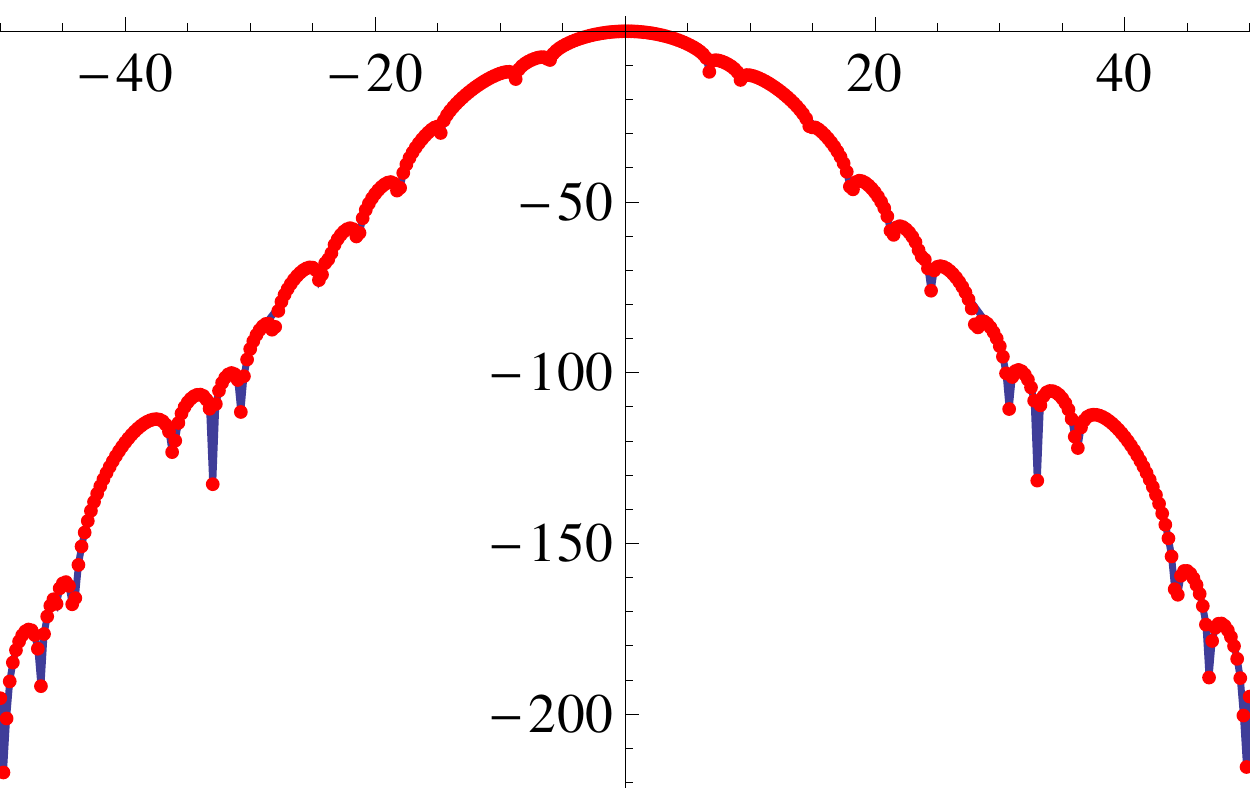} \;\; \includegraphics[height=1.27in]{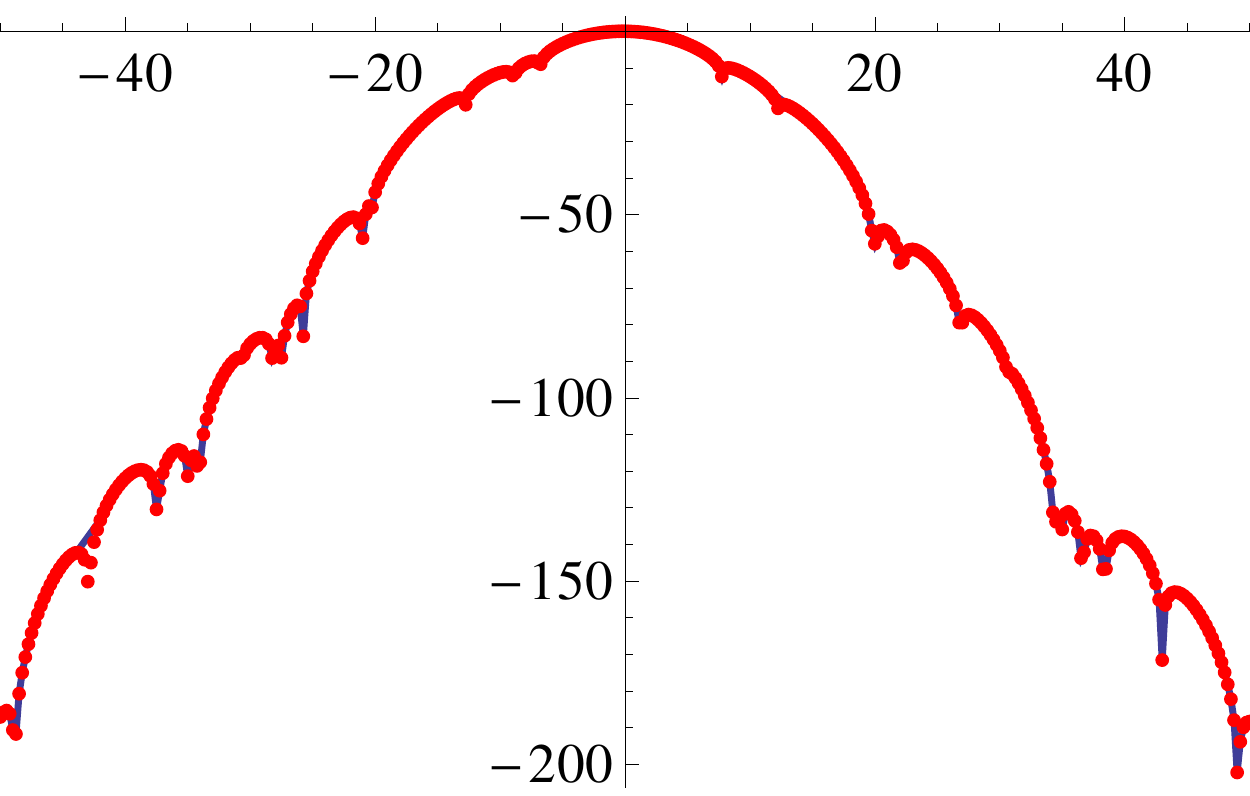}  }
\caption{Plot of $\log |\Psi_{HH} (c_k) |$ for $N=2$ for the first nine spherical harmonics mapped to $\mathbb{R}^3$ given in (\ref{sphericalharmonicR3}) using $l_{max} = 45$. Notice that only the zeroth harmonic is non-normalizable in the negative $c_0$ direction.}\label{harmonicsPlots}
\end{center}
\end{figure}

Postponing a more systematic study for the future, here we simply consider a modified version of the single Gaussian deformation given in (\ref{gauss}). Previously, we had found that as the overall coefficient of the profile becomes large and negative, the partition function diverges. The new profile we will study is obtained by mapping the Gaussian profile to the three-sphere, subtracting off its zero mode, and mapping it back to $\mathbb{R}^3$. Notice that the Gaussian profile mapped to the three-sphere is constructed from an infinite number of harmonics, and here we are subtracting the piece that seems problematic from our analysis of harmonics and finite linear combinations thereof.
Since $F_0(\psi)=c_0$ is simply a constant, the condition to be met is:
\begin{equation}\label{sphericalharmonic}
\int_0^\pi d\psi \left(\frac{r(\psi)^2+1}{2}\right)^2 \hat m(r(\psi))~ \sin^2 \psi = 0~,
\end{equation}
where $r(\psi)=\text{tan}(\frac{\psi}{2})$ and $\hat m(r)=A e^{-r^2}-4 a_1/(r^2+1)^2$. In this case the integral can be done explicitly and we can solve for the coefficient $a_1$ analytically. The final form of the single Gaussian radial deformation orthogonal to the zero mode of the three-sphere becomes:
\begin{equation}\label{normalizedGaussian}
\hat m(r)=A ~e^{-r^2}-\frac{8 A \left(1-e \sqrt{\pi }~ \text{Erfc}[1]\right)}{\sqrt{\pi } \left(1+r^2\right)^2}~.
\end{equation}
\begin{figure}
\begin{center}
{ \includegraphics[height=1.7in]{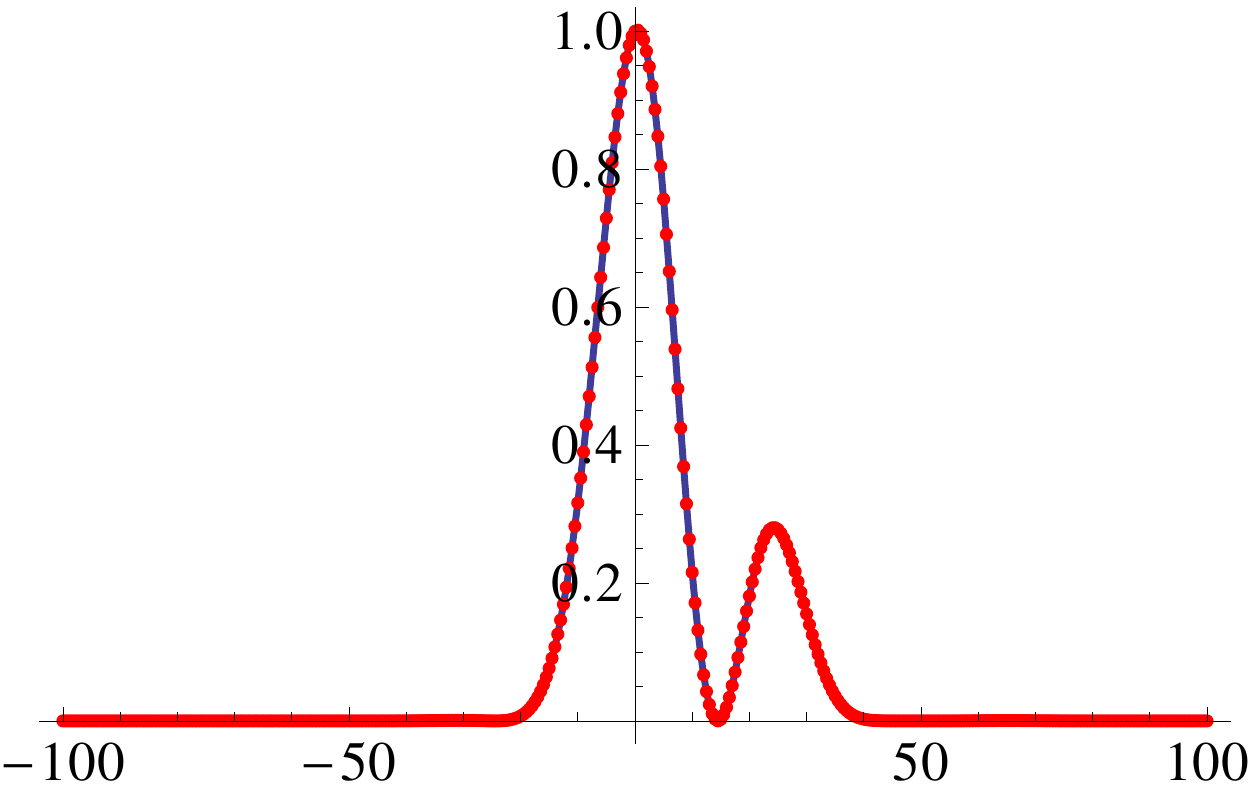}  \includegraphics[height=1.7in]{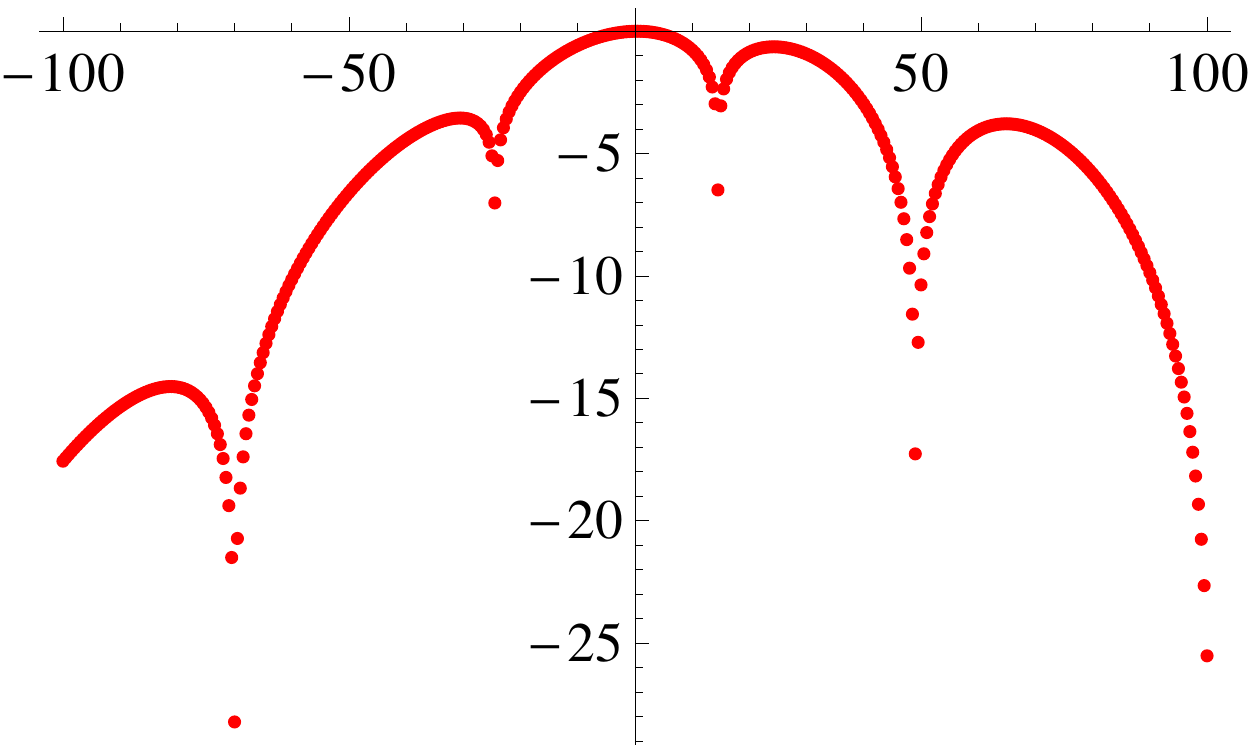}}
\caption{$|\Psi_{HH}(A)|^2$ (left) and $\text{log}|\Psi_{HH}(A)|$ (right) for $N=2$ as a function of $A$, the overall size of the radial deformation in (\ref{normalizedGaussian}) which is constructed to be orthogonal to the zero mode of the three-sphere ($l_{max}=45$).}\label{normGauss}
\end{center}
\end{figure}We plot the functional determinant as a function of $A$ in figure \ref{normGauss}. Interestingly, the partition function is once again well-behaved for large values of $A$. An analogous analysis for the balloon geometries, where we fix the zero-mode on the conformally related three-sphere, also results in the boundedness of the partition function in the $\zeta$  direction ($\zeta$ defined in (\ref{balloonfunction})).

The above results motivate the conjecture: 
\newline\newline
{\it The partition function of any $SO(3)$ symmetric ``radial" deformation for which the three-sphere zero mode harmonic is fixed is bounded}. 
\newline

Notice that the three-sphere is chosen as the geometry in the conformal class on which to fix the uniform profile. For example, in the case of the peanut geometries one can show that fixing the uniform profile on the peanut is \emph{not} sufficient to ensure that the partition function is bounded (though fixing some non-uniform profile would suffice). This is simply because keeping the uniform mass profile fixed to some negative value while taking $\zeta$ large and negative, which corresponds to the conformally related sphere getting fatter at the waist, implies that the uniform profile on the sphere is getting large and negative. Consistent with the rest of our observations, we see that the partition function is unbounded in this direction. Furthermore, the next section will provide evidence for a more general conjecture that would extend beyond the conformal class of the sphere.

\section{Three-sphere squashed and massed}\label{secthree}

In this section, we would like to briefly revisit and extend some of the observations of \cite{Anninos:2012ft} for a constant mass deformation on an $S^3$. There, it was observed that in the $\hat{\sigma}$-basis the wavefunction on an $S^3$ with a uniform mass deformation oscillated and diverged at large negative $m_{S^3}$. We explore in this section a new direction which is the squashing parameter of the round metric on the three-sphere (in the presence of a non-zero uniform scalar profile) and its effect on the zeroes and maxima. Unlike the previous $SO(3)$ preserving deformations which were inhomogeneous, this $SO(3) \times U(1)$ preserving deformation is homogeneous yet anisotropic. Furthermore, the squashed three-sphere is not conformal to the ordinary round three-sphere. In fact, squashed spheres with different values of the squashing parameter belong to distinct conformal classes.

Part of our motivation is to provide further evidence that the zeroes of the wavefunction are extended and that the local maxima of the wavefunction (other than the pure de Sitter one) will no longer necessarily peak about homogeneous and isotropic geometries. Additionally, and in the same spirit as the observations made in section \ref{secconj}, we find that upon fixing the value of the uniform profile over the whole squashed three-sphere the wavefunction is normalizable in the squashing direction.

\subsection{Squashed and massed}

\begin{figure}
\begin{center}
{\includegraphics{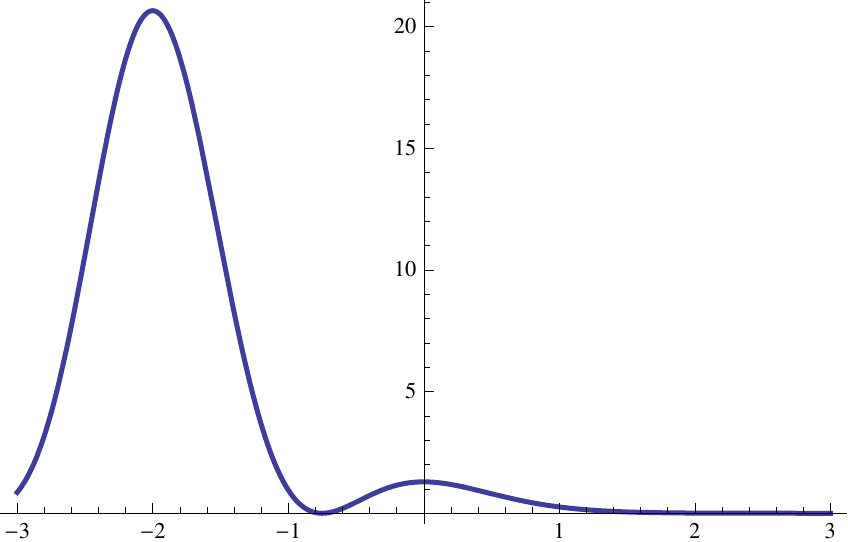}}
\end{center}
\caption{Plot of $|\Psi_{HH}[{m_{S^3}}]|^2$ given by expression (\ref{s3z}) for $N = 2$.}\label{sigmafig}
\end{figure}

Consider turning on a constant mass $m_{S^3}$ for the free $Sp(N)$ model on the round metric on an $S^3$ whose radius $a$ is fixed to one unless otherwise specified. The partition function is given by \cite{Anninos:2012ft}:
\begin{equation}\label{s3z}
N^{-1} \log Z_{free} [m_{S^3}] = \frac{1}{16} \left({\log} \; 4-\frac{3 {\zeta}(3)}{\pi ^2}\right) -\frac{\pi}{8} \int^{m_{S^3}}_0 d \sigma \sqrt{1 - 4\sigma} \cot \left(\frac{\pi}{2}\sqrt{1-4\sigma}  \right)~.
\end{equation}
The analyticity of $Z_{free}[m_{S^3}]$ in the complex $m_{S^3}$-plane is ensured by the uniform convergence of the (regularized) infinite product of analytic eigenvalues which defines it. For physical applications we restrict to real $m_{S^3}$ and note that this implies that the Taylor series expanded about any $m_{S^3} = m^0_{S^3}$ converges to the partition function above. Furthermore, the partition function has a zero if and only if one of the product eigenvalues in the functional determinant vanishes, which can only happen for $m_{S^3}  < 0$. As shown in figure \ref{sigmafig}, this wavefunction grows exponentially and oscillates in the negative $m_{S^3}$ direction.\footnote{Divergences of the Hartle-Hawking wavefunctional have been discussed in other circumstances such as Einstein gravity coupled to a scalar field with a quadratic scalar potential and vanishing cosmological constant \cite{Hawking:1983hj} or the wavefunction of dS$_3$ on a toroidal boundary \cite{Castro:2012gc}. Their physical interpretation remains to be understood. For example, it might be a result of very sharp conditioning or an indication of an instability.} It is worth noting that the $m_{S^3}$ dependent part of the phase of the wavefunction vanishes. The real part of $\log Z_{free}/N$ goes as $|m_{S^3}|$ for large negative $m_{S^3}$ and $-(m_{S^3})^{3/2}$ for large positive $m_{S^3}$.  

\begin{figure}
\begin{center}
{\includegraphics[width=2.8in]{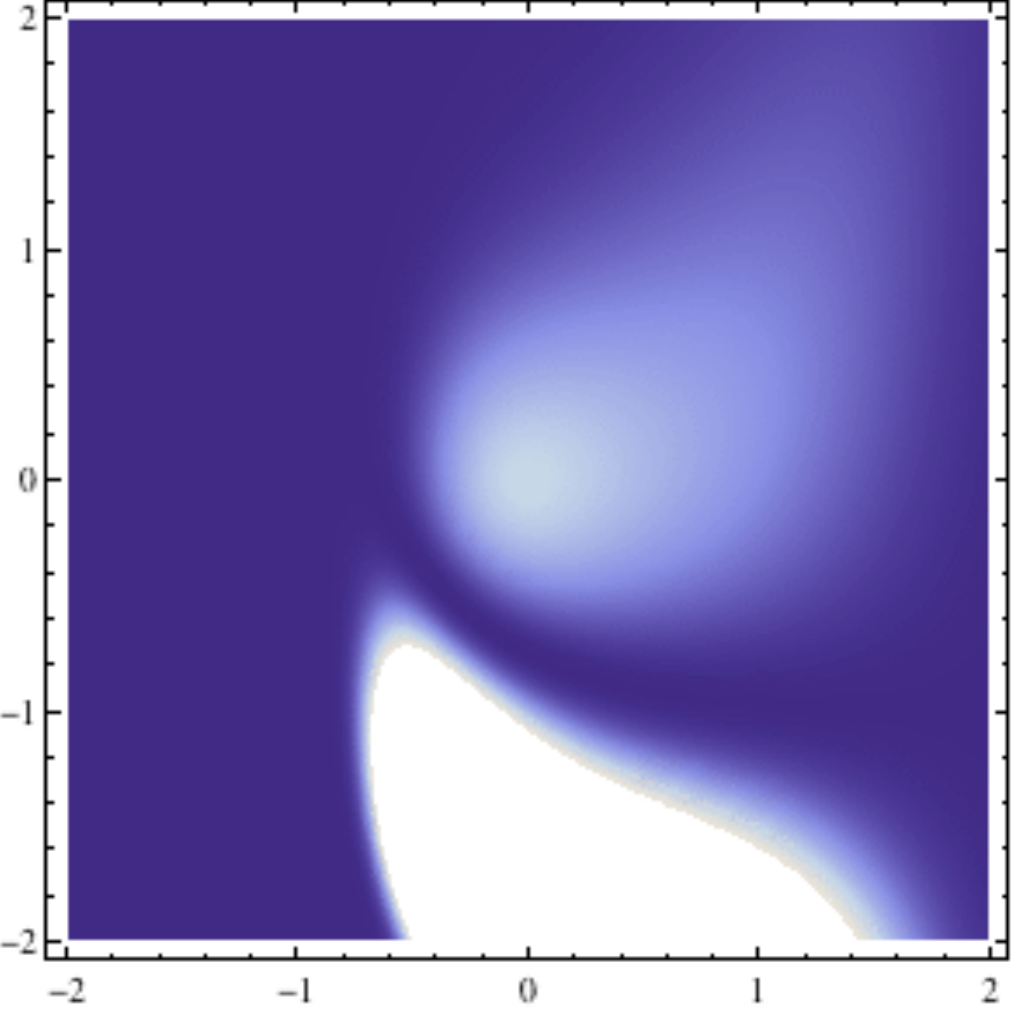} \;  \includegraphics[width=3.5in]{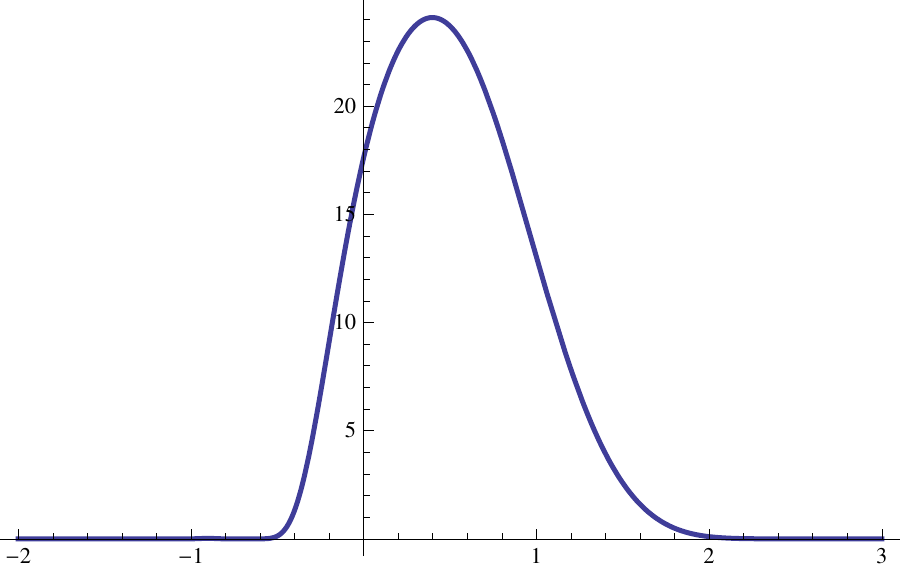}}
\caption{Left: Density plot of $|\Psi_{HH}[\rho,m_{S^3}]|^2$ for $N=2$. The fainter peak centered at the origin reproduces perturbation theory in the empty de Sitter vacuum. Horizontal lines are lines of constant $m_{S^3}$ and vertical lines are lines of constant $\rho$. Right: Plot of $|\Psi_{HH}[\rho,-2.25]|^2$ for $N=2$. Notice that it peaks away from $\rho = 0$.}\label{sqM}
\end{center}
\end{figure}

It is worth studying what happens to the zeroes and local maxima of $Z_{free}$ in the presence of an additional deformation. A computationally convenient deformation is to squash the round metric on $S^3$ into that of a squashed sphere, which is a homogeneous yet anisotropic geometry. In this sense, this deformation is complementary to the inhomogeneous deformations we have been studying so far. We review the metric and eigenvalues of the squashed sphere with squashing parameter $\rho$ in appendix \ref{squashed}. Our method of regularization is a straightforward extension of heat kernel techniques used in section 3.2 of \cite{Anninos:2012ft}, and details can be found therein. In figure \ref{sqM} we present a plot of the wavefunction as a function of the mass $m_{S^3}$ and the squashing parameter $\rho$ (the round metric on $S^3$ occurs at $\rho = 0$). 

We find that the local maxima are in general pushed away from $\rho = 0$ and the zeroes of the wavefunction are extended to enclose the local maxima. The fact that the zeroes are extended (codimension one in the $(\rho, m_{S^3})$ plane) is perhaps not surprising given that the zeroes in figure \ref{sigmafig} arise from $\Psi_{HH}[{m_{S^3}}]$ (which is purely real) changing sign. This feature should not disappear, at least for small perturbations to the equation which determines $\Psi_{HH}[{m_{S^3}}]$. Furthermore, the maxima of figure \ref{sigmafig} are pushed even higher when squashing is allowed. This can be seen in the right hand side of figure \ref{sqM}, where one notices that the second local maximum of $\Psi_{HH}[{m_{S^3}}]$ at $\rho = 0$ is pushed further up to some $\rho >0$. This observation strongly suggests that away from the origin, and under such extreme conditioning of the late time profiles of the bulk fields, the wavefunction peaks in regions where the metric and perhaps even the higher spin fields are highly excited.  On the other hand, the perturbative de Sitter saddle centered at the origin remains a true local maximum. 
\begin{figure}
\begin{center}
{\includegraphics[width=3.1in]{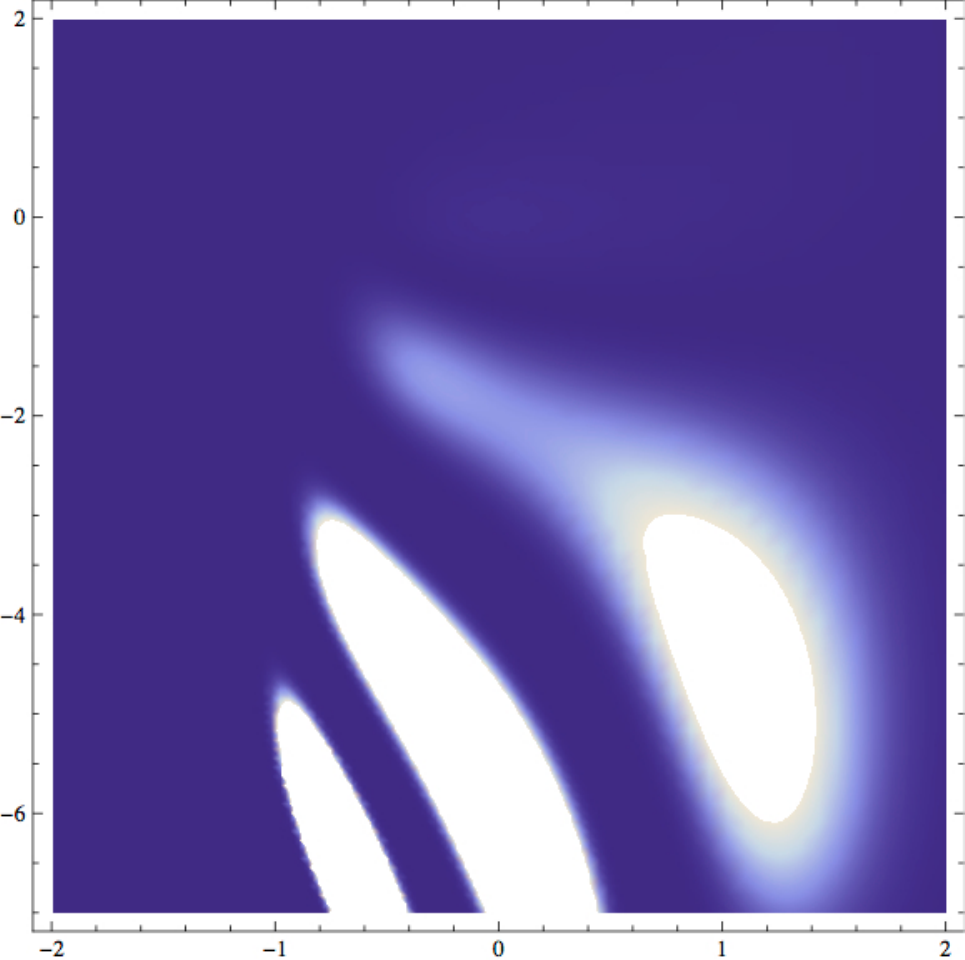}  \quad \includegraphics[width=3.1in]{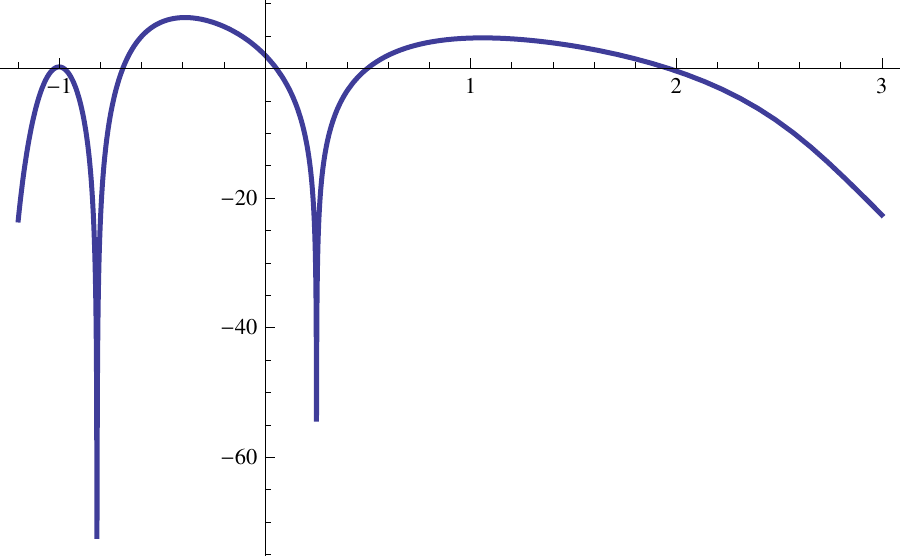} }
\caption{Left: Density plot of $|\Psi_{HH}[\rho,m_{S^3}]|^2$ for $N=2$ for a slightly larger range. Notice the local de Sitter maximum visible in figure \ref{sqM} is already too faint to be seen. Horizontal lines are lines of constant $m_{S^3}$ and vertical lines are lines of constant $\rho$. Right: Plot of $2 \log |\Psi_{HH}[\rho,-4.5]|$ for $N=2$.}\label{sqMfull}
\end{center}
\end{figure}

\section{Basis change, critical $Sp(N)$ model, double trace deformations} \label{secthreehalf}

In this section we further discuss the transformation to the field basis and clarify the role of double trace deformations. We would like to comment on the transform to the bulk field basis, which gives us the wavefunctional as a functional of $\nu \equiv \sqrt{N} \tilde{\sigma}$ (with $\nu$ defined in (\ref{fg}) and $\tilde{\sigma}$ being related the source of the single-trace operator dual to the bulk scalar), in the large $N$ limit. 

As discussed in  \cite{Anninos:2012ft}, one has to consider the free theory deformed by a relevant double trace operator $f (\chi \cdot \chi)^2/(8N)$. We also keep a source $- i f \tilde{\sigma}$ turned on for the single-trace $\chi \cdot \chi$ operator. The parameter $f \in \mathbb{C}$ has units of energy.
Performing a Hubbard-Stratonovich transformation by introducing an auxiliary scalar field $\sigma$ we find:
\begin{equation}
S^{(f)} = \frac{1}{2} \int d^3x \sqrt{g} \left( \Omega_{AB} \left( \partial_i \chi^A \partial_j \chi^B g^{ij} - i f \tilde{\sigma} \chi^A \chi^B\right)   - \left( \frac{N \sigma^2}{f} - \sigma \; \Omega_{AB} \chi^A \chi^B  \right) \right)~.
\end{equation}
Integrating out the $\chi^A$ fields, the partition function becomes:
\begin{equation}\label{critz}
Z^{(f)}_{crit} [\tilde{\sigma}] = e^{-\frac{N f}{2}  \int d^3 x \sqrt{g} \; \tilde{\sigma}^2 }\int \mathcal{D} \sigma \; \exp \left[ N \int d^3 x \sqrt{g} \left(  \frac{ \sigma^2}{2 f} + i \tilde{\sigma} {\sigma} \right) \right] Z_{free}  [\sigma]~,
\end{equation}
where we have transformed variables $\sigma \to ( \sigma - i f \tilde{\sigma})$. Having interpreted $Z_{free}[\sigma] = \langle \sigma | E \rangle = \Psi_{HH}[\sigma]$, we see that the above expression is a Fourier type transform of the $\hat{\sigma}$-basis wavefunction. In order for this to become the actual basis changing transform (\ref{tranqminv}) to the eigenbasis of the field operator, the constant $f$ must be taken to infinity. 
The $f \to \infty$ limit (where we are keeping the size of the three-sphere fixed) corresponds to sending the ultraviolet cutoff (of the infrared fixed point theory) to infinity, in the same sense as \cite{Wilson:1972cf}, or roughly speaking it corresponds to the late time limit in the bulk.\footnote{Another way to think about this is keeping $f$ fixed and  taking the large size limit of the three-sphere that the CFT lives on. This is because the dimensionless quantity is $|f| a$ where $a$ is the size of the sphere. Indeed at late times the three-sphere grows large.} 

We begin by performing a perturbative analysis for infinitesimal deformations of the free $Sp(N)$ theory at large $N$ on an $\mathbb{R}^3$.


\subsection{Perturbative analysis on $\mathbb{R}^3$}

For the sake of simplicity, we will put the theory on the flat metric on $\mathbb{R}^3$, akin to studying perturbations in a small piece of $\mathcal{I}^+$.\footnote{The parallel story in anti-de Sitter space has been studied extensively \cite{Witten:2001ua,Mueck:2002gm,Gubser:2002zh,Gubser:2002vv}. From the bulk perspective in planar AdS$_4$: $ds^2 = \ell_A^2 (dz^2 + d\vec{x}^2)/z^2$, at least perturbatively about the empty AdS$_4$ vacuum, the finite $f_A \in \mathbb{R}$ double trace deformed theory computes correlation functions of the bulk scalar quantized with mixed boundary conditions. Near the boundary $z \to 0$ of AdS$_4$, the bulk scalar with mass $m^2\ell_{A}^2 = -2$ behaves as $\phi(z,\vec{x}) \sim \alpha(\vec{x}) z^2 + \beta(\vec{x}) z$ with $\alpha(\vec{x}) = f_A \beta(\vec{x})$. This boundary condition is different from the conformally invariant one which sets either $\alpha(\vec{x})$ or $\beta(\vec{x})$ to zero, corresponding to the free or critical $O(N)$ models respectively. In de Sitter space we would consider a wavefunction of a scalar of mass $m^2\ell^2 = +2$ computed by imposing future boundary conditions \cite{Maldacena:2002vr,Harlow:2011ke,Anninos:2011jp} $\phi(\eta,\vec{x}) \sim \alpha(\vec{x}) \eta^2 + \beta(\vec{x}) \eta$ (with $\alpha(\vec{x}) = f_D \beta(\vec{x})$) and Bunch-Davies conditions $\phi \sim e^{i k \eta}$ for $k |\eta| \gg 1$. At the level of perturbation theory this is computed by continuing the Euclidean AdS$_4$ partition function by $z \to -i\eta$, $\ell_{A} \to i \ell$ and $f _A \to i f_D$ (where $f_D \in \mathbb{R}$ for the `normalizable' profile of the scalar field to be real).} For reasons that will be clear momentarily we choose $f= i |f|$ to be pure imaginary. 
 
We are interested in setting up a perturbative expansion about the $\sigma \sim 0$ Gaussian peak of $Z_{free}[\sigma]$. From (\ref{critz}) 
we can compute the two point function of $\mathcal{O} \equiv \chi \cdot \chi$ at large $N$ in the double trace deformed $Sp(N)$ theory on an $\mathbb{R}^3$. We do this by taking two variational derivatives of the logarithm of (\ref{critz}) with respect to $\tilde{\sigma}(x^i)$ and evaluating at $\tilde{\sigma}(x^i)=0$:
\begin{equation}
\langle \mathcal{O}_{\vec{k}} \mathcal{O}_{-\vec{k}} \rangle_f = {N} \left( \frac{|f|^2 G(k)}{N+2 i |f| G(k)} \right)~, \quad G(k) \equiv \langle \mathcal{O}_{\vec{k}} \mathcal{O}_{-\vec{k}} \rangle_{f=0} = - \frac{N}{k}~,
\end{equation}
where $k$ is the magnitude of the momentum. For $k \gg |f|$ this becomes the two-point function of the free $Sp(N)$ model, whereas for $k \ll |f|$ this becomes the two-point function of the critical $Sp(N)$ model. Expanding for large $|f|$ we find:
\begin{equation}\label{2pf}
\langle \mathcal{O}_{\vec{k}} \mathcal{O}_{-\vec{k}} \rangle_f = - i \frac{N |f|}{2} - \frac{N}{4} k + \ldots
\end{equation}
Notice that the real part of the two-point function (\ref{2pf}) is negative. Recalling (\ref{pertdscft}), this is the Gaussian suppression of the Hartle-Hawking wavefunction near the de Sitter vacuum. Also, the local momentum independent term has become a phase for pure imaginary $f$. We can compare this to the bulk Hartle-Hawking wavefunction for a free $m^2\ell^2 = +2$ scalar in planar coordinates computed in (\ref{planarwf}) of the appendix. This allows us to (roughly) identify $|f|^{-1}$ with the late time cutoff $|\eta_c|$ at large $|f|$. In a similar way we can compute the rest of the perturbative correlators of the critical $Sp(N)$ theory \cite{LeClair:2007iy}.

Beyond such perturbative analyses, we must resort to a saddle point approximation which we now proceed to.

\subsection{Large $N$ saddles for uniform $S^3$ profiles} 

We now put the theory on the round metric on $S^3$. At large $N$, we can evaluate (\ref{critz}) by solving the saddle point equation (for $\sigma = \sigma(\tilde{\sigma})$ and $\tilde{\sigma}$ uniform over the whole three-sphere):
\begin{equation}\label{saddle}
\frac{16\pi \sigma}{f} + 16 \pi i \tilde{\sigma} = \sqrt{1 - 4 \sigma} \cot \left( \frac{\pi}{2} \sqrt{1 - 4 \sigma} \right)~.
\end{equation} 
For a given solution $\Sigma_i$ of (\ref{saddle}), we can evaluate $Z_{crit}[\tilde{\sigma},\Sigma_i]$. For example, there is a solution $\Sigma_0$ where $\sigma \sim 0$ when $\tilde{\sigma} \sim 0$ (with $f \to \infty$). It is the piece of the wavefunctional evaluated from this saddle near $\tilde{\sigma} = 0$ that reproduces the dS invariant perturbation theory in the Bunch-Davies state (about the pure de Sitter vacuum). 

\begin{figure}
\begin{center}
{\includegraphics[width=2.1in]{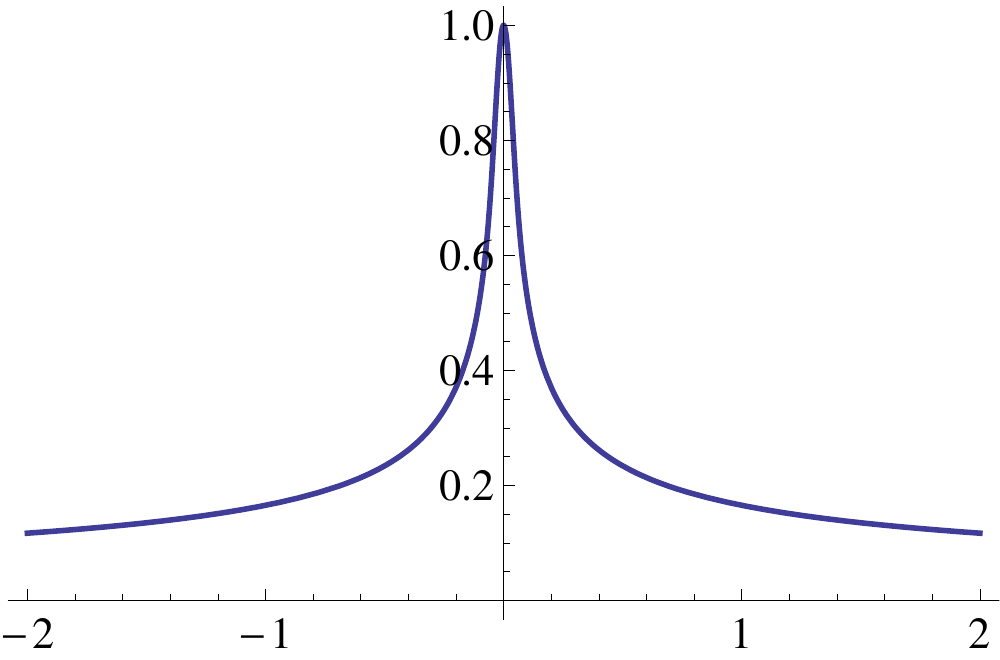} \includegraphics[width=2.1in]{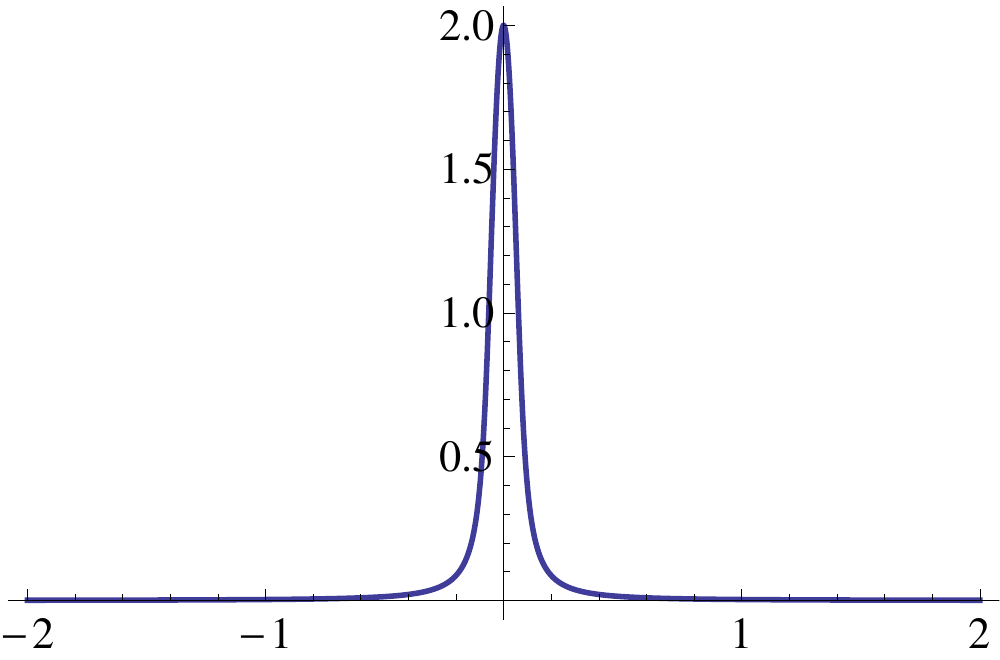} \includegraphics[width=2.1in]{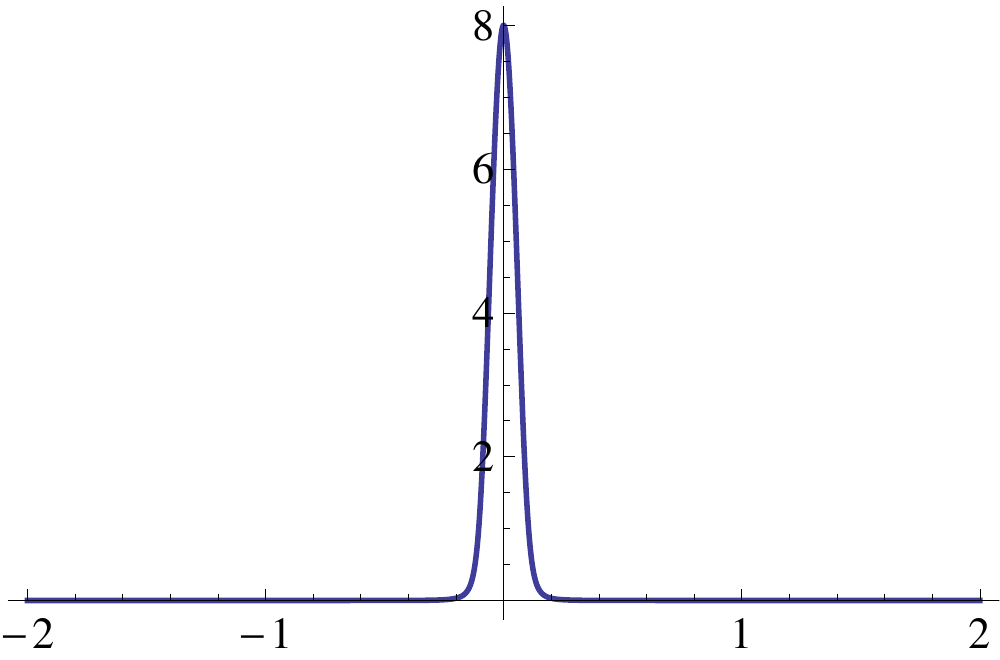}}
\caption{Plots of the $N$th root of $|\Psi_{HH}[\tilde{\sigma},\Sigma_0]|$, $|\Psi_{HH}[\tilde{\sigma},\Sigma_1]|$, $|\Psi_{HH}[\tilde{\sigma},\Sigma_2]|$ at large $f$ from left to right. Notice that the higher saddles dominate near $\tilde{\sigma}=0$ but fall off faster for large $\tilde{\sigma}$.}\label{saddlesfig}
\end{center}
\end{figure}

In figure \ref{saddlesfig} we plot $Z_{crit}$ for the first few $\Sigma_i$ at large $f = i |f|$.
These have $\tilde{\sigma} = 0$ near the subsequent zeroes of $\cot \left( \frac{\pi}{2} \sqrt{1 - 4 \sigma} \right)$ in (\ref{saddle}). Notice that for all large $N$ saddles $Z_{crit}[\tilde{\sigma}]$ is peaked at $\tilde{\sigma} = 0$ but the saddles coming from the more negative $\sigma$ peaks contribute more near $\tilde{\sigma} = 0$. 

Away from the large $N$ limit we must compute the integral in (\ref{critz}) without resorting to a saddle point approximation. If we restrict to uniform $\sigma$ and $\tilde{\sigma}$, we note that as we increase $f=i|f|$ more and more of the growing negative $\sigma$ peaks in $Z_{free}[\sigma]$ contribute to the integral before it is cutoff by the rapid oscillations due to the $i N \sigma^2/|f|$ piece. One can check that the integral grows for large and negative $\tilde{\sigma}$ upon fixing $|f|$.

Some of the saddles in the large $N$ limit should correspond to classical (complex) bulk solutions with a uniform late time profile of the scalar on the round metric of the three-sphere. Some of these solutions labelled by a continuous parameter, which only involve the bulk metric and scalar, were found by Sezgin and Sundell in \cite{Sezgin:2005hf}. It is worth noting that though it may sound confusing that there are bulk solutions that have only the metric and scalar turned on (since all higher spin fields interact on an equal footing), this is natural from the CFT since turning on an $SO(4)$ symmetric source for $J^{(0)} = \chi \cdot \chi$ need not source the traceless higher spin currents due to symmetry reasons. The metric is non-vanishing since in effect we have also turned on a source for it by having the round metric on $S^3$ at the boundary.
At finite $N$, all these saddles mix quantum mechanically.  Each $\Psi_{HH}[\tilde{\sigma},\Sigma_i]$ comes with a phase, so one should be careful when summing contributions from different saddles. Finally, it would be extremely interesting to understand the Lorentzian cosmologies associated to the wavefunction using the ideas developed in \cite{Hartle:2008ng,Hartle:2007gi} (see also \cite{Sarangi:2005cs}). In order to have a classical cosmology (or an ensemble of such cosmologies) we must ensure that the wavefunction takes a WKB form with a phase oscillating much more rapidly than its absolute value. At least in the large $N$ limit, and for large $|f|$, this is ensured by the first term in (\ref{critz}).

\subsection{Double trace deformations as convolutions}

We can also consider keeping $f$ finite and real. This defines a double-trace deformed field theory, in and of its own right, whose partition function $Z^{(f)} [{\sigma}']$ can be computed in the large $N$ limit. We also keep a uniform (on $S^3$) source ${\sigma}'$ turned on for the single-trace $\chi \cdot \chi$ operator. 
This partition function is no longer computing overlaps between the Bunch-Davies vacuum and some late time field configuration which is an eigenstate of the field operator $\hat{\phi}$. Given that $Z_{free}[\sigma] = \langle \sigma | E \rangle = \Psi_{HH}[\sigma]$, we see that $Z^{(f)} [{\sigma}']$ is computing instead a convolution of the wavefunction in the $\hat{\sigma}$-basis:\footnote{Though we will not do so here, we can study higher multitrace deformations of the $Sp(N)$ theory and arrive at similar expressions. Thus, the fact that multi-trace operators are irrelevant seems far less threatening than having low-spin, single-trace operators which are irrelevant and correspond to bulk tachyons.}
\begin{equation}
Z^{(f)} [{\sigma}'] = \int \mathcal{D} \sigma \exp \left[ N  \int d\Omega_3 \; \frac{\left({\sigma}' - \sigma \right)^2}{2f}   \right] \Psi_{HH}[\sigma]~.
\end{equation}
We can also view $Z^{(f)} [{\sigma}']$ as computing the overlap of the Hartle-Hawking state with the state:
\begin{equation}
| f \rangle \equiv \int \mathcal{D} \sigma \exp \left[  N  \int d\Omega_3 \; \frac{\left({\sigma}' - \sigma \right)^2}{2f}   \right] | \sigma \rangle~.
\end{equation}
Notice that though the integral itself is convergent for finite $f$, the resulting function $Z^{(f)} [{\sigma}']$ will grow exponentially at large negative ${\sigma}'$. One could also consider more generally a complex valued $f \in \mathbb{C}$ which would correspond to a kind of windowed Fourier transform.

\subsection{Euclidean AdS$_4$ with an $S^3$ boundary}

\begin{figure}
\begin{center}
{\includegraphics[width=3.3in]{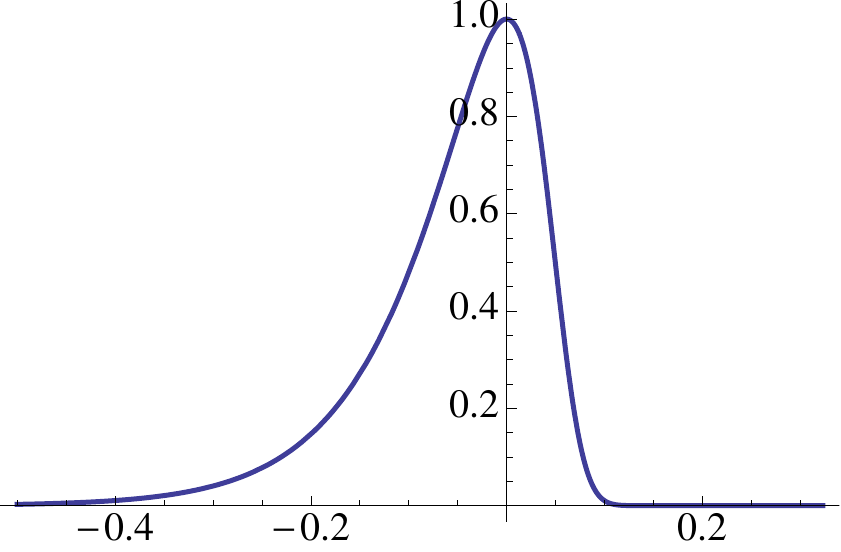}  }
\caption{Plot of the $N^{th}$ root of the finite part of $Z^{O(N)}_{crit}[{\sigma}_A]$ for a uniform source ${\sigma}_A$ over the whole three-sphere. We have normalized such that $Z^{O(N)}_{crit}[{\sigma}_A] = 1$ at ${\sigma}_A=0$.}\label{oncrit}
\end{center}
\end{figure}

The situation can be contrasted with the case of Euclidean AdS$_4$ (with an $S^3$ boundary) \cite{Gubser:2002vv}. The partition function $Z^{O(N)}_{crit}[{\sigma}_A]$ of the critical $O(N)$ model on an $S^3$, dual to Euclidean AdS$_4$ in higher spin gravity, is obtained again from the free $O(N)$ model by a double trace deformation in the limit $f_A \to  + \infty$:
\begin{equation}\label{critzon}
Z^{O(N)}_{crit} [{\sigma}_A] = \lim_{f_A \to\infty} e^{\frac{N f_A}{2}  \int d\Omega_3 {\sigma}_A^2 }\int \mathcal{D} \sigma \; \exp \left[ N \int d\Omega_3 \left( {\frac{ \sigma^2}{2 f_A} - {{\sigma}_A {\sigma}} }\right) \right] Z^{O(N)}_{free}  [\sigma]~, \quad f_A \in \mathbb{R}~.
\end{equation}
The first term in front of the integral is local in the limit $f_A \to +\infty$ and we can remove it by adding a counterterm. To ensure convergence of the integral we must choose an appropriate contour, which in this case is given by $\sigma$ running along the imaginary axis (see for example \cite{Anninos:2012ft}). $Z^{O(N)}_{free}[\sigma]$ is the partition function of the free $O(N)$ model and is related to the free $Sp(N)$ partition function by $N \to -N$. Note that $Z^{O(N)}_{free}$ has poles precisely at the values where the wavefunction in the $\hat{\sigma}$-basis (\ref{s3z}) vanishes. At large $N$ the integral (\ref{critzon}) can be evaluated by a saddle point approximation. In figure \ref{oncrit} we display $Z^{O(N)}_{crit} [{\sigma}_A]$ for the case of a uniform source $f_A {\sigma}_A$ over the whole $S^3$.
In principle, this plot should be reproducible by computing the regularized on-shell Vasiliev action on the asymptotically Euclidean AdS$_4$ Sezgin-Sundell solution \cite{Sezgin:2005hf}.





\section{Extensions of higher spin de Sitter holography?}\label{secsix}

So far, our discussion was restricted to the minimal bosonic higher spin theory. A natural question that arises, particularly given possible interpretational issues of the wavefunction such as its (non)-normalizability, is whether this theory is part of a larger framework. We briefly discuss possible extensions of higher spin de Sitter holography, inspired by the analogous situation in  anti-de Sitter space.

\subsection{AdS$_4$}

There exist parity violating deformations of the bulk equations of motion which deform the original Vasiliev equations (in anti-de Sitter space) to a one-parameter family \cite{Vasiliev:1992av,Vasiliev:1995dn,Vasiliev:1999ba}. It was proposed that the dual description is given by coupling the theory to a Chern-Simons theory with level $k$, at least for simple enough topologies. The new parameter is given by the 't Hooft coupling $\lambda = N/k$ which is small when the dual is higher spin gravity. Such a theory was shown in the infinite $N$ limit with fixed small $\lambda$ \cite{Aharony:2011jz,Chang:2012kt,Giombi:2011kc} to have a spectrum of single-trace operators which is precisely that of the free $U(N)$ model, namely a tower of higher spin currents which are conserved (up to $\mathcal{O}(1/\sqrt{N})$ corrections), in accordance with a bulk higher spin theory. 

As discussed in \cite{Chang:2012kt,Giombi:2012ms} in the context of anti-de Sitter higher spin gravity, one can also endow the bulk higher spin fields with $U(M)$ Chan-Paton factors, such that the fields all lie in the adjoint representation of the $U(M)$. In the dual theory, this corresponds to adding a $U(M)$ flavor symmetry to the $U(N)$ model. If the flavor symmetry is weakly gauged, which can be achieved through the procedure described in \cite{Witten:2003ya}, then one can form single-trace operators of the form $\text{Tr} \left( A B A B \ldots A B\right)$. The fields $A$ and $B$ transform in the $(\square,\overline{\square})$ and $(\overline{\square},\square)$ of the $U(N) \times U(M)$ gauge group respectively. As $M/N$ increases the `glue' between each $\text{Tr} AB$ (which are dual to higher spin fields in the bulk) becomes stronger. From the bulk point of view, it was suggested that the higher spin fields, now endowed with additional $U(M)$ interactions, form bound states with binding energy that increases as we crank up $M/N$. These bound states would correspond to the $\text{Tr} \left( A B A B \ldots A B\right)$ operators. An appropriately supersymmetrized version of this story \cite{Chang:2012kt,Giombi:2012ms} was conjectured to connect the higher spin (supersymmetric) theory to the ABJ model \cite{Aharony:2008gk} where such long string operators are dual to bulk strings. 

\subsection{dS$_4$}

It is convenient in our discussion for the bulk to contain a spin-one gauge field in its spectrum. We thus consider the non-minimal higher spin model with even and odd spins whose dual (at least at the level of correlation functions on $\mathbb{R}^3$) is a free/critical $U(N)$ theory with $N$ anti-commuting scalars transforming as $U(N)$ vectors. We refer to this as the $\tilde{U}(N)$ model. 

Given that the parity violating deformations of the Vasiliev equations in AdS retain the original field content and reality conditions, it seems natural that they are present for the de Sitter theory as well. Thus, one might consider that such theories are dual to parity violating extensions of the $\tilde{U}(N)$ theory obtained by adding a level $k$ Chern-Simons term to the free anti-commuting complex scalars. The Lagrangian of this theory is:
\begin{equation}\label{csds}
S_{CSM} = - \frac{i k}{4\pi} \int d^3 x  \left( \epsilon_{ijk} \; A^a_i \partial_{i} A^a_{k}  + \frac{1}{3}\epsilon_{ijk} f^{abc} A^a_i A^b_j A^c_k \right) + \int d^3 x \; \nabla_i \chi^{A} \nabla^i \bar{\chi}^{\bar{A}}~.
\end{equation}
The fields $A_i^a$ are (possibly complexified) $U(N)$ gauge fields, the $\chi^{A}$ fields are anti-commuting complex scalars transforming in the fundamental of the $U(N)$ gauge symmetry, and the $\nabla_i$ derivative is covariant with respect to $U(N)$ gauge transformations. 

Though classically this theory is conformally invariant, this need not be the case when we include loops, as the $\beta$-function might be non-vanishing. The ordinary $U(N)$ Chern-Simons theory coupled to a vector of charged scalars
has two exactly marginal deformations at infinite $N$ as in \cite{Aharony:2011jz,Maldacena:2011jn,Maldacena:2012sf}. These are the 't Hooft coupling $\lambda \equiv N/k$ and the coupling constant $\lambda_6$ of the triple trace interaction $(\chi^A \bar{\chi}^{\bar{A}})^3$. 
Though the Chern-Simons-$\tilde{U}(N)$ theory (\ref{csds}) is non-unitary, it is conceivable that it is also has a vanishing $\beta$-function at large $N$ \cite{workinprogress}.
This theory would also have an unchanged spectrum of single-trace operators at large $N$ and small $\lambda$ by the same arguments as those in \cite{Aharony:2011jz,Girardello:2002pp}.\footnote{At finite $k$, the partition function on a non-trivial topology such as $\mathcal{M}  = S^1 \times \text{Riemm}_g$ will grow as \cite{Banerjee:2012gh} (see also \cite{Radicevic:2012in}): $Z_{CFT} \sim \exp\left({(g-1)N^2 \log{k}+\mathcal{O}(N)}\right)$ where $g$ is the genus of the $\mathcal{M}$. This drastically favors higher topologies if interpreted as a probability, but it is unclear whether and how one should compare topologies and what the correct normalization for $\Psi_{HH}$ is. It is interesting that at finite $k$ one might also encounter monopole operators. In ABJM, such operators are dual to D0-branes in the bulk. It is unclear how they should be understood in the context of de Sitter space and higher spin gravity. For instance, they have a conformal weight that goes like $k$, which might suggest taking $k$ to be complex or imaginary \cite{Witten:2010cx} in the de Sitter case. It is also worth noting that the potentially infinite wealth of topological data at $\mathcal{I}^+$ might be at odds with the finiteness of de Sitter entropy \cite{Banks:2005bm}. In Einstein gravity adding too much topology at $\mathcal{I}^+$ often results in bulk singularities \cite{Andersson:2002nr}. The issue of topology in the context of dS/CFT is further discussed in \cite{foam}.}

One can also consider adding $U(M)$ Chan-Paton factors to the bulk higher spin de Sitter theory. This merely requires tensoring the $*$ algebra with that of $M\times M$ matrices. Once again, this will not affect the reality conditions on the higher spin fields and they will all transform in the adjoint of the $U(M)$. This corresponds to adding a $U(M)$ flavor symmetry to the $\tilde{U}(N)$ vector model, which can be weakly gauged (see appendix \ref{u1appendix} for a discussion). The single-trace operators $\text{Tr} \left( A B A B \ldots A B\right)$ have increasingly real conformal weight. From the point of view of dS/CFT this would imply that the bulk theory has a tower of tachyonic bulk fields since the conformal weight of a bulk field goes as $\Delta_\pm \sim 3/2 \pm \sqrt{9/4-m^2\ell^2}$. One might suspect that these will be the continuations of the higher spin bound states previously discussed for the anti-de Sitter case. Thus we see that even though the fundamental constituents (i.e. the higher spin particles) of such an extension of higher spin de Sitter gravity are not pathological (at least at the level of perturbation theory), they may form configurations which resemble tachyonic fields in de Sitter space. It is of interest to understand whether the late time behavior of such a theory can ever be asymptotically de Sitter \cite{workinprogress}. From the CFT point of view these are highly irrelevant operators which are {\it not} conserved currents. For there to be a late time de Sitter phase, one would require that turning on such irrelevant deformations can flow the theory to a UV fixed point. 

\section*{Acknowledgements}

It is a great pleasure to thank Tarek Anous, Shamik Banerjee, Gerald Dunne, Daniel Harlow, Jim Hartle, Tom Hartman, Sean Hartnoll, Simeon Hellerman, Thomas Hertog, Diego Hofman, Juan Maldacena, Alex Maloney, Steve Shenker, Eva Silverstein, Andy Strominger and Lenny Susskind for useful discussions. The authors are also grateful to the ``Cosmology and Complexity" conference in Hydra for their hospitality while this work was in progress. This work has been partially funded by DOE grant DE-FG02-91ER40654 and by a grant of the John
Templeton Foundation. The opinions expressed in this publication are those of the authors and do not necessarily reflect the views of the John Templeton Foundation.

\appendix

\section{Conformal  Transformation from round $S^3$ to flat $\mathbb{R}^3$}\label{r3s3}

In this appendix, we remind the reader of the conformal relation between the three-sphere and the three-plane. 

\subsection{Coordinate transformation and Weyl rescaling}

Consider the metric of the three-sphere:
\begin{equation}
ds^2 = d\psi^2 + \sin^2 \psi \; d\Omega_2^2~.
\end{equation}
Upon a coordinate transformation, $\psi(r) = 2 \cot^{-1} r^{-1}$, the above metric maps to:
\begin{equation}
ds^2 = \left( \frac{2}{ r^2 + 1 } \right)^2 \left( dr^2 + r^2  d\Omega_2^2 \right)~,
\end{equation}
which is conformally equivalent to the flat metric on $\mathbb{R}^3$. According to our discussion in section \ref{wf}, a constant mass source $m_{S^3}$ in the free $Sp(N)$ theory on a three-sphere corresponds to the free $Sp(N)$ theory on the flat metric on $\mathbb{R}^3$ with the following source for the $\chi \cdot \chi$ operators:
\begin{equation}\label{ms3}
m_{\mathbb{R}^3}(r) =  \left( \frac{2}{r^2 + 1} \right)^2 m_{S^3}~.
\end{equation}

\subsection{Numerical Error}

\begin{figure}
\begin{center}
{\includegraphics[height=2in]{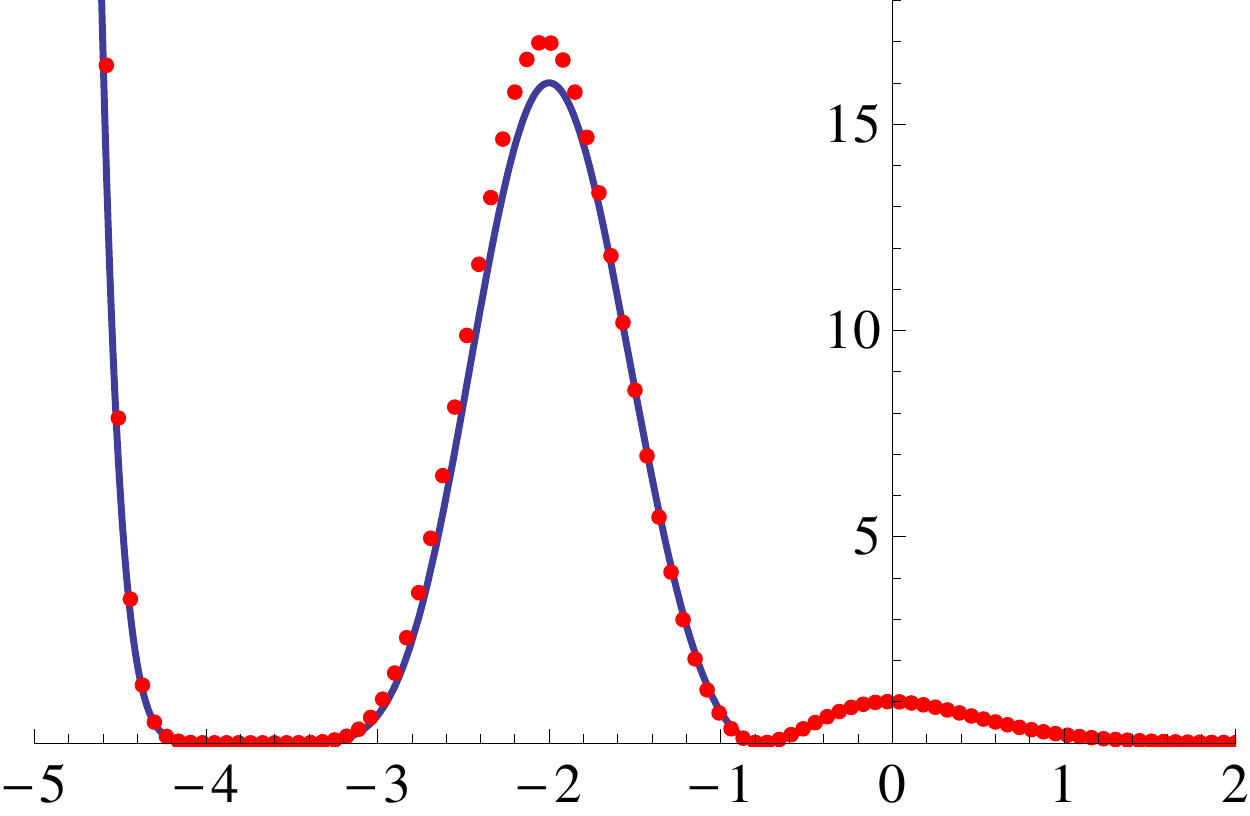} \quad {\includegraphics[height=2in]{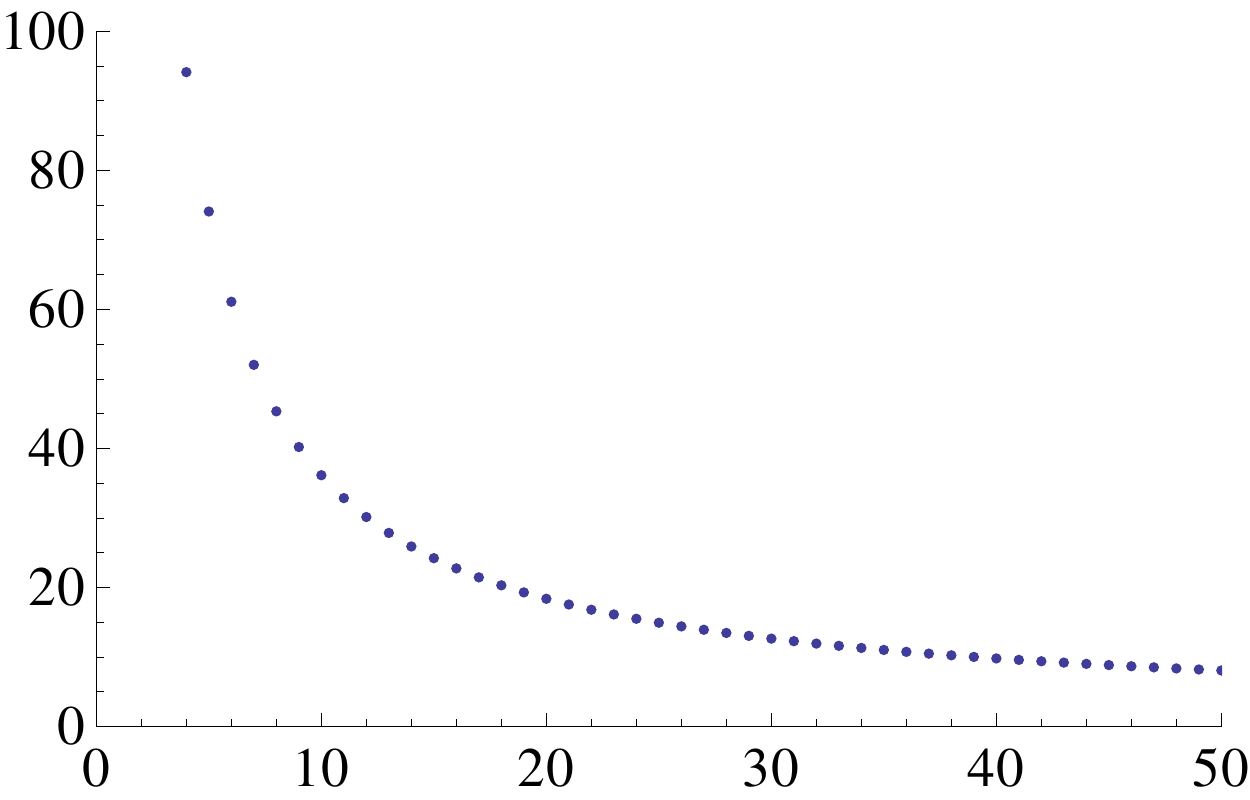}}}
\caption{Left: Comparison of $|\Psi_{HH}(m_{S^3})|^2$ for $N=2$ as obtained by calculating ${Z}_{CFT}[m_{S^3}]$ analytically (blue line) and given in equation (\ref{s3z}), and by numerically evaluating the functional determinant using the Dunne-Kirsten regularization method with $l_{max}=45$ (red dots). We have normalized such that $|\Psi_{HH}(m_{S^3})|^2 = 1$ at $m_{S^3}=0$. Right: Plot of the percentage error as a function of the numerical cutoff (at $m_{S^3}=-2.2$).}\label{ms3plot}
\end{center}
\end{figure}

As a check on our numerics, we display in figure \ref{ms3plot} a plot of the functional determinant using the Dunne-Kirsten method for the radial mass given in (\ref{ms3}) laid over the analytic result for the three sphere written in (\ref{s3z}). In the right hand side of the figure we display the error between the two as a function of the cutoff value of $l$ for the sum in (\ref{dunne}). For a maximum cutoff $l_{max}=200$ the partition function is within 3 percent of the exact answer at $m_{S^3}=-2.2$.

\subsection{Balloon Geometries}\label{balloonapp}

We can use the same conformal transformation between the round metric on $S^3$ and the flat metric on $\mathbb{R}^3$ to show the conformal equivalence of the balloon geometry on an $\mathbb{R}^3$ topology discussed in section \ref{balloonsec} and a different geometry on an $S^3$ topology:
\begin{align}
ds^2 = dr^2 + r^2 f_\zeta(r)^2 d\Omega^2_2 =  \left(2 \cos^2(\psi/2)\right)^{-2}(d\psi^2 + \sin^2 \psi \; f_\zeta(\psi)^2 d\Omega_2^2)~.
\end{align}
Upon a conformal rescaling, we can see that this is just a deformed three-sphere:
\begin{equation}\label{balloontopeanut}
d\tilde{s}\,^2\equiv\left(2\cos^2(\psi/2)\right)^2 ds^2=d\psi^2 + \sin^2 \psi \; f_\zeta(\psi)^2 d\Omega_2^2~.
\end{equation}
The function $f_\zeta(\psi)$ serves to add a waist to the three-spheres along the $\psi$ direction, with the parameter $\zeta$ controlling whether the sphere tapers or bulges. Thus, the mass deformation chosen in section \ref{balloonsec} for the balloon geometry corresponds to a constant mass deformation on its conformally related deformed sphere. When mapping the balloon geometry to flat space to apply the Dunne-Kirsten method for computing the partition function, we should understand that the answer we recover is also the answer for the partition function on this deformed sphere with a \emph{constant} mass. The pinching limit of the balloon geometry maps into a pinch that tears the peanut-like geometry into two warped spheres.

\section{Review of the squashed sphere}\label{squashed}

The metric of the squashed sphere, which is a homogeneous yet anisotropic space on an $S^3$ topology, is given by:
\begin{equation}
ds^2 = \frac{1}{4} \left( d\theta^2 + \cos^2\theta d\phi^2 + \frac{1}{1+\alpha}\left( d\psi + \sin\theta d\phi \right)^2 \right)~,
\end{equation}
with $\psi \sim \psi + 4\pi$. The geometry is an $S^1$ fiber over $S^2$ and consequently has an $SO(3) \times U(1)$ isometry group. The constant Ricci scalar and volume are given by:
\begin{equation}
R = \frac{2(3+4\alpha)}{(1+\alpha)}~,  \quad V = \frac{2\pi^2}{\sqrt{1+\alpha}}~.
\end{equation}
At $\alpha = 0$ we recover the $S^3$ with enhanced $SO(4)$ isometry group. It is convenient to parametrize the squashing parameter by $\rho$ such that $\alpha = e^{2\rho} - 1$, and we will display our results using this parameter. The eigenvalues of the conformal Laplacian in the presence of a uniform mass term $\sigma$ are given by:
\begin{equation}
\lambda_{n,q} = \left[ n^2 - \frac{1}{4(1+\alpha)} + \alpha \left( n - 1 - 2 q \right)^2 + \sigma \right]~, \; q = 0,1,\ldots,n-1~, \; n = 1,2,\ldots~,
\end{equation}
with degeneracy $n$. Knowing the eigenvalues analytically allows for easier computation of the functional determinant.

\section{Perturbative Bunch-Davies modes for $m^2 \ell^2 = + 2$}\label{bdapp}

We briefly review the Bunch-Davies zero modes of a free conformally coupled scalar with $m^2\ell^2 = +2$ in a fixed global dS$_4$ background:
\begin{equation}
ds^2 = -d\tau^2 + \ell^2 \cosh^2 \frac{\tau}{\ell} \; d\Omega^2_3~.
\end{equation}
The action of the scalar is given by:
\begin{equation}
S_\phi = - \frac{1}{2} \int d^4x \sqrt{-g} \left( g^{\mu\nu} \partial_\mu \phi \partial_\nu \phi + \frac{2}{\ell^2} \phi^2 \right)~.
\end{equation}
Assuming $\phi = \phi(\tau)$ does not depend on the three-sphere coordinates, it is governed by the equation of motion:
\begin{equation}
\frac{2}{\ell} \; \phi (\tau) + {3 \tanh \frac{\tau }{\ell} \; \phi '(\tau )} + \ell \; \phi ''(\tau) = 0~.
\end{equation}
The general solution to the above equation is given by:
\begin{equation}
\phi(\tau) = \sech^2 \frac{\tau }{\ell}\left(c_2+{c_1} \sinh \frac{\tau }{\ell}\right)~.
\end{equation}
The positive frequency modes of the Bunch-Davies vacuum are those which are regular on the lower Euclidean hemisphere $\theta \in [-\pi/2,0]$ obtained by continuing the global metric by $\frac{\tau}{\ell} \to - i \theta$.\footnote{Notice that the continuation $\tau/\ell\to i\theta$ with the same regularity condition would lead to a wavefunctional \ref{wavefunctional} that diverges with the late-time profile of the scalar. It is thus an inappropriate choice for a perturbatively stable vacuum state.} This fixes $c_1 = i c_2$ resulting in:
\begin{equation}\label{bdsol}
\phi_{BD}(\tau) = c_2 \; \sech^2 \frac{\tau }{\ell}\left(1+ i \; \sinh \frac{\tau }{\ell}\right)~.
\end{equation}
We can expand the Bunch-Davies solution at late times to find:
\begin{equation}
\phi_{BD}(\tau) \sim 2 \; c_2 \left( 2 \; e^{-2\tau/\ell} + i \; e^{-\tau/\ell} \right)~.
\end{equation}

\subsection{Continuation from Euclidean AdS$_4$}

Another way to see this is by continuing the perturbative solutions of a conformally coupled free scalar with mass $m^2 \ell^2_{A} = -2$ in a fixed Euclidean AdS$_4$ background:
\begin{equation}
ds^2 = d\rho^2 + \ell_{A}^2 \sinh^2 \frac{\rho}{\ell_{A}} d\Omega_3^2~.
\end{equation}
which are smooth in the interior. These scalars obey the equation:
\begin{equation}
\frac{2 }{\ell_{A}}\;\phi (\rho)+3 \coth \frac{\rho }{\ell_{A}} \; \phi' (\rho)+ \ell_{A} \; \phi''(\rho) = 0~.
\end{equation}
The smooth solution in the interior is given by:
\begin{equation}
\phi_{smooth}(\rho) = \tilde{c}_2 \; \text{csch}^2 \frac{\rho }{\ell_A} \left(-1+ \cosh \frac{\rho}{\ell_A} \right)~.
\end{equation}
Near the boundary of Euclidean AdS the smooth solution behaves as:
\begin{equation}
\phi_{smooth}(\rho) \sim - 2 \; \tilde{c}_2  \left( 2 \; e^{-2\rho/\ell_A} - e^{-\rho/\ell_A}  \right)~.
\end{equation}
Notice that the Euclidean AdS and global dS metrics map onto each other under the transformation $\ell_A \to i \ell$ and $\rho \to i \tau+\ell\pi/2$. Under this analytic continuation the smooth solution maps to the Bunch-Davies solution. 

\subsection{Wavefunctional}

We can compute the Hartle-Hawking wavefunctional using the complex solutions in (\ref{bdsol}). We choose $c_2$ such that the solution at some late time cutoff $\tau = \tau_c \gg \ell$ has the real profile $\phi_0$. Evaluating the on-shell action of a free scalar field with mass $m^2\ell^2 = 2$ at late times and computing $e^{i S_\phi}$ gives:
\begin{equation}\label{wavefunctional}
\Psi_{HH} [\phi_0,\tau_c] \sim \exp \left[ - \frac{\ell^2 \pi^2}{4} \phi_0^2  \left(\;  e^{2\tau_c/\ell} + \ldots \right) \right] \; \exp \left[- i \frac{\ell^2 \pi^2}{8}  \phi^2_0 \left( \; e^{3\tau_c/\ell} + \ldots \right) \right]~, 
\end{equation}
where we have separated the phase from the magnitude and expanded in powers of $e^{\tau/\ell}$. Notice that the wavefunction is Gaussian suppressed and the phase diverges at late times. Picking a convenient overall normalization for $\hat{\sigma}$, we define $\hat{\sigma}=e^{2\tau_c/\ell}\,\left(\hat{\phi}_0-\alpha \, \hat{\pi_{\phi}}\right)$ and fix $\alpha = (e^{-3\tau_c/\ell})/8$ so that the late-time behavior of $\sigma$ is purely fast-falling (this is the choice for which the late-time correlators in the bulk are computed by a CFT). We use the same convention for $\hat{\pi_{\phi}}$ as in (\ref{convention2}). For $\tau_c \gg \ell$ we find:

\begin{equation}
\Psi_{HH}[\sigma,\tau_c] \sim \exp \left[ - \frac{\ell^2 \pi^2}{16} \sigma^2 + \ldots \right]~. 
\end{equation}
Notice that at the level of Gaussian wavefunctionals, the phase vanishes in the $\hat{\sigma}$-basis. 

For the sake of completeness we also include below the wavefunctional of the $m^2\ell^2=+2$ free scalar in planar coordinates $ds^2 = \ell^2 (-d\eta^2 + d\vec{x}^2)/\eta^2$. If its late time profile at some cuttoff $\eta = \eta_c$ is given by $\phi_{\vec{k}}$ then we have:
\begin{equation}\label{planarwf}
\log \Psi_{HH} [\phi_{\vec{k}},\eta_c] \sim   -\frac{\ell^2}{2} \int \frac{d^3 {k}}{(2\pi)^3}  \left(\frac{k}{\eta_c^2} - \frac{i}{\eta_c^3}   \right) \; \phi_{-\vec{k}} \; \phi_{\vec{k}}~.
\end{equation}
Transforming to the $\hat{\sigma}=\eta_c^{-2}\left(\hat{\phi}-\eta_c^3 \,\hat{\pi_{\phi}}\right)$ basis as before, we find:
\begin{equation}
\log \Psi_{HH} [\sigma_{\vec{k}},\eta_c] \sim -\frac{\ell^2}{2} \int \frac{d^3 {k}}{(2\pi)^3} \; \frac{1}{k}\;\sigma_{-\vec{k}}\; \sigma_{\vec{k}}~.
\end{equation}
This leads to a correlation function in position space given by:
\begin{equation}
\langle \mathcal{O}(x) \mathcal{O}(y) \rangle \sim \frac{1}{|x-y|^2}\;,
\end{equation}
which is obtained by differentiating twice the logarithm of the wavefunction. This is the appropriate answer for an operator of conformal weight $\Delta=1$, which for the $Sp(N)$ theory corresponds to $\chi \cdot \chi$. 

\section{Wavefunctionals for bulk gauge fields}\label{u1appendix}

We consider here the perturbative Hartle-Hawking wavefunctional for a bulk $U(1)$ gauge field with action:
\begin{equation}
S = - \frac{1}{4g^2} \int d^4 x \sqrt{-g}\, F_{\mu\nu} F^{\mu\nu}~, \quad F_{\mu\nu} = \partial_\mu A_{\nu} - \partial_\nu A_\mu~,
\end{equation}
in a fixed de Sitter background $ds^2 = \ell^2/\eta^2 \left(  - d\eta^2 + d\vec{x}^2 \right)$. The putative dual CFT would have a $U(1)$ global symmetry.\footnote{As an example, the non-minimal bosonic Vasiliev theory, which includes all non-negative integer spins, contains a massless bulk $U(1)$ gauge field in its spectrum. The bulk gauged $U(1)$ symmetry is dual to the global $U(1)$ flavor symmetry of the anti-commuting complex scalars in the CFT.} Working in the gauge $A_\eta = 0$, the on-shell action is simply given by a boundary piece at $\eta = \eta_c$:
\begin{equation}
S_{on-shell} =  \frac{1}{2g^2} \int d^3 x A_i \partial_\eta A_i |_{\eta=\eta_c}~,
\end{equation}
which is related by an analytic continuation $\eta = i z$ to the on-shell action in Euclidean AdS$_4$ (see for example section 5.3 of \cite{Chang:2012kt}). As a simple example, consider a solution in the $A_\eta = 0$ gauge with $k_y=k_z=0$. Then we can consistently set $A_x = A_y = 0$ and remain with the following solution satisfying the Bunch-Davies condition:
\begin{equation}
A_z = \int \frac{d k_x}{(2\pi)} \tilde{A}^{(k_x)}_z e^{- i |k_x| (\eta - \eta_c) + i k_x x}~, \quad \quad |\eta_c| \ll 1~.
\end{equation}
Reality of the profile at $\eta = \eta_c$ requires $\left( \tilde{A}^{(k_x)}_z \right)^* = \tilde{A}^{(-k_x)}_z$. The on-shell action at $\eta = \eta_c$ on the above complex solution is then:
\begin{equation}
i \; S_{on-shell} = - \frac{1}{2g^2} \int \frac{d k_x}{2\pi} \tilde{A}^{(k_x)}_z (\tilde{A}^{(k_x)}_z)^* |k_x|~.
\end{equation}
The wavefunction is Gaussian suppressed as expected and the single power of $k$ is in accordance with the conformal weight $\Delta = 2$ of the current operator dual to the bulk $U(1)$ gauge field.

One can also consider adding a $\theta$-term to the bulk action:
\begin{equation}
i \; S_\theta = i \; \frac{\theta}{16\pi} \int_{\mathcal{M}} d^4 x \sqrt{-g} \; \epsilon^{\mu\nu\rho\sigma} F_{\mu\nu} F_{\rho\sigma}~, \quad \theta \in \mathbb{R}~.
\end{equation}
This term is independent of the metric. It is a total derivative and is equivalent to a Chern-Simons term at $\eta = \eta_c$:
\begin{equation}
i \; S_{\theta} = i \; \frac{\theta}{4\pi} \int_{\partial \mathcal{M}} d^3 {x} \; \epsilon_{ijk} A_i \partial_j A_k |_{\eta=\eta_c}~.
\end{equation}
Since the profile of the gauge field is real at $\eta = \eta_c$, the above is pure imaginary.

Using the on-shell action we can construct the Bunch-Davies wavefunctional:
\begin{equation}
\Psi^{(\theta)}_{HH}[A_i,\eta_c] \sim e^{{i S_{on-shell}} + i S_{\theta}}~.
\end{equation}
Notice that the absolute value squared $|\Psi^{(\theta)}_{HH}[A_i,\eta_c] |^2$ is independent of $\theta$. Although this would seem to suggest that the $\theta$-term plays no role in the cosmological correlators obtained from $|\Psi^{(\theta)}_{HH}[A_i,\eta_c]|^2$, it can still appear in computing observables involving the conjugate momentum of the gauge field, i.e. the electric field $\vec{E} = \partial_\eta \vec{A}\;$. Indeed, in the $A_\eta = 0$ gauge, the wavefunction at zero $\vec{E}$-field is given by:
\begin{equation}
\tilde{\Psi}^{(\theta)}_{HH}[\vec{E}=0,\eta_c] = \int \mathcal{D} A_i \; \Psi^{(\theta)}_{HH}[A_i,\eta_c]~.
\end{equation}
Notice that we are now performing a path integral over a functional of $A_i$ which includes the Chern-Simons term. Thus, the gauge field becomes dynamical \cite{Witten:2003ya}. 

Much of our above discussion follows mostly unchanged when the $U(1)$ gauge field is replaced with a non-Abelian gauge field. It would be interesting to understand how the topological dependence of the Chern-Simons term manifests itself in terms of cosmological expectation values.


\begin{thebibliography}{99}

\bibitem{Hartle:1983ai} 
  J.~B.~Hartle and S.~W.~Hawking,
  ``Wave Function of the Universe,''
  Phys.\ Rev.\ D {\bf 28}, 2960 (1983).

\bibitem{Bunch:1978yq} 
  T.~S.~Bunch and P.~C.~W.~Davies,
  ``Quantum Field Theory in de Sitter Space: Renormalization by Point Splitting,''
  Proc.\ Roy.\ Soc.\ Lond.\ A {\bf 360}, 117 (1978).

\bibitem{Mottola:1984ar} 
  E.~Mottola,
  ``Particle Creation in de Sitter Space,''
  Phys.\ Rev.\ D {\bf 31}, 754 (1985).

\bibitem{Allen:1985ux} 
  B.~Allen,
  ``Vacuum States in de Sitter Space,''
  Phys.\ Rev.\ D {\bf 32}, 3136 (1985).

\bibitem{Chernikov:1968zm} 
  N.~A.~Chernikov and E.~A.~Tagirov,
  ``Quantum theory of scalar fields in de Sitter space-time,''
  Annales Poincare Phys.\ Theor.\ A {\bf 9}, 109 (1968).

\bibitem{Schomblond:1976xc} 
  C.~Schomblond and P.~Spindel,
  ``Unicity Conditions of the Scalar Field Propagator Delta(1) (x,y) in de Sitter Universe,''
  Annales Poincare Phys.\ Theor.\  {\bf 25}, 67 (1976).

\bibitem{Sasaki:1994yt} 
  M.~Sasaki, T.~Tanaka and K.~Yamamoto,
  ``Euclidean vacuum mode functions for a scalar field on open de Sitter space,''
  Phys.\ Rev.\ D {\bf 51}, 2979 (1995)
  [gr-qc/9412025].

\bibitem{Dowker:1975tf} 
  J.~S.~Dowker and R.~Critchley,
  ``Effective Lagrangian and Energy Momentum Tensor in de Sitter Space,''
  Phys.\ Rev.\ D {\bf 13}, 3224 (1976).

\bibitem{Candelas:1975du} 
  P.~Candelas and D.~J.~Raine,
  ``General Relativistic Quantum Field Theory-An Exactly Soluble Model,''
  Phys.\ Rev.\ D {\bf 12}, 965 (1975).

\bibitem{Boerner:1969ff} 
  G.~Boerner and H.~P.~Duerr,
  ``Classical and quantum fields in de sitter space,''
  Nuovo Cim.\ A {\bf 64}, 669 (1969).

\bibitem{Maldacena:2002vr}
  J.~M.~Maldacena,
  ``Non-Gaussian features of primordial fluctuations in single field
  inflationary models,''
  JHEP {\bf 0305}, 013 (2003)
  [arXiv:astro-ph/0210603].

\bibitem{Strominger:2001pn}
  A.~Strominger,
  ``The dS / CFT correspondence,''
  JHEP {\bf 0110}, 034 (2001)
  [arXiv:hep-th/0106113].
  
\bibitem{Witten:2001kn}
  E.~Witten,
  ``Quantum gravity in de Sitter space,''
  arXiv:hep-th/0106109.

\bibitem{Anninos:2012qw} 
  D.~Anninos,
  ``De Sitter Musings,''
  Int.\ J.\ Mod.\ Phys.\ A {\bf 27}, 1230013 (2012)
  [arXiv:1205.3855 [hep-th]].
  
\bibitem{Alishahiha:2004md} 
  M.~Alishahiha, A.~Karch, E.~Silverstein and D.~Tong,
  ``The dS/dS correspondence,''
  AIP Conf.\ Proc.\  {\bf 743}, 393 (2005)
  [hep-th/0407125].
  
\bibitem{Dong:2010pm} 
  X.~Dong, B.~Horn, E.~Silverstein and G.~Torroba,
  ``Micromanaging de Sitter holography,''
  Class.\ Quant.\ Grav.\  {\bf 27}, 245020 (2010)
  [arXiv:1005.5403 [hep-th]].
  
\bibitem{Dong:2011uf} 
  X.~Dong, B.~Horn, S.~Matsuura, E.~Silverstein and G.~Torroba,
  ``FRW solutions and holography from uplifted AdS/CFT,''
  Phys.\ Rev.\ D {\bf 85}, 104035 (2012)
  [arXiv:1108.5732 [hep-th]].
  
\bibitem{McFadden:2009fg} 
  P.~McFadden and K.~Skenderis,
  ``Holography for Cosmology,''
  Phys.\ Rev.\ D {\bf 81}, 021301 (2010)
  [arXiv:0907.5542 [hep-th]].
  
\bibitem{Banks:2011qf} 
  T.~Banks and W.~Fischler,
  ``Holographic Theories of Inflation and Fluctuations,''
  arXiv:1111.4948 [hep-th].
  
\bibitem{Banks:2008ep} 
  T.~Banks,
  ``Holographic Space-time from the Big Bang to the de Sitter era,''
  J.\ Phys.\ A A {\bf 42}, 304002 (2009)
  [arXiv:0809.3951 [hep-th]].
  
\bibitem{Freivogel:2006xu} 
  B.~Freivogel, Y.~Sekino, L.~Susskind and C.~-P.~Yeh,
  ``A Holographic framework for eternal inflation,''
  Phys.\ Rev.\ D {\bf 74}, 086003 (2006)
  [hep-th/0606204].
  
\bibitem{Anninos:2011kh} 
  D.~Anninos and F.~Denef,
  ``Cosmic Clustering,''
  arXiv:1111.6061 [hep-th].
  
\bibitem{Roberts:2012jw} 
  D.~A.~Roberts and D.~Stanford,
  ``On memory in exponentially expanding spaces,''
  arXiv:1210.5238 [hep-th].
  
\bibitem{Harlow:2012dd} 
  D.~Harlow, S.~H.~Shenker, D.~Stanford and L.~Susskind,
  ``The Three Faces of a Fixed Point,''
  arXiv:1203.5802 [hep-th].
  
\bibitem{Garriga:2008ks} 
  J.~Garriga and A.~Vilenkin,
  ``Holographic Multiverse,''
  JCAP {\bf 0901}, 021 (2009)
  [arXiv:0809.4257 [hep-th]].
  
\bibitem{Parikh:2004wh} 
  M.~K.~Parikh and E.~P.~Verlinde,
  ``De Sitter holography with a finite number of states,''
  JHEP {\bf 0501}, 054 (2005)
  [hep-th/0410227].
  
\bibitem{Anninos:2009yc} 
  D.~Anninos and T.~Hartman,
  ``Holography at an Extremal De Sitter Horizon,''
  JHEP {\bf 1003}, 096 (2010)
  [arXiv:0910.4587 [hep-th]].
  
\bibitem{Anninos:2010gh} 
  D.~Anninos and T.~Anous,
  ``A de Sitter Hoedown,''
  JHEP {\bf 1008}, 131 (2010)
  [arXiv:1002.1717 [hep-th]].
  
\bibitem{Anninos:2011zn} 
  D.~Anninos, T.~Anous, I.~Bredberg and G.~S.~Ng,
  ``Incompressible Fluids of the de Sitter Horizon and Beyond,''
  JHEP {\bf 1205}, 107 (2012)
  [arXiv:1110.3792 [hep-th]].
  
\bibitem{Anninos:2011af} 
  D.~Anninos, S.~A.~Hartnoll and D.~M.~Hofman,
  ``Static Patch Solipsism: Conformal Symmetry of the de Sitter Worldline,''
  Class.\ Quant.\ Grav.\  {\bf 29}, 075002 (2012)
  [arXiv:1109.4942 [hep-th]].
  
\bibitem{Hertog:2011ky} 
  T.~Hertog and J.~Hartle,
  ``Holographic No-Boundary Measure,''
  JHEP {\bf 1205}, 095 (2012)
  [arXiv:1111.6090 [hep-th]].
  
\bibitem{Bzowski:2012ih} 
  A.~Bzowski, P.~McFadden and K.~Skenderis,
  ``Holography for inflation using conformal perturbation theory,''
  arXiv:1211.4550 [hep-th].

\bibitem{Vasiliev:1990en}
  M.~A.~Vasiliev,
  ``Consistent equation for interacting gauge fields of all spins in (3+1)-dimensions,''
  Phys.\ Lett.\  {\bf B243}, 378-382 (1990).  
  
\bibitem{Vasiliev:1999ba}
  M.~A.~Vasiliev,
  ``Higher spin gauge theories: Star product and AdS space,''
  arXiv:hep-th/9910096.

\bibitem{Iazeolla:2007wt}
  C.~Iazeolla, E.~Sezgin, P.~Sundell,
  ``Real forms of complex higher spin field equations and new exact solutions,''
  Nucl.\ Phys.\  {\bf B791}, 231-264 (2008).
  [arXiv:0706.2983 [hep-th]].    
 
\bibitem{Vasiliev:1986td}
  M.~A.~Vasiliev,
 ``Free Massless Fields Of Arbitrary Spin In The De Sitter Space And Initial Data For A Higher Spin Superalgebra,''
  Fortsch.\ Phys.\  {\bf 35}, 741-770 (1987).

\bibitem{Vasiliev:1992av} 
  M.~A.~Vasiliev,
  ``More on equations of motion for interacting massless fields of all spins in (3+1)-dimensions,''
  Phys.\ Lett.\ B {\bf 285}, 225 (1992).
  
\bibitem{Vasiliev:1995dn} 
  M.~A.~Vasiliev,
  ``Higher spin gauge theories in four-dimensions, three-dimensions, and two-dimensions,''
  Int.\ J.\ Mod.\ Phys.\ D {\bf 5}, 763 (1996)
  [hep-th/9611024].

\bibitem{arXiv:1108.5735} 
  D.~Anninos, T.~Hartman and A.~Strominger,
  ``Higher Spin Realization of the dS/CFT Correspondence,''
  arXiv:1108.5735 [hep-th].
  
\bibitem{Das:2012dt} 
  D.~Das, S.~R.~Das, A.~Jevicki and Q.~Ye,
  ``Bi-local Construction of Sp(2N)/dS Higher Spin Correspondence,''
  JHEP {\bf 1301}, 107 (2013)
  [arXiv:1205.5776 [hep-th]].
  
\bibitem{Ng:2012xp} 
  G.~S.~Ng and A.~Strominger,
  ``State/Operator Correspondence in Higher-Spin dS/CFT,''
  arXiv:1204.1057 [hep-th].


\bibitem{LeClair:2007iy}
  A.~LeClair, M.~Neubert,
  ``Semi-Lorentz invariance, unitarity, and critical exponents of symplectic fermion models,''
  JHEP {\bf 0710}, 027 (2007).
  [arXiv:0705.4657 [hep-th]].


\bibitem{Anninos:2012ft} 
  D.~Anninos, F.~Denef and D.~Harlow,
  ``The Wave Function of Vasiliev's Universe - A Few Slices Thereof,''
  arXiv:1207.5517 [hep-th].

\bibitem{Sezgin:2005hf} 
  E.~Sezgin and P.~Sundell,
  ``On an exact cosmological solution of higher spin gauge theory,''
  hep-th/0511296.

\bibitem{Dunne:2006ct} 
  G.~V.~Dunne and K.~Kirsten,
  ``Functional determinants for radial operators,''
  J.\ Phys.\ A {\bf 39}, 11915 (2006)
  [hep-th/0607066].
  
 \bibitem{Klebanov:2002ja}
 I.R.~Klebanov and A.M.~Polyakov,
``AdS dual of the critical O(N) vector model,"
Phys.\ Lett.\  B {\bf 550} (2002) 213
{\tt [arXiv:hep-th/0210114]}.

%

\bibitem{Sezgin:2002rt}
E.~Sezgin and P.~Sundell,
``Massless higher spins and holography,"
Nucl.\ Phys.\  B {\bf 644} (2002) 303
[Erratum-ibid.\  B {\bf 660} (2003) 403]
{\tt [arXiv:hep-th/0205131]}.

\bibitem{hep-th/0103247} 
  B.~Sundborg,
  ``Stringy gravity, interacting tensionless strings and massless higher spins,''
  Nucl.\ Phys.\ Proc.\ Suppl.\ \ {\bf 102}, 113  (2001)
  [hep-th/0103247].
  
\bibitem{arXiv:0912.3462} 
  S.~Giombi and X.~Yin,
  ``Higher Spin Gauge Theory and Holography: The Three-Point Functions,''
  JHEP\ {\bf 1009}, 115  (2010)
  [arXiv:0912.3462 [hep-th]].
  
  
\bibitem{Starobinsky:1982mr} 
  A.~A.~Starobinsky,
  ``Isotropization of arbitrary cosmological expansion given an effective cosmological constant,''
  JETP Lett.\  {\bf 37}, 66 (1983).

\bibitem{fg} 
  C. Fefferman and C. R. Graham, 
  ``Conformal Invariants," 
  Elie Cartan et les Mathematiques dÃ•Aujoudhui (Asterisque, 1985) 95.

\bibitem{Anninos:2010zf} 
  D.~Anninos, G.~S.~Ng and A.~Strominger,
  ``Asymptotic Symmetries and Charges in De Sitter Space,''
  Class.\ Quant.\ Grav.\  {\bf 28}, 175019 (2011)
  [arXiv:1009.4730 [gr-qc]].
  
\bibitem{Gelfand:1959nq} 
  I.~M.~Gelfand and A.~M.~Yaglom,
  ``Integration in functional spaces and it applications in quantum physics,''
  J.\ Math.\ Phys.\  {\bf 1}, 48 (1960).
  
\bibitem{Bordag:1996fw} 
  M.~Bordag, K.~Kirsten and J.~S.~Dowker,
  ``Heat kernels and functional determinants on the generalized cone,''
  Commun.\ Math.\ Phys.\  {\bf 182}, 371 (1996)
  [hep-th/9602089].
  
\bibitem{Bousso:1998bn} 
  R.~Bousso,
  ``Proliferation of de Sitter space,''
  Phys.\ Rev.\ D {\bf 58}, 083511 (1998)
  [hep-th/9805081].
  
\bibitem{Hawking:1983hj} 
  S.~W.~Hawking,
  ``The Quantum State of the Universe,''
  Nucl.\ Phys.\ B {\bf 239}, 257 (1984).

\bibitem{Castro:2012gc} 
  A.~Castro and A.~Maloney,
  ``The Wave Function of Quantum de Sitter,''
  JHEP {\bf 1211}, 096 (2012)
  [arXiv:1209.5757 [hep-th]].
  
  
\bibitem{Wilson:1972cf} 
  K.~G.~Wilson,
  ``Quantum field theory models in less than four-dimensions,''
  Phys.\ Rev.\ D {\bf 7}, 2911 (1973).

\bibitem{Witten:2001ua} 
  E.~Witten,
  ``Multitrace operators, boundary conditions, and AdS / CFT correspondence,''
  hep-th/0112258.

\bibitem{Mueck:2002gm} 
  W.~Mueck,
  ``An Improved correspondence formula for AdS / CFT with multitrace operators,''
  Phys.\ Lett.\ B {\bf 531}, 301 (2002)
  [hep-th/0201100].

\bibitem{Gubser:2002zh} 
  S.~S.~Gubser and I.~Mitra,
  ``Double trace operators and one loop vacuum energy in AdS / CFT,''
  Phys.\ Rev.\ D {\bf 67}, 064018 (2003)
  [hep-th/0210093].

\bibitem{Gubser:2002vv} 
  S.~S.~Gubser and I.~R.~Klebanov,
  ``A Universal result on central charges in the presence of double trace deformations,''
  Nucl.\ Phys.\ B {\bf 656}, 23 (2003)
  [hep-th/0212138].


\bibitem{Harlow:2011ke}
  D.~Harlow, D.~Stanford,
  ``Operator Dictionaries and Wave Functions in AdS/CFT and dS/CFT,''
  [arXiv:1104.2621 [hep-th]]. 

\bibitem{Anninos:2011jp} 
  D.~Anninos, G.~S.~Ng and A.~Strominger,
  ``Future Boundary Conditions in De Sitter Space,''
  JHEP {\bf 1202}, 032 (2012)
  [arXiv:1106.1175 [hep-th]].
  
\bibitem{Hartle:2008ng} 
  J.~B.~Hartle, S.~W.~Hawking and T.~Hertog,
  ``The Classical Universes of the No-Boundary Quantum State,''
  Phys.\ Rev.\ D {\bf 77}, 123537 (2008)
  [arXiv:0803.1663 [hep-th]].
  
\bibitem{Hartle:2007gi} 
  J.~B.~Hartle, S.~W.~Hawking and T.~Hertog,
  ``No-Boundary Measure of the Universe,''
  Phys.\ Rev.\ Lett.\  {\bf 100}, 201301 (2008)
  [arXiv:0711.4630 [hep-th]].

\bibitem{Sarangi:2005cs} 
  S.~Sarangi and S.~-H.~H.~Tye,
  ``The Boundedness of Euclidean gravity and the wavefunction of the universe,''
  hep-th/0505104.
  

  








%
%

  

  
















\bibitem{Aharony:2011jz} 
  O.~Aharony, G.~Gur-Ari and R.~Yacoby,
  ``d=3 Bosonic Vector Models Coupled to Chern-Simons Gauge Theories,''
  JHEP {\bf 1203}, 037 (2012)
  [arXiv:1110.4382 [hep-th]].

\bibitem{Chang:2012kt} 
  C.~-M.~Chang, S.~Minwalla, T.~Sharma and X.~Yin,
  ``ABJ Triality: from Higher Spin Fields to Strings,''
  arXiv:1207.4485 [hep-th].

\bibitem{Giombi:2011kc} 
  S.~Giombi, S.~Minwalla, S.~Prakash, S.~P.~Trivedi, S.~R.~Wadia and X.~Yin,
  ``Chern-Simons Theory with Vector Fermion Matter,''
  Eur.\ Phys.\ J.\ C {\bf 72}, 2112 (2012)
  [arXiv:1110.4386 [hep-th]].

\bibitem{Giombi:2012ms} 
  S.~Giombi and X.~Yin,
  ``The Higher Spin/Vector Model Duality,''
  arXiv:1208.4036 [hep-th].

\bibitem{Witten:2003ya} 
  E.~Witten,
  ``SL(2,Z) action on three-dimensional conformal field theories with Abelian symmetry,''
  In *Shifman, M. (ed.) et al.: From fields to strings, vol. 2* 1173-1200
  [hep-th/0307041].
  

  


  
\bibitem{Aharony:2008gk} 
  O.~Aharony, O.~Bergman and D.~L.~Jafferis,
  ``Fractional M2-branes,''
  JHEP {\bf 0811}, 043 (2008)
  [arXiv:0807.4924 [hep-th]].
  
\bibitem{Maldacena:2011jn} 
  J.~Maldacena and A.~Zhiboedov,
  ``Constraining Conformal Field Theories with A Higher Spin Symmetry,''
  arXiv:1112.1016 [hep-th].
  
\bibitem{Maldacena:2012sf} 
  J.~Maldacena and A.~Zhiboedov,
  ``Constraining conformal field theories with a slightly broken higher spin symmetry,''
  arXiv:1204.3882 [hep-th].

\bibitem{workinprogress}
 work in progress





\bibitem{Girardello:2002pp} 
  L.~Girardello, M.~Porrati and A.~Zaffaroni,
  ``3-D interacting CFTs and generalized Higgs phenomenon in higher spin theories on AdS,''
  Phys.\ Lett.\ B {\bf 561}, 289 (2003)
  [hep-th/0212181].

\bibitem{Banerjee:2012gh} 
  S.~Banerjee, S.~Hellerman, J.~Maltz and S.~H.~Shenker,
  ``Light States in Chern-Simons Theory Coupled to Fundamental Matter,''
  arXiv:1207.4195 [hep-th].
  
\bibitem{Radicevic:2012in} 
  D.~Radicevic,
  ``Singlet Vector Models on Lens Spaces,''
  arXiv:1210.0255 [hep-th].

\bibitem{Witten:2010cx} 
  E.~Witten,
  ``Analytic Continuation Of Chern-Simons Theory,''
  arXiv:1001.2933 [hep-th].



  
\bibitem{Banks:2005bm} 
  T.~Banks,
  ``Some thoughts on the quantum theory of stable de Sitter space,''
  hep-th/0503066.
  
\bibitem{Andersson:2002nr} 
  L.~Andersson and G.~J.~Galloway,
  ``DS / CFT and space-time topology,''
  Adv.\ Theor.\ Math.\ Phys.\  {\bf 6}, 307 (2003)
  [hep-th/0202161].

\bibitem{foam}
 Shamik Banerjee, Alexandre Belin, Simeon Hellerman, Arnaud Lepage-Jutier, Alexander Maloney, Dorde Radicevic, Stephen Shenker,
 ``Topology of future infinity in dS/CFT,"
  to appear.

\end{thebibliography}
\end{document}